\documentclass[a4paper,11pt]{article}
\pdfoutput=1 % if the images are in pdf, png or jpg format
\usepackage{jinstpub}
\usepackage{textcomp,gensymb} % symbols
\usepackage{multirow}
\hypersetup{breaklinks,colorlinks,allcolors=blue}
\graphicspath{{./Figures/}}

\title{\boldmath A detector system for \lq{}absolute\rq{} measurements of fission 
cross sections at n\_TOF in the energy range below 200~MeV}

\author[a,1]{E.~Pirovano,\note{Corresponding author.}}
\author[b,c]{A.~Manna}
\author[d]  {O.~Aberle}
\author[e]  {S.~Amaducci}
\author[f]  {N.~Colonna}
\author[g]  {P.~Console Camprini}
\author[e]  {L.~Cosentino}
\author[a]  {M.~Dietz}
\author[a]  {Q.~Ducasse}
\author[e]  {P.~Finocchiaro}
\author[b,c]{C.~Massimi}
\author[g]  {A.~Mengoni}
\author[a]  {R.~Nolte}
\author[a]  {D.~Radeck}
\author[h]  {L.~Tassan-Got}
\author[i]  {N.~Terranova}
\author[c]  {G.~Vannini}
\affiliation[a]{Physikalisch-Technische Bundesanstalt (PTB), Braunschweig, Germany}
\affiliation[b]{Dipartimento di Fisica e Astronomia, Universit\`{a} di Bologna, Italy}
\affiliation[c]{Istituto Nazionale di Fisica Nucleare (INFN), Sezione di Bologna, Italy}
\affiliation[d]{European Laboratory for Particle Physics (CERN), Geneva, Switzerland}
\affiliation[e]{INFN Laboratori Nazionali del Sud, Catania, Italy}
\affiliation[f]{Istituto Nazionale di Fisica Nucleare (INFN), Sezione di Bari, Italy}
\affiliation[g]{Agenzia nazionale per le nuove tecnologie (ENEA), Bologna, Italy}
\affiliation[h]{Institut de Physique Nucl\'{e}aire (IPN), Orsay, France}
\affiliation[i]{Agenzia nazionale per le nuove tecnologie (ENEA), Frascati, Italy}

\collaboration[c]{on behalf of the n\_TOF collaboration}

\emailAdd{elisa.pirovano@ptb.de}

\abstract{A new measurement of the \textsuperscript{235}U(n,f) cross section was
performed at the neutron time-of-flight facility n\_TOF at CERN. The experiment
focused on neutron energies from 20~MeV to several hundred MeV, and was 
normalized to neutron scattering on hydrogen.
This is a measurement first of its kind at this facility, in an energy range 
that was until now not often explored, so the detector development phase was 
crucial for its success. 
Two detectors are presented, a parallel plate fission chamber (PPFC)
and a recoil proton telescope (RPT), both dedicated  to perform measurements 
in the incident neutron energy range from 30~MeV to 200~MeV. 
The experiment was designed to minimize statistical uncertainties in the 
allocated run time. Several efforts were made to ensure that the systematic
effects were understood and under control. The results show that the detectors
are suited for measurements at n\_TOF above 30~MeV, and indicate the path for
possible future lines of development.}


\begin{document}
\maketitle
\flushbottom

\section{Introduction}
\label{sec:Intro}

   The cross section for neutron-induced fission at neutron energies above 
20~MeV is of interest for basic nuclear physics \cite{LOM15} as well as for 
nuclear technology, especially for improving the predictive capability of 
neutron transport codes \cite{DAV15}. The emission of two kinematically 
correlated fission fragments with kinetic energies around 80~MeV is a very 
distinctive signature which makes fission an ideal reference reaction for the 
measurements of cross sections for other neutron-induced reactions \cite{COL20}.
For this reason, the cross sections for neutron-induced fission of 
\textsuperscript{235}U and \textsuperscript{238}U belong to the set of standard 
cross sections \cite{CAR18}. 

   While most of the other reference cross sections are standards for selected 
neutron energy ranges below 20~MeV, the cross sections for 
\textsuperscript{235,238}U(n,f) are standards up to 200~MeV. 
The only other standard in this energy range is the neutron-proton (n-p) 
scattering cross section. The evaluated uncertainty of the integral and 
differential cross sections for n-p scattering is smaller than those of all 
other cross section standards \cite{CAR18}. Therefore, measurements of the 
fission cross section standards relative to the neutron-proton scattering cross
section are of particular importance for reducing the uncertainties of neutron
measurements, and there is a constant request for more data from such experiments
\cite{MAR15}.
   Despite its importance for neutron measurements, there is only one 
experimental data set for the \textsuperscript{235}U(n,f) cross section at 
energies above 20~MeV which was determined relative to n-p scattering and has a 
continuous energy coverage \cite{LIS91}. These data were measured in the 1980s 
at the Weapons Neutron Research (WNR) spallation neutron source and determine 
the evaluated cross section in this energy range until today \cite{CAR18}. 

   To cross-check these results and reduce the uncertainty on the evaluation, 
while also covering the largest possible neutron energy interval, and also 
to minimize the systematic uncertainties, two 
detection setups were designed to perform a new measurement of the 
\textsuperscript{235}U(n,f) cross section at the n\_TOF neutron source of 
CERN \cite{BOR03}. Except for measurements of fission cross section ratios 
\cite{TAR14}, the energy range above 20~MeV was not explored very much at 
n\_TOF so far. Therefore, the present measurement was the first of this kind at 
this neutron source.
One setup was developed aiming at improving the existing standard, 
with focus on the neutron energy range below 200~MeV. The other was designed
to cover the largest possible neutron energy range, with the goal 
of extending the measurement to energies higher than 200~MeV.
In this paper, they are also referred as the \lq low-energy\rq\ 
and the \lq high-energy\rq\ experiment. 

   The two experiments were carried out simultaneously, using the same neutron 
beam. In both setups, recoil proton telescopes (RPTs) of different 
designs were employed to measure the incident neutron energy distribution 
relative to the differential n-p scattering cross section. The fission detector 
for the low-energy experiment was a parallel-plate fission chamber (PPFC), while
parallel-plate avalanche counters (PPAC) were used for the high-energy 
experiment. Such a measurement is usually termed \lq absolute\rq, although it is
actually a measurement relative to the  n-p scattering cross section. Above thermal 
neutron energies and below the pion production threshold, however, elastic 
neutron-proton scattering is the only open reaction channel in the 
neutron-proton system. In principle, the differential scattering cross section 
can be determined from a relative measurement of the angular distribution of 
recoil protons and a determination of the total cross section, which only 
requires a relative measurement as well. Although experimental details are a 
bit more complicated, this may justify calling the present experiment an 
absolute measurement.

In the energy range below 20~MeV, two measurements of fission cross sections 
\cite{MAT17,BEL22} were carried recently using monoenergetic neutrons and 
primary reference instruments for neutron fluence measurements. These 
instruments are compared to each other in regular key comparisons \cite{GRE14}.
Hence, their uncertainty budget is well known and small remaining uncertainties 
can be achieved. A similar achievement is expected from the planned experiment 
by using two simultaneous but independent measurements to check for consistency
and identify sources of systematic uncertainties.

   The present publication reports about the technical design of the low-energy 
experiment and the properties of the detectors used. The decision to optimize 
the setup for the energy region from 30~MeV to 200~MeV was motivated by several 
reasons. Nuclear model codes such as GNASH, EMPIRE and TALYS used for producing 
the data libraries for neutron transport codes extend up to this energy while 
the intranuclear cascade (INC) model becomes applicable above this energy. From 
the practical point of view, the neutron energy range of interest in radiation 
protection and technological applications such as the design of 
accelerator-driven neutron sources extends to about 200~MeV because neutron 
energy distributions produced by bombardment of stopping-length targets with 
light ion beams exhibit a broad maximum around 100~MeV in addition to the 
evaporation peak in the MeV region.

The data shown in this paper were mostly acquired during detector tests. Not all
configurations investigated in these runs could finally be used for the cross 
section measurement. The design of the high-energy experiment is reported in a 
second paper \cite{SIS} and the results of the measurements will be reported in forthcoming publications.

%-------------------------------------------------------------------------------

\section{Experimental setup}
\label{sec:Experimental-setup}

The measurements were carried out at n\_TOF, the neutron time-of-flight facility 
of CERN. The pulsed white neutron beam of n\_TOF is produced via spallation
reactions by protons impinging on a massive lead target with a momentum of 20~GeV/$c$. 
The protons are extracted from the Proton Synchrotron (PS) accelerator 
ring at a repetition rate of less than 1~Hz. They are grouped in 7~ns wide 
bunches formed by about $7\cdot10^{12}$ (the so-called \lq dedicated\rq\ pulses) 
or $3\cdot10^{12}$ (\lq parasitic\rq\ pulses) particles. 
The interaction of the PS proton beam with the n\_TOF target produces about
300 neutrons per proton which then travel along evacuated tubes to reach the 
experimental areas where the detection setups are installed.
A detailed description of the facility can be found for example in 
\cite{BOR03} while the characteristics of the lead target and neutron beam
are described in \cite{BAR13, GUE13}. 

\begin{figure}[htb]
\centering
\includegraphics[width=10cm]{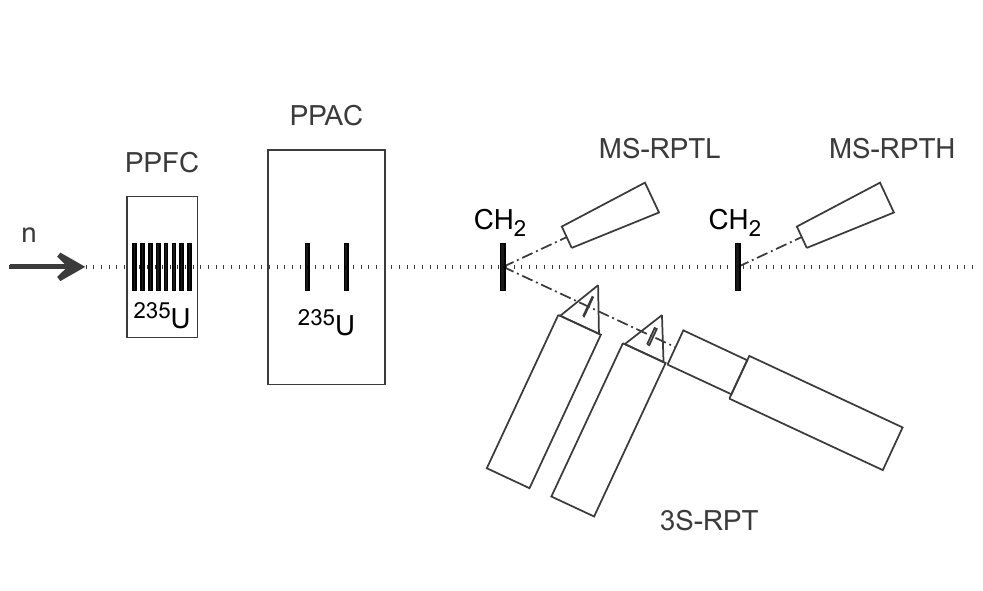}
\caption{\label{fig:Setup} Schematic layout of the two experiments. 
The low-energy experiment (incident neutron energy below 200~MeV) used the 
recoil telescope 3S-RPT (\lq triple-stage\rq\ RPT) and the PPFC. 
The PPAC and the two telescopes MS-RPTL and MS-RPTH (\lq multiple-stage\rq\ 
RPTs) belong to the high-energy experiment. 
The RPT of the low-energy experiment and the first telescope of the 
high-energy experiment used the same polyethylene and graphite
targets.}
\end{figure}

The detectors for the \textsuperscript{235}U(n,f) cross section measurement
were installed in the Experimental Area 1 (EAR1) at a distance of 183~m from 
the spallation target, where the neutron energy distribution extends 
from thermal energies up to several hundred MeV.
The setup included ten \textsuperscript{235}U samples, mounted in two 
reaction chambers, and two polyethylene (PE) samples.
The fission detectors were a parallel-plate fission chamber (PPFC) developed
by PTB (Physikalisch-Technische Bundesanstalt) and a reaction
chamber based on parallel plate avalanche counters (PPACs) developed by IPN 
(Institut de Physique Nucl\'{e}aire d'Orsay). 
The light charged particles emitted from the PE samples were counted using three 
Recoil Proton Telescopes (RPTs), two designed by INFN (Istituto Nazionale di 
Fisica Nucleare) and one by PTB. 
The layout is also shown schematically in figure~\ref{fig:Setup}. 

The \lq low-energy experiment\rq\ consisted of the PPFC and the 3S-RPT, a 
triple-stage telescope composed of two transmission detectors and one stop 
detector. 
The reasons behind the choice of these detectors and their design are introduced 
in section~\ref{sec:Concept}. The technical details of the fission chamber and 
characteristics of the uranium samples are reported in section~\ref{sec:FC}, 
while those of the polyethylene samples and of the RPT are in 
section~\ref{sec:RPT}. 

The fission chamber was installed at about 183~m from the spallation target,
the first polyethylene sample was placed about 1~m downstream and the second
sample at an additional distance of 50~cm .
The triple-stage telescope was mounted at an angle of 25.2(1)\degree\ with respect to 
the direction of the beam. The distance from the center of the radiator and the 
center of the transmission detectors, and the front face of stop detector, were 
206.0(15)~mm, 355.0(15)~mm, and 410(2)~mm, respectively. 

The distances between PE samples and detectors were measured with sub-millimeter 
precision with laser trackers and reflectors positioned at strategic positions,
as for example the center of the sample support and the center of the 
transmission detectors, or the front face of the stop detector, determining at 
the same time also the angles between beam and telescope axes. The position 
and direction of the neutron beam was known from previous surveys and 
measurements with Medipix detectors. 
The survey was carried out once, at the beginning of the measurement campaign,
as it is very time consuming. Every time the setup was modified, e.g. when 
one of the RPT detectors was replaced, the measurements were controlled again 
using a rigid ruler with 0.5-mm precision. The frame supporting samples and 
telescopes was never moved, and whenever possible, the RPT detectors were 
replaced without removing the PMT from the frame. This resulted in a spread in 
the repeated measurements that was no more than 2~mm, which was used also to 
determine the uncertainties on distances and angle.

The alignment of the chambers and the PE samples was checked with
Gafchromic films, and it was found that both were slightly off-center with 
respect to the beam. In the reference system with origin at the 
PPFC center, with the neutron beam pointing perpendicularly towards it, the beam 
center was positioned at (-2.5(5)~mm, -2.2(5)~mm) in the horizontal/vertical 
plane. As for PE sample, using the same reference system but with origin on 
the sample center, the beam spot was at (-4.4(5)~mm, 16.9(5)~mm).
Considering the homogeneity of the PE samples and the large transversal size 
compared to the beam diameter (10~cm by 10~cm compared to a FWHM of less than 
2~cm), the main effect of the misalignment for the RPTs was in the calculation 
of the angle relative to the beam axis, which was anyway determined 
independently during the laser tracking-system survey.
For the fission chamber, this mainly affected the determination of the 
mass of uranium effectively irradiated with the neutron beam, as discussed in
section~\ref{subsec:FC-Targets}.

The beam profile and its energy dependence were measured with the PPAC. 
The details are presented in \cite{SIS}, but to summarize briefly, 
the beam profile followed a bell-shaped \lq flat top\rq\ function, with the 
flat top becoming larger with increasing energy, and a FWHM ranging from 
1.6~cm to 1.7~cm.

The data acquisition system (DAQ) at n\_TOF is based on SP~Devices 
ADQ412-DC and ADQ14-DC-4C digitizers, with 12-bit and 14-bit resolution and 
variable sampling rates, set to 1~GS/s for this measurement. 
The DAQ is synchronized with the operation of PS: the n\_TOF proton 
extraction signal is distributed to the cards as trigger and the detector 
signals are recorded over a time window that for EAR1 is about 100~ms long,
which covers events from the detection of the gamma flash to that
of thermal neutrons \cite{MAS17}.
The digital waveforms are processed using the Pulse Shape Analysis (PSA) 
routine described in \cite{ZUG16}. The PSA includes algorithms e.g. for pulse 
recognition, baseline subtraction, and determination of the arrival time
of gamma flash and neutron-induced events, and produces files in which 
quantities such as timestamp, pulse height, or pulse integral are listed for 
each detected signal.

%-------------------------------------------------------------------------------

\section{Conceptual considerations}
\label{sec:Concept}

   The conceptual design for the low-energy experiment was guided by the goal of
achieving sufficiently small statistical uncertainties with the granted number 
of $4 \cdot 10^{18}$ protons on the n\_TOF spallation target, which in EAR1
translates to $5.3 \cdot 10^{10}$ neutrons with energies between 20~MeV and 200~MeV
at the target position, over the beam's cross-sectional area.  
With respect to the statistical uncertainty, the fission arm of the experiment 
sets the main constraint because the efficiency of any fission detector is given
by the number of fissile nuclei intercepted by the neutron beam. Usually, this 
number is limited by the availability of samples of sufficient quality. 

   For the low-energy experiment, a parallel-plate fission ionization chamber 
was selected as the detector for fission events. The main advantage of this 
detector is simplicity. Its efficiency for the detection of fission fragments is
close to unity and can be calculated easily if the composition, thickness and 
homogeneity of the fissile layer are well characterized. The eight available 
samples had an average \textsuperscript{235}U mass per unit area 
$\bar{m}_{\rm U}$ of about 300~\micro g/cm\textsuperscript{2}. 
For neutron beams smaller than the sample diameter, this detector has a neutron
detection efficiency
\begin{equation}
\label{eq:FC-Efficiency}
\epsilon_{\rm n} = \frac{N_{\rm f}}{ N_{\rm n}} 
   = \epsilon_{\rm f} \, 
     \biggl( \frac{N_{\rm A}}{\mu_{\rm U}} \biggr) \, 
     n_{\rm U} \, \bar{ m}_{\rm U} \sigma_{\rm f}  
\end{equation}
ranging between $7.4 \cdot 10^{-6}$ and $5.1 \cdot 10^{-6}$ for neutron energies
between 20~MeV and~200 MeV. Here, $N_{\rm n}$ denotes the number of neutrons in
a given small energy interval, $\epsilon_{\rm f} \approx 1$ the zero-bias 
fission fragment detection efficiency and $N_{\rm f}$ the number of detected 
fission events, respectively. 
The number of fissile layers is denoted by $n_{\rm U}$, the atomic mass of 
\textsuperscript{235}U by $\mu_{\rm U}$, and $N_{\rm A}$ is the Avogadro number.

   The neutron detection efficiency 
\begin{equation}
\label{eq:RPT-Efficiency}
\epsilon_{\rm n} = \frac{N_{\rm p}}{ N_{\rm n}} 
   = x_{\rm H} \, 
     \biggl( \frac{N_{\rm A}}{\mu_{\rm PE}} \biggr) \, 
     m_{\rm PE} \, 
     \biggl( \frac{{\rm d} \sigma_{\rm np}}{{\rm d}\Omega_{\rm p}} \biggr)(\Theta_{\rm p}) \, 
     \Delta\Omega_{\rm p} 
\end{equation}
of the recoil proton telescope (RPT) in the n-p scattering arm is determined by 
the mass $m_{\rm PE}$ per unit area of the polyethylene target (H/C ratio 
$x_{\rm H} \approx 2$), the differential proton emission cross section  
$({\rm d}\sigma_{\rm np}/{\rm d}\Omega_{\rm p})(\Theta_{\rm p})$ for n-p 
scattering at the proton emission angle $\Theta_{\rm p}$ and the solid angle 
$\Delta \Omega_{\rm p}$ covered by the RPT. 
The neutron detection efficiency of the RPT should match the neutron detection 
efficiency of the detector in the fission arm of the experiment detector to 
minimize the overall statistical uncertainty of the measured cross section 
ratio. 

By combining equations~\ref{eq:FC-Efficiency} and \ref{eq:RPT-Efficiency}, the 
fission cross section is obtained as
\begin{equation}
\label{eq:Cross-section}
\sigma_{\rm f} =  \biggl( \frac{N_{\rm f}}{N_{\rm p}} \biggr)  
                  \biggl( \frac{x_{\rm H}}{\epsilon_{\rm f}} \biggr)
                  \biggl( \frac{\mu_{\rm U}}{ \mu_{\rm PE}} \biggr)
                  \biggl( \frac{m_{\rm PE}}{n_{\rm U} \, \bar{m}_{\rm U}} \biggr)
                  \biggl( \frac{{\rm d} \sigma_{\rm np}}{{\rm d}\Omega_{\rm p}}\biggr)(\Theta_{\rm p})
                  \, \Delta\Omega_{\rm p}.
\end{equation}

   To reduce background and provide clear signatures for recoil proton events, 
triple-stage telescopes with two transmission ($\Delta E$) detectors and one 
stop ($E$) detector were used to identify recoil protons emitted from a 
polyethylene target, the so-called radiator. The neutron energy range covered by
a RPT is limited by the thickness of the transmission detectors and the 
thickness of the radiator. The energy loss of the recoil protons in these 
elements should be small enough to avoid excessive energy and angular straggling
of the protons. It should also leave enough kinetic energy to produce a clear 
signal in the stop detector. On the other hand, the signal in the $\Delta E$ 
detectors must be large enough to avoid losing valid recoil proton events below 
the noise level. 
The thickness of the radiator should increase with neutron energy to compensate 
for the decrease in the n-p scattering cross section and keep the recoil proton 
yield almost constant. These design constraints made it necessary to construct 
three different RPTs for overlapping energy ranges from 
30~MeV to 80~MeV, 50~MeV to 100~MeV and 80~MeV to 150~MeV with neutron detection
efficiencies ranging between $8.4\cdot10^{-6}$ and $1.1\cdot10^{-5}$.

Neutron-induced reactions on the carbon nuclei in the PE targets of the RPTs 
will produce a background of protons and deuterons from 
\textsuperscript{12}C(n,px) and \textsuperscript{12}C(n,dx) reactions. 
While deuterons can be discriminated using the differential energy losses in the
transmission detectors, protons have to be subtracted by supplementary 
measurements using matched graphite targets anyway. Hence, the demand on the 
particle discrimination properties of the RPTs are not very stringent.

   The geometrical arrangement of the RPT, especially the emission angle of the 
recoil protons, is an important parameter for the uncertainty of neutron fluence
measurements relative to the differential n-p scattering cross section. With 
respect to the uncertainty of the reference cross section, a proton emission 
angle of zero degree would be desirable because the majority of the data used 
for the phase shift analyses are for backward scattering of neutrons in the 
center-of-mass system. Hence, the uncertainty of the phase shift solutions is 
expected to be smaller than for smaller neutron scattering angles, although a 
precise statement about the uncertainty of the phase shift solutions is still 
lacking \cite{ARN91}. 

\begin{figure}[htb]
\centering 
\includegraphics[width=7cm]{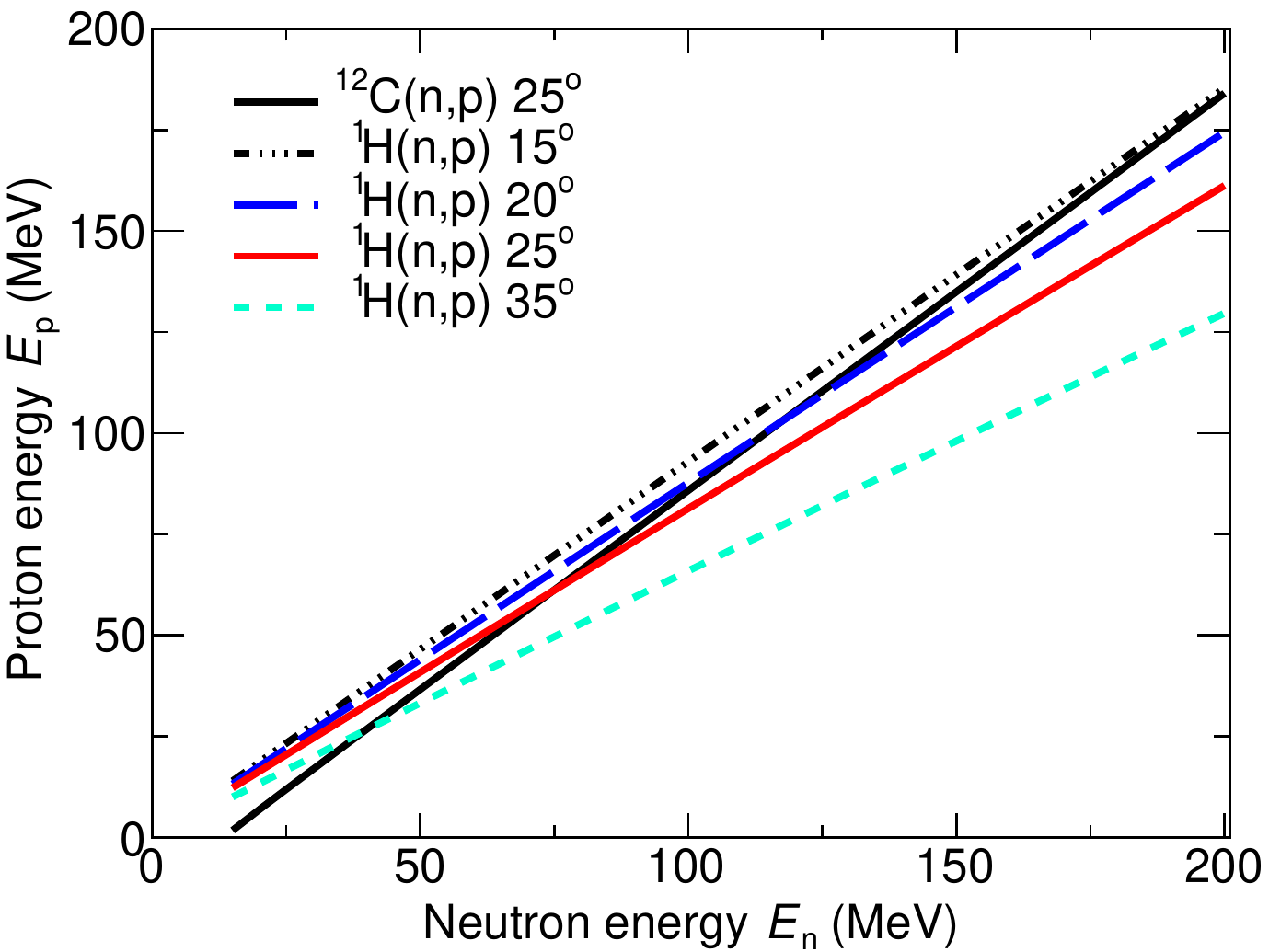}
\caption{\label{fig:RPT-Angles} Maximum energy of protons from n-p scattering, 
\textsuperscript{1}H(n,p)n, and from the 
\textsuperscript{12}C(n,p)\textsuperscript{12}B reaction as a function of the 
neutron energy for several proton emission angles $\Theta_{\rm p}$.}
\end{figure}

   In the past, RPT configurations at zero degree were used at 
quasi-monoenergetic sources up to 100~MeV \cite{SHI10} and at spallation sources
for energies up to 30~MeV \cite{CAR88}. In such configurations, an annular 
radiator is used and the RPT detectors must be shielded behind a massive shadow
bar. This was not considered a viable option for an experiment at EAR1 of 
n\_TOF because background problems were expected, and the simultaneous use of 
two telescopes would have been impossible. Moreover, the size of the available 
\textsuperscript{235}U samples (see below) made it necessary to use a collimator that 
limited the beam diameter to about 16~mm (FWHM). It is almost impossible to 
realize a zero-degree RPT in such a narrow beam. Therefore, a proton emission 
angle of 25~degrees in the laboratory system was selected for the present 
experiment, which allowed the RPT detectors to be placed outside of the 
neutron beam. 
An important aspect for the selection of the proton emission angle is the 
overlap of protons from reactions in carbon nuclei with the recoil proton. This 
is illustrated in figure~\ref{fig:RPT-Angles} which shows the 
maximum proton energy from \textsuperscript{12}C(n,p)\textsuperscript{12}B 
and n-p scattering for various emission angles. Only for emission angles 
smaller than 15~degrees it is possible to avoid interference of the two 
proton energy distributions for neutron energies up to about 200~MeV. 
For the emission angle of 25~degrees selected in the present experiment the 
recoil protons are only separated from the \textsuperscript{12}C(n,p) 
contribution up to a neutron energy of 70~MeV.

Because of a lack of space in the experimental area, it was impossible to use 
the three RPTs simultaneously. Instead, only one RPT could be used together with
the PPFC at a time and the reduced number of neutrons available for one RPT had
to be compensated by an increased efficiency, i.e. a larger solid angle 
$\Delta\Omega_{\rm p}$ covered by the telescope. 

\begin{figure}[htb]
\centering 
\includegraphics[width=7cm]{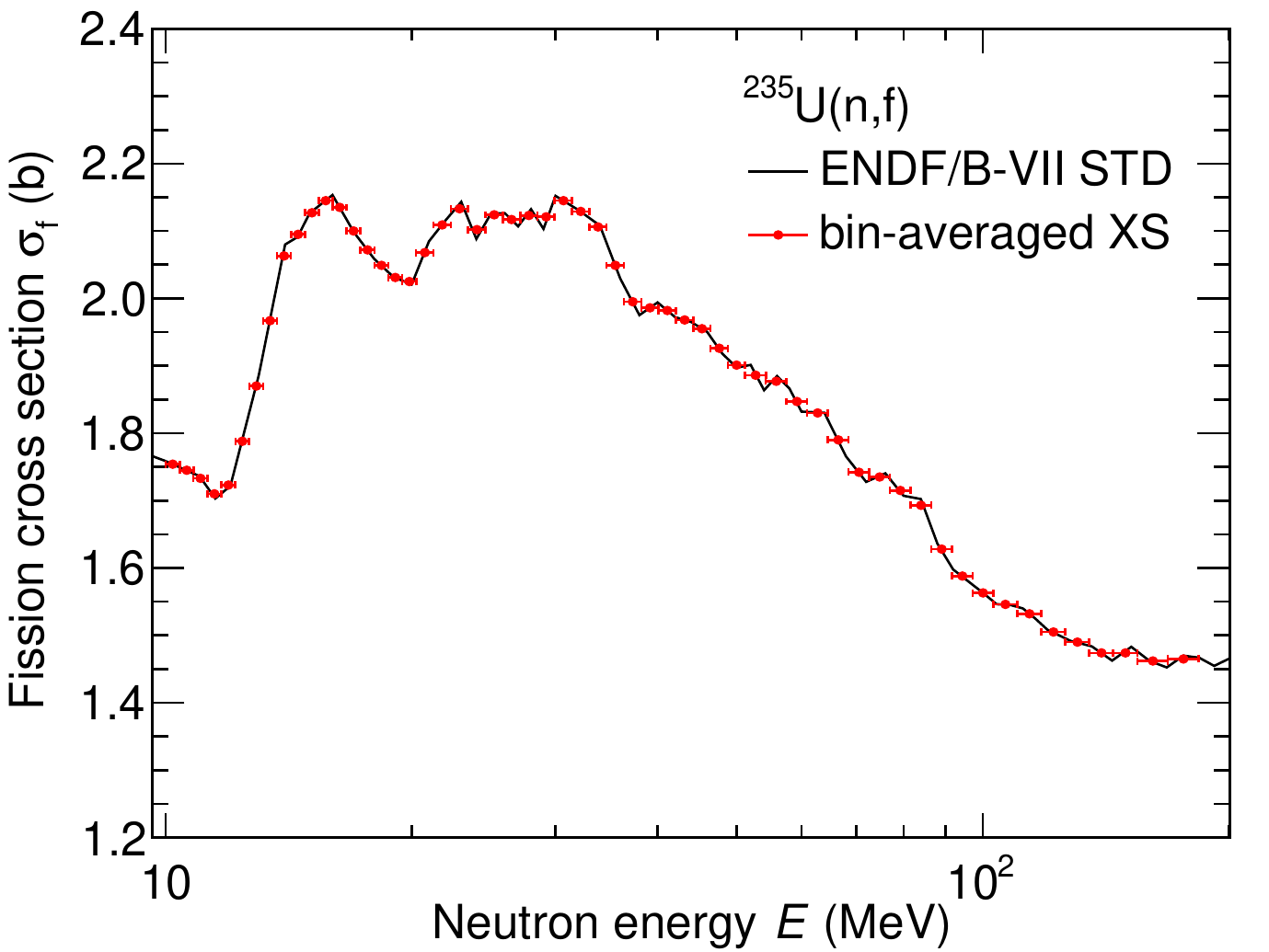}
\caption{\label{fig:Energy-grid} The solid line shows the 
\textsuperscript{235}U(n,f) reference cross section. The red error bars indicate
the energy bin structure used for the present experiment. The symbols show the 
bin-averaged cross section.}
\end{figure}

   The energy resolution required for the present experiment is determined by 
the energy dependence of the \textsuperscript{235}U(n,f) cross section while the
n-p scattering cross section is smooth and does not pose tight restrictions. 
As shown in figure~\ref{fig:Energy-grid}, a five percent energy resolution is 
sufficient to resolve the structures for neutron energies between 20~MeV and 
50~MeV. At higher energies, the cross section is rather smooth which allows 
increasing the required resolution to seven percent at 200~MeV.  With this 
binning and for beam pulses with $7\cdot10^{12}$ protons, the number of neutrons
per energy bin varies between $9.8\cdot10^2$ per beam pulse at 20~MeV and 
$2.4\cdot10^3$ at 200~MeV. It should be noted that the energy resolution 
required by the \textsuperscript{235}U cross section is still larger than the 
energy resolution of the neutron beam for a given flight time which is 
determined by the duration of the proton beam pulse of about 7~ns. At a neutron
energy of 200~MeV this translates into an energy resolution of about 5~MeV, 
i.e. 2.5~\%. At lower energies, the energy resolution of the neutron beam 
becomes even smaller.

   The energy range covered by the low-energy experiment is limited by the 
maximum proton energy which could be stopped in the last detector.
For a given maximum length of the stop detector, the energy range could be 
extended by using larger scattering angles which would reduce the energy of the
recoil protons at a given neutron energy. However, this would require separate 
measurements with this configuration, which was incompatible with the allocated 
run time, or a second dedicated telescope, which was ruled out by the available 
space. As shown in figure~\ref{fig:RPT-Angles}, a larger scattering angle would
also increase the interference of recoil protons and protons from 
\textsuperscript{12}C(n,p) reactions.

   As for every bremsstrahlung or spallation neutron source, the high-intensity 
prompt gamma flash determines the maximum neutron energy which can be handled 
with a given detector technology without significant saturation problems. The 
distance between the spallation target and the measurement position at EAR1 of 
n\_TOF is about 183~m. At this distance, a neutron energy of 200~MeV corresponds
to a time difference between the arrival of the gamma flash and the neutrons 
of about 400~ns. Hence, all detectors must recover from saturation effects 
induced by the gamma flash within this time. This requires low-mass designs and 
the use of low-$Z$ materials to reduce the photon detection probability.

%-------------------------------------------------------------------------------

\section{Parallel-plate fission chamber}
\label{sec:FC}

\subsection{Design}
\label{subsec:FC-Design}

\begin{figure}[htb]
\centering
\includegraphics[width=8cm,trim=4cm 2cm 4cm 2.5cm,clip]{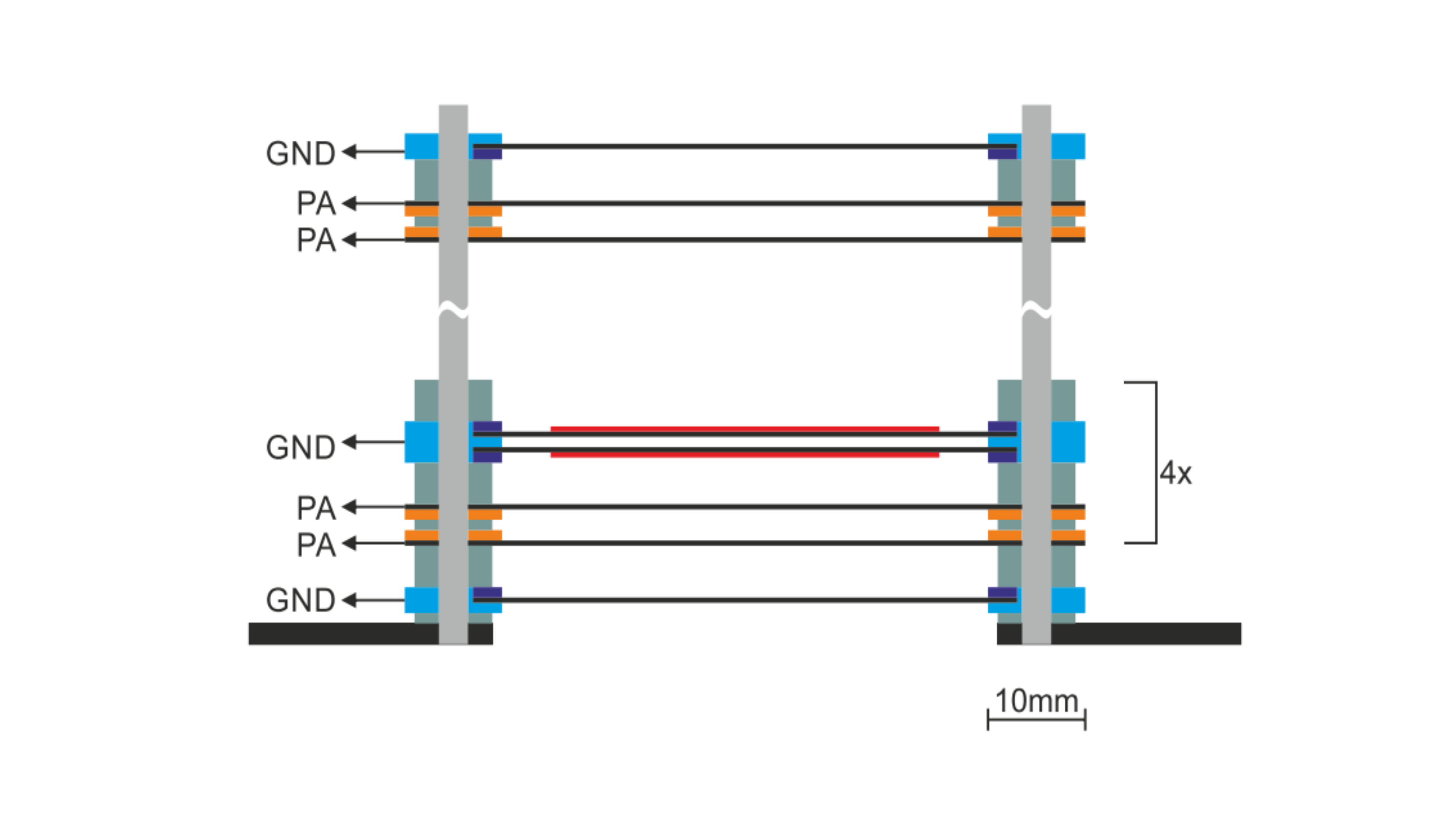}
\caption{\label{fig:FC-Stack}
Schematic cross-sectional view of the PPFC electrode stack.
The eight \textsuperscript{235}U samples consisted of 30 {\micro m} thick 
aluminum foils (black) coated with layers containing \textsuperscript{235}U 
(red). The samples were mounted back-to-back in aluminum frames and used as 
cathodes on ground (GND) potential. At the bottom and top of the stack two blank
samples were placed. The opposite read-out anodes were made of 20~{\micro m} 
aluminum foils. They were biased at +230 V and connected to individual 
Canberra~2006 preamplifiers (PA) followed by ORTEC~474 timing filter amplifiers. 
The distances between samples and read-out electrodes were 5~mm. Note that the 
thicknesses of the foils and fissile layers are not to scale. The counting gas 
was Ar/CF\textsubscript{4}(10\,vol\%) at ambient pressure. The electrode stack 
was mounted in a stainless steel housing equipped with Kapton entrance and 
exit windows for the neutron beam.}
\end{figure}

   Figure~\ref{fig:FC-Stack} shows the layout of the PPFC. The focus of the PPFC
design was to keep the mass intercepting the neutron beam at minimum to 
avoid saturation effects induced by the gamma flash and the high instantaneous 
neutron rate at high energies. The detector consists of a stack of eight fissile 
samples with a diameter of 42.00(3)~mm and \textsuperscript{235}U masses per unit area ranging between 
264~{\micro g}/cm\textsuperscript{2} and 
372~{\micro g}/cm\textsuperscript{2}. The backings of these samples are aluminum foils 30~{\micro m} in 
thickness which is much larger than the range of fission fragments in aluminum, 
i.e. the backings are \lq thick\rq . The foils are glued to 1~mm thick 
aluminum rings with inner and outer diameters of 49\,mm and 52\,mm, 
respectively.
Two samples were inserted back-to-back in an aluminum holder ring 70~mm in 
diameter and 3~mm in thickness. The holder rings were connected to ground 
potential, i.e. the samples acted as cathodes. The anodes facing the samples 
consisted of 20~{\micro m} aluminum foils glued to rings made of 
fiber-reinforced plastic material (G10), 1~mm in thickness and with inner and 
outer diameters of 49~mm and 70~mm respectively. The thickness of the anode 
foils was sufficient to just stop fission fragments emitted from the adjacent 
fissile layer. The cathode and anode foils were held at a distance of 5~mm by 
six PEEK spacers. Four stacks containing two cathodes and two anodes each were 
stacked on top of each other with spacers between the adjacent anode such that 
the distance between the anode foils was 3~mm.  Below and above the stack two 
cathodes with blank samples and the corresponding anodes were mounted. The 
whole stack was held together by six 3~mm Trovidur rods and plastic nuts. 
 
   The stack was mounted in a stainless steel vacuum chamber with Kapton windows
60~mm in diameter and 50~{\micro m} in thickness. The fission chamber was 
operated with Ar/CF\textsubscript{4}(10\,vol\%) at ambient pressure and a 
regulated gas flow rate of about 15~ml/min. Including the counting gas, the 
total mass per unit area intercepted by the neutron beam amounted to 
147~mg/cm\textsuperscript{2}. 
The ten anodes of the PPFC were biased at +230~V to maximize the drift 
velocity of the electrons \cite{CHR79}. 
Each anode was connected to a separate Canberra~2006 charge-sensitive 
preamplifier with a sensitivity of 235~mV per $10^6$ ion pairs. Hence, assuming
a mean energy of 24~eV required to create an electron-ion pair, about 
14~fission fragments could be detected simultaneously (within a time interval of
about 50~ms, i.e. the decay time constant of the preamplifier)  without 
saturating the preamplifier. 
This is sufficient for the expected number of fission events induced per proton 
beam pulse by neutrons with energies above 10~MeV. The preamplifier signals 
were fed into ORTEC~474 timing filter amplifiers, shaped to pulses with a 
width of about 130~ns (FWHM) and then digitized using the n\_TOF data
acquisition system. 

   The response of the PPFC to the n\_TOF gamma flash was calculated with MCNPX 
\cite{PEL11} using the energy distribution of prompt photons from \cite{GUE13} 
as input. In these simulations photon and electron transport was switched on 
and the production of secondary bremsstrahlung was sampled in the analog mode. 
The energy deposition in each cathode - anode gap was calculated and compared 
with the maximum energy deposition resulting from fission fragments. The results
showed that for beam pulses with the maximum number of $7\cdot10^{12}$ protons 
the signal induced by the gamma flash amounts to about 30~\% of the maximum 
signal expected for a fission fragment. The number of photofission events 
induced by the gamma flash in one fissile layer is about $1.3\cdot10^{-2}$ per 
proton beam pulse of $7\cdot10^{12}$ protons. 
In summary, these results demonstrate that the energy deposition induced by the
n\_TOF gamma flash does not pose a significant challenge for the PPFC setup for
this experiment.

\subsection{Characterization of the uranium targets}
\label{subsec:FC-Targets}

The uranium deposits were provided by the EC Joint Research Centre in Geel 
(JRC-Geel). They were prepared via molecular plating 
from a solution of UO\textsubscript{2}(NO\textsubscript{3})\textsubscript{2} 
in isopropanol \cite{SIB18}. The enrichment of the isotope 
\textsuperscript{235}U was 99.9336(14)\,\%.
A mask 42.00(3)~mm in diameter was used to define the size of the samples.
The mass of each deposit was determined from the activity, measured via
alpha counting, and the isotopic composition was measured by mass spectrometry.
The diameter of the deposits was not measured directly, 
so for the average mass per unit area the diameter of the mask was used.

In \cite{SIB18}, the chemical composition of deposits produced
with the same technique is discussed; from their analysis it resulted that 
the samples obtained by molecular plating are not pure oxide, but contain 
traces of other elements, for example carbon. The morphology was also studied;
it was found that deposits produced by molecular plating present a maze-like
cracking pattern, with huge fluctuations on the local density for scales of the
order of 10~{\micro m}.
All these information were included in our considerations for the determination
of the detection efficiency of fission fragments in 
section~\ref{subsec:FC-Efficiency}.

\begin{figure}[htb]
\centering
\includegraphics[width=7cm]{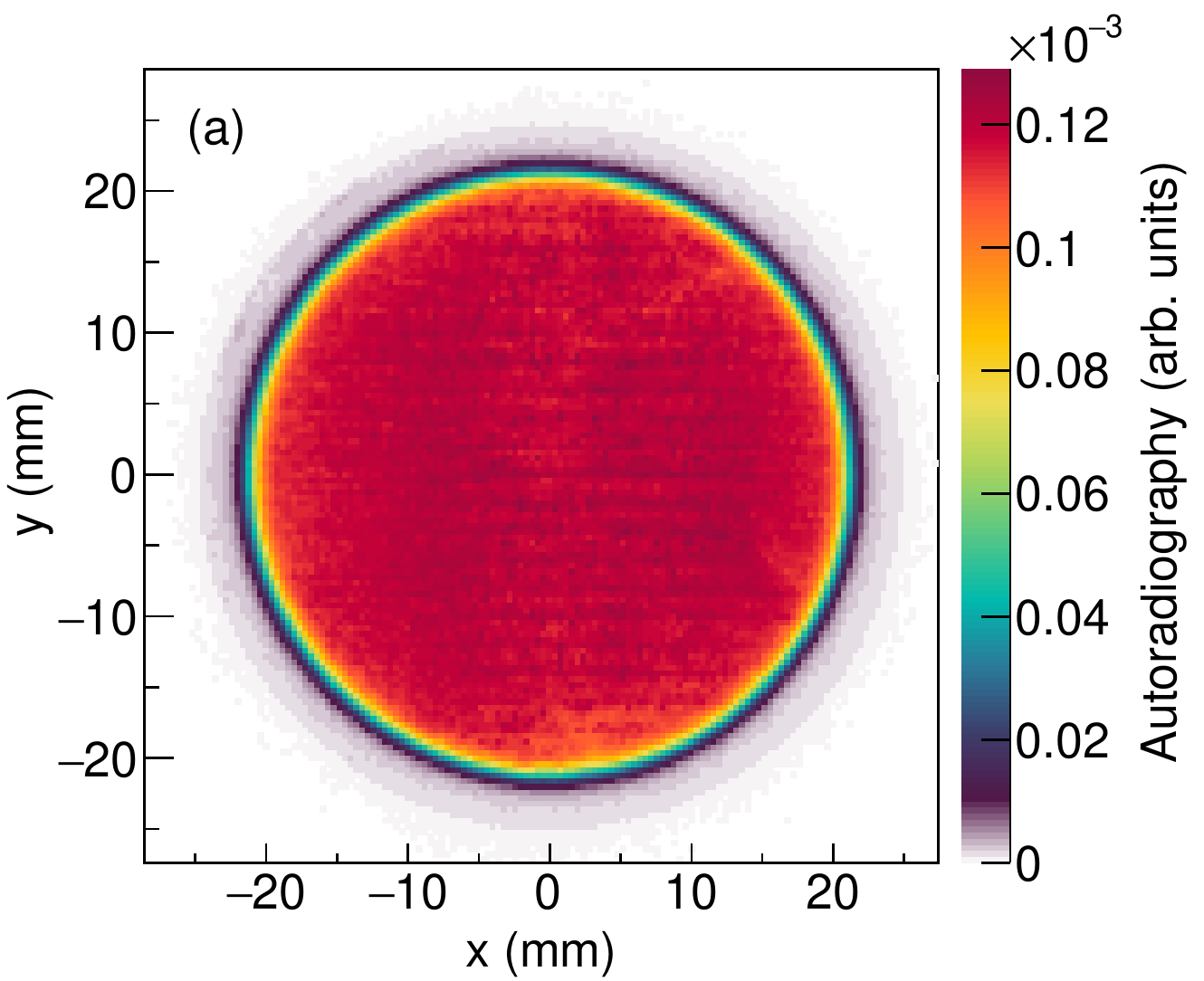}\qquad
\includegraphics[width=7cm]{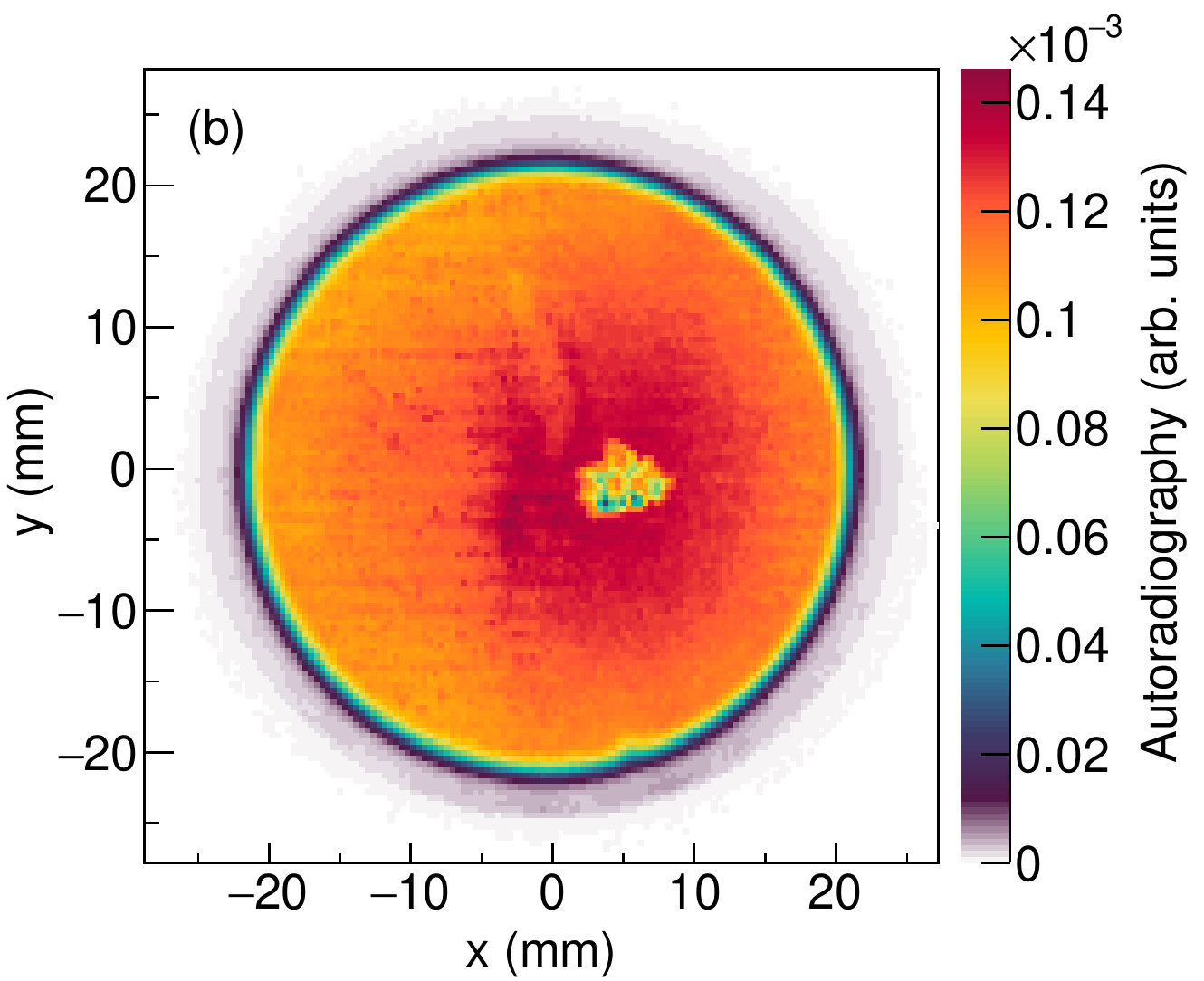}
\caption{Results of the autoradiography of two uranium samples, selected to show
the variability in the uniformity of the deposit layers: (a) is characterized
by a smooth profile, while (b) has a \lq dent\rq\ and it is overall less regular.}
\label{fig:FC-autorad}
\end{figure}

To determine the uniformity of the layers of fissile material, autoradiographies 
of all samples were produced by placing the samples on top a FUJIFILM BAS-IP MS
storage phosphor screen at a distance of 1~mm. In figure~\ref{fig:FC-autorad},
the digital scan of two autoradiographies is shown, selected to represent the 
variability among samples. As expected, the samples are not perfectly uniform: 
they are thinner at the borders and thicker in the center, the width of the 
borders varying from sample to sample. In some cases, irregular structures which
were not  evident by visual inspection were also revealed; that was the case for 
example for the sample shown in figure~\ref{fig:FC-autorad}(b).

Considering the non-uniformity of the samples and the beam profile, the amount
of fissile material effectively irradiated by the neutron beam cannot be simply 
determined from the average mass per unit area. 
A correction factor $k_{\rm U}$ was therefore calculated as the convolution of the 
autoradiography images $I(x,y)$ with the beam profile $\phi(x,y)$ measured
for neutron energies above 30~MeV:
\begin{equation}
\label{eq:FC-kU}
    k_{\rm U} = \pi r_{\rm U}^2 \frac
    { \int_0^{r_{\rm U}} I(x,y) \phi(x,y) \, {\rm d}x {\rm d}y }
    { \int_0^{r_{\rm U}} I(x,y) \, {\rm d}x {\rm d}y \int_0^{r_{\rm U}} \phi(x,y) \, {\rm d}x {\rm d}y}
\end{equation}
where $r_{\rm U}$ is the nominal radius of the deposits. The neutron beam 
profile $\phi(x,y)$ was measured with the PPAC detector \cite{SIS}, while the 
alignment of the samples with respect to the axis of the beam was carried out
using Gafchromic films. The uncertainty on 
$k_{\rm U}$ was determined by varying the position of the neutron beam relative 
to the sample, and trying different parametrizations for $\phi(x,y)$
(as for example the parametrizations obtained for neutron energies below 
30~MeV). The results for each sample are reported in table~\ref{tab:FC-Targets}.
It was found that, globally, the non-uniformity of the samples and the beam
resulted in a correction factor of 1.075 on the \textsuperscript{235}U
areal density.

\begin{table}[htb]
\centering
\caption{\label{tab:FC-Targets} Mass per unit area of 
\textsuperscript{235}U $m_{\rm U}$, provided by JRC-Geel, 
and correction factor $k_{\rm U}$ that accounts for the amount of uranium 
effectively irradiated by the neutron beam.
The effective mass $m_{\rm U,eff}$ accounts for the stoichiometry and
microscopic structure of the layers and was used only to calculate the zero-bias 
fragment detection efficiency $\epsilon_{\rm f}$ for thermal neutrons.}
\smallskip
\begin{tabular}{|l|c|c|c|c|}
\hline
Sample & $m_{\rm U}$       & $k_{\rm U}$ & $m_{\rm U,eff}$  & $\epsilon_{\rm f}({\rm 25.4\,meV})$ \\
\quad  & \micro g/cm$^2$   & \quad       & \micro g/cm$^2$  & \quad                               \\
\hline                                                                                      
U1     & 337.6(21)         & 1.058(21)   & 503(16)          & 0.901(3)    \\
U2     & 316.5(20)         & 1.033(9)    & 510(18)          & 0.900(4)    \\
U3     & 264.1(17)         & 1.044(9)    & 344(14)          & 0.9325(27)  \\
U4     & 282.9(18)         & 1.094(26)   & 295(15)          & 0.9418(29)  \\
U5     & 289.4(18)         & 1.086(19)   & 402(20)          & 0.921(4)    \\
U6     & 279.1(18)         & 1.083(14)   & 415(17)          & 0.918(4)    \\
U7     & 280.5(18)         & 1.105(26)   & 560(18)          & 0.889(4)    \\
U8     & 307.3(20)         & 1.101(17)   & 550(21)          & 0.892(4)    \\
\hline
\end{tabular}
\end{table}

\subsection{Fragment detection efficiency}
\label{subsec:FC-Efficiency}

   The main advantage of a PPFC for the measurements of fission cross sections 
compared with a parallel-plate avalanche counter is the large detection 
efficiency for fission fragment which is usually above 90~\%. In a PPFC, only 
fragments emitted almost parallel to the fissile layer lose such a large 
fraction of their kinetic energy in the layer that the signal produced by the 
drifting secondary electrons falls below the detection threshold which is given 
by the maximum signal produced by alpha particles, recoil nuclei from 
neutron-induced reaction in the backing or electromagnetic noise. 

   The calculation of the fragment detection efficiency can be carried out by 
analytical formulas \cite{CAR74} or using the Monte Carlo method. In addition to
energy-loss data for the fissile layer, this requires knowledge of the 
kinematic properties of the fission fragments, i.e. the angular distribution 
of the fission fragments in center-of-mass system and the partial transfer of 
linear momentum from the incident neutron to the fissioning compound nucleus. 

   The analytical formulas given in \cite{CAR74} are sufficient for calculating 
the probability that at least one of the two fragments deposits energy in the 
counting gas, the so-called zero-bias efficiency. 
The key assumption behind these formulas is that of a homogeneous fissile layer 
with well-determined stoichiometry. As discussed above, this assumption is not 
valid for the samples used for the present experiment. Moreover, the zero-bias 
efficiency $\epsilon_{\rm f}$ is always several percent larger than 
the efficiency for a finite 
pulse-height threshold imposed by the experimental constraints  
which requires an extrapolation of the experimental pulse-height distribution to
zero pulse height.

   For the calculation of the zero-bias efficiency the effective 
mass $m_{\rm U,eff}$ was treated as an adjustable parameter that
was determined from a
comparison of experimental pulse-height distributions with predictions of a 
dedicated Monte Carlo model of a PPFC. Mass, charge and kinetic energy 
distributions of fission fragments were obtained from the GEF code \cite{SCH16} 
for neutron energies between 25.4~meV to 100~MeV, the maximum neutron energy in 
GEF. Above this energy the data for 100~MeV were used. 
In the energy range between 20~MeV and 200~MeV, experimental data for the 
angular distribution of fission fragments are available for 
\textsuperscript{235}U(n,f) \cite{VOR17,GEP19,LEA16}. Data for the effect of the 
partial transfer of linear momentum were taken from results for 
\textsuperscript{235}U(p,f) \cite{FAT85}. The energy loss of the fission 
fragments in the fissile layer and the counting gas were calculated using 
energy-loss or range data from the SRIM2013 code \cite{ZIE13}. The effective 
stoichiometric composition of the layers was an average of the results reported 
in \cite{SIB18}.

   Correlated pairs of fragments were started at a random depth in the fissile 
layer and transported into the backing or into the gas volume.
The charge $Q$ induced on the anode by the drifting electrons was
calculated in the usual way by applying the Shockley-Ramo theorem \cite{ZHO01} 
and assuming a homogeneous electrical field is 
\begin{equation}
\label{eq:FC-U}
Q = \frac{e_0}{W} 
     \int_0^{R(E_0)} S(E) \; \bigl( 1-\frac{r}{d}\cos{(\Theta}) \bigr) \; {\rm d} r, 
\end{equation}
where $W$, $d$ and $e_0$ denote the mean energy required to produce an
electron-ion pair, the anode-cathode distance, and 
the elementary charge, respectively. $S(E)$ denotes the differential energy loss
in the counting gas of a fission fragment with kinetic energy $E$ and $R(E_0)$ 
is its range after leaving the fissile layer with kinetic energy $E_0$. The angle 
of the fragment relative to the normal on the cathode and the distance along the
fragment trajectory are denoted by $\Theta$ and $r$, respectively.  
Eq.~\ref{eq:FC-U} applies to fragments stopped in the counting gas. Wall effects
are accounted for by restricting the integration to the distance traveled in the
counting gas. The ionization quenching observed for fission fragments in
Ar/CF\textsubscript{4}(10\,vol\%) gas \cite{PEC19} was included in the 
simulation. The electronic resolution of the whole signal processing chain is 
accounted for by folding the simulated pulse-height distributions with a 
Gaussian of constant relative width. The relative widths were adjusted 
individually for each signal-processing chain.

   The adjustment of the effective thickness of the fissile layers was carried 
out for a neutron energy window around the thermal point at 25.4~meV, i.e. 
about 83~ms after the gamma flash, because electronic distortions induced by 
the gamma flash vanished at this large time of flight. Moreover, the angular 
distribution of the fission fragments is isotropic at thermal energies. 
 
\begin{figure}[htb]
\centering
\includegraphics[width=7cm]{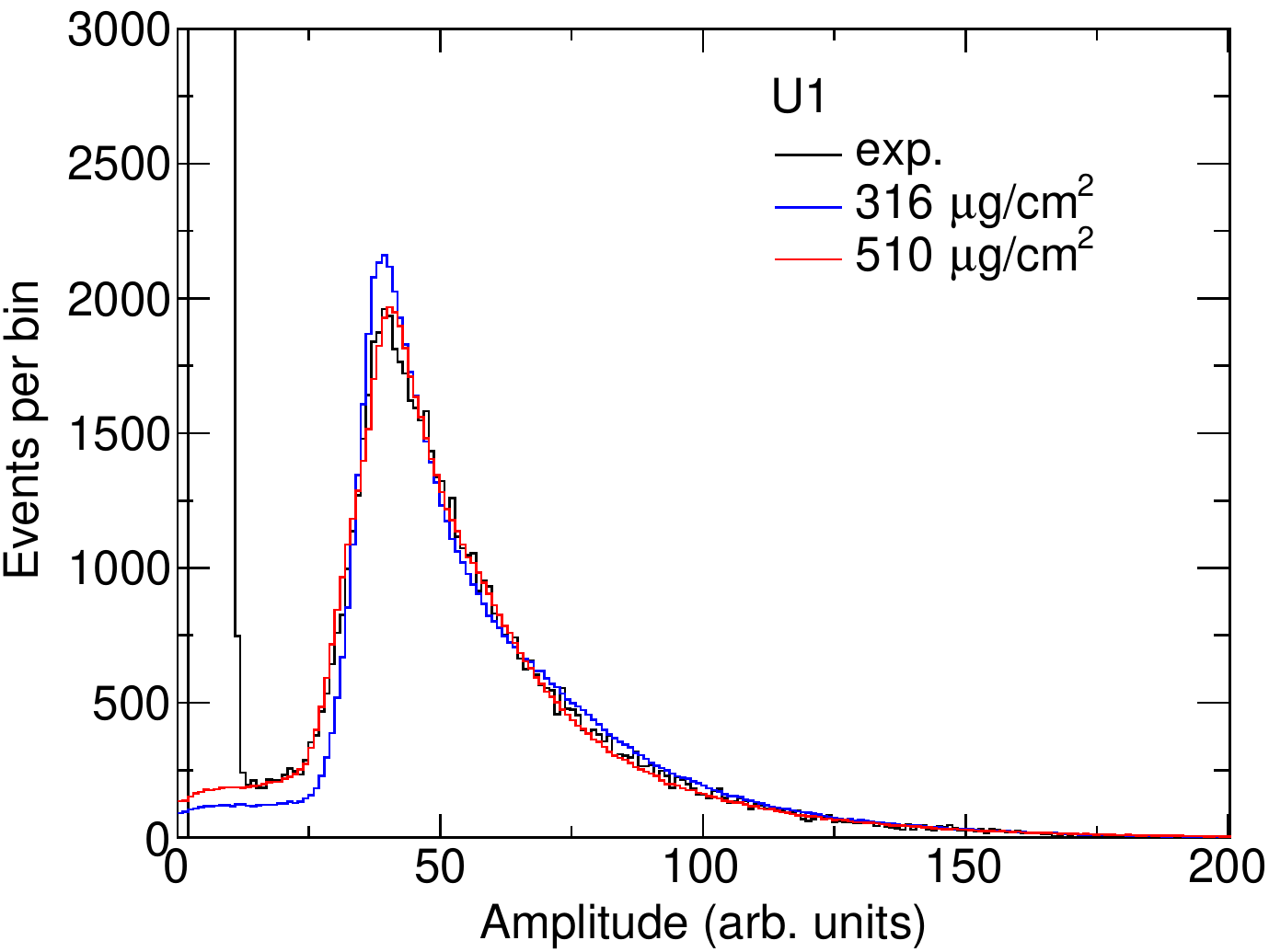}\qquad
\includegraphics[width=7cm]{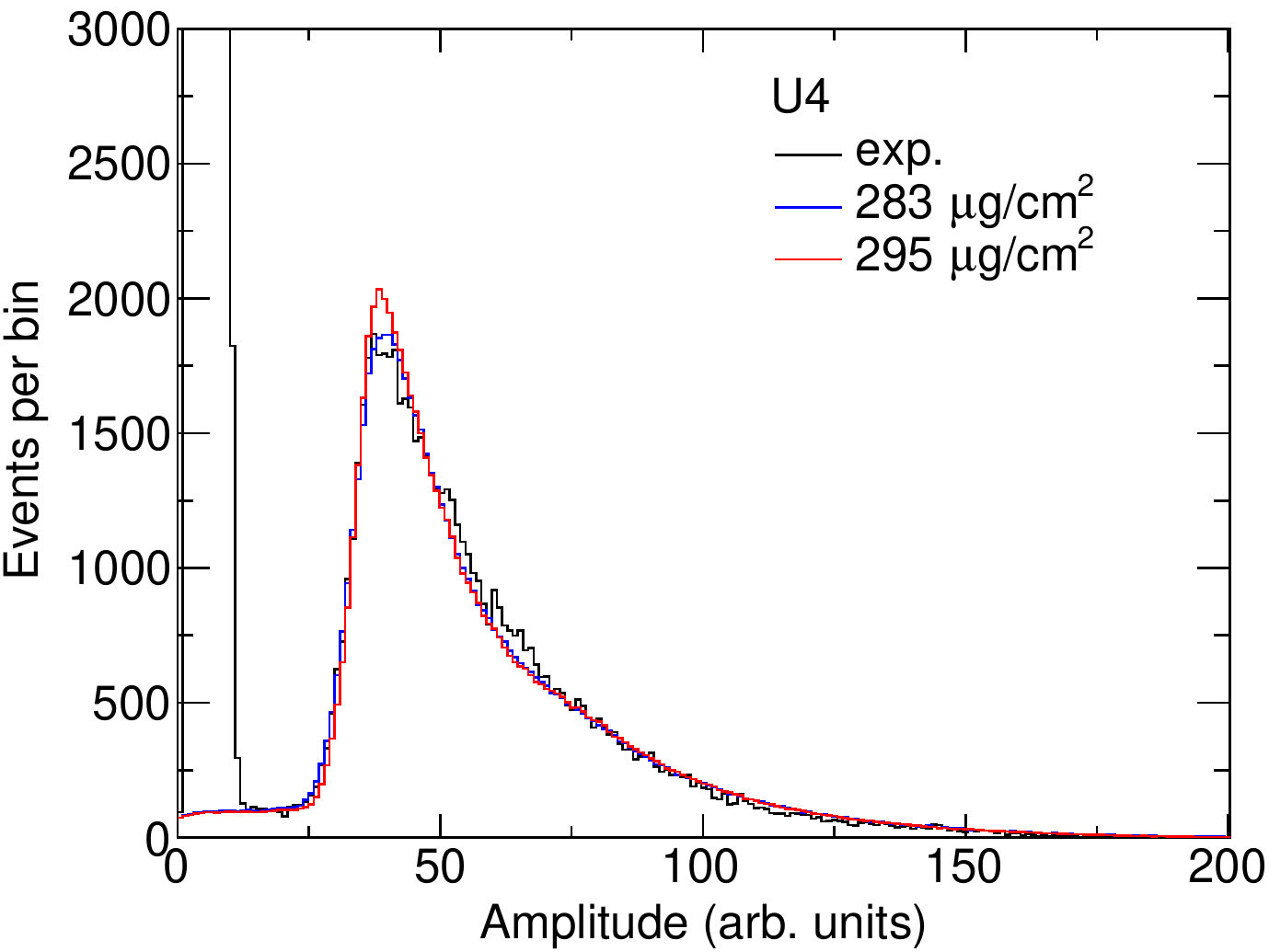}
\caption{\label{fig:FC-MC} Experimental and calculated pulse-height 
distributions for the two samples U1 (left panel) and U4 (right panel). 
The black histograms are the experimental data. The blue histograms were 
calculated for the nominal mean masses per unit area $m_{\rm U}$, calculated 
from the measured \textsuperscript{235}U mass per unit area and the effective 
stoichiometry taken from \cite{SIB18}. 
For the red histograms, the effective masses per unit area $m_{\rm U,eff}$ were
adjusted such that the pedestals below the maximums of the fragment pulse-height
distributions and the noise were reproduced.}
\end{figure}

As shown in figure~\ref{fig:FC-MC}, an adjustment of the mass per unit area 
leads to a satisfactory agreement over the full pulse-height range, 
including the region above the maximum which contains fragments leaving the 
fissile layer at angles $\Theta$ close to 90\textdegree. 
The effective \textsuperscript{235}U masses per unit area $m_{\rm U,eff}$ and 
the corresponding zero-bias fragment detection efficiencies $\epsilon_{\rm f}$ 
of the PPFC for thermal neutrons are indicated in table~\ref{tab:FC-Targets}.
The comparison of experimental and calculated pulse height distribution also 
confirms that the usual constant extrapolation of the pulse-height distribution 
in the region dominated by alpha particles and noise can also be carried out 
for the present samples (see section~\ref{subsec:FC-Analysis}). 

   For the sample shown on the left side of figure~\ref{fig:FC-MC} the effective
mass exceeds the nominal mass by about 60~\%. 
In contrast, there is only a slight deviation between the effective and the
nominal mass for the sample shown on the right side. 
This shows that the \lq cracky\rq\ microstructure of the layers obviously varies 
from sample to sample. 
It should be noted that this analysis is only sensitive to the cumulative effect
of the fissile mass per unit area, the stoichiometric composition of the fissile
layer and the range data for fragments in the fissile layer. 
However, range data for fission fragments calculated using the earlier version 
SRIM2008 were found to exceed experimental data by not more than 10\,\%-20\,\% 
\cite{FIL10} which is insufficient to explain the observed discrepancies. 
Therefore, the majority of the observed deviations must be due to the 
imperfections of the surface structure.   

\begin{figure}[htb]
\centering 
\includegraphics[width=7cm]{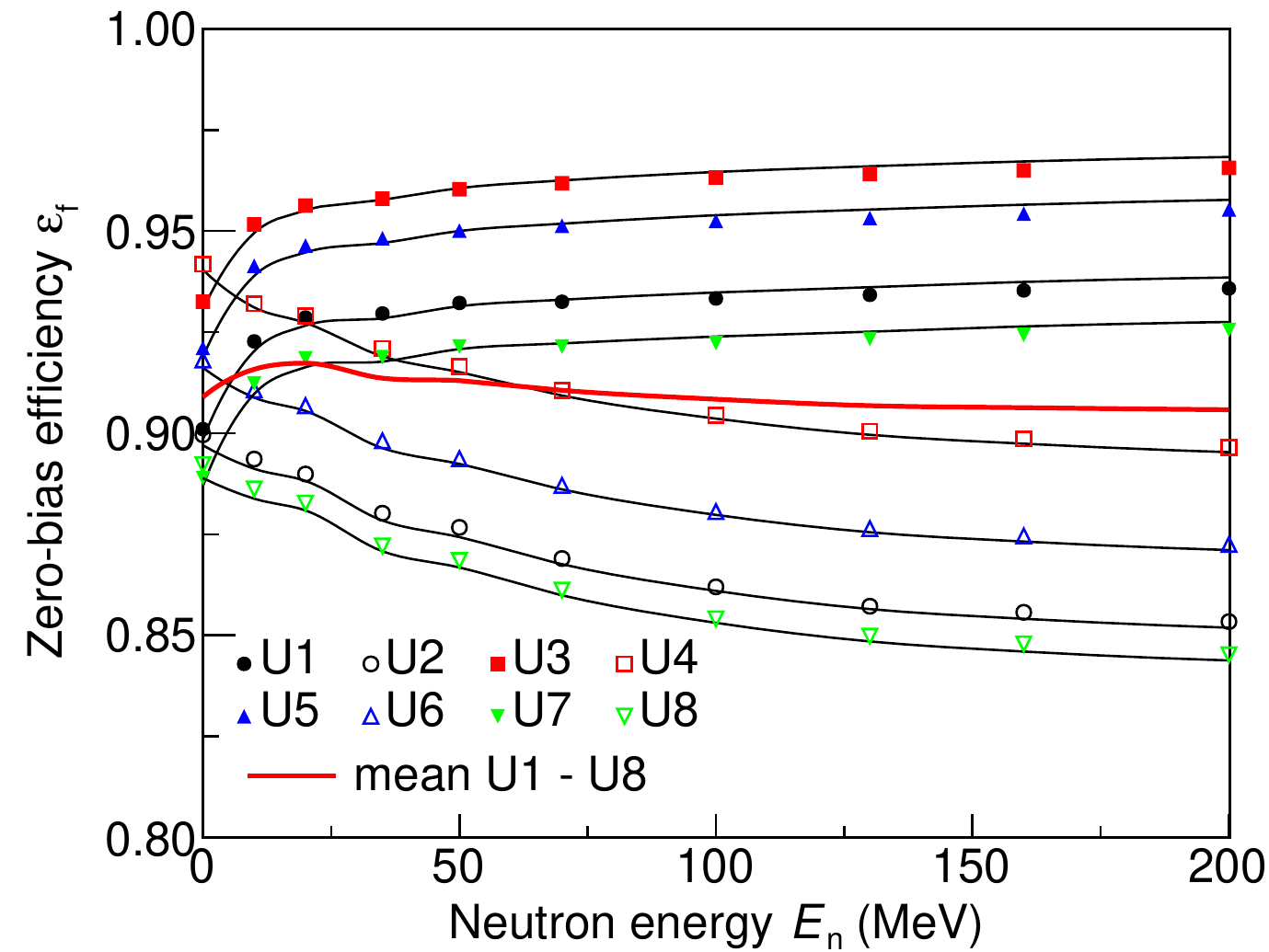} 
\caption{\label{fig:FC-ZBEfficiency} Zero-bias fission fragment detection 
efficiency $\epsilon_{\rm f}$ for the \textsuperscript{235}U layers produced by 
electroplating. The effective mass $m_{\rm U,eff}$ per unit area 
(see table~\ref{tab:FC-Targets}) was adjusted to match the experimental 
pulse-height distributions for thermal neutrons using the effective 
stoichiometric composition of layers \cite{SIB18}. The solid lines show the 
result of the Carlson's analytical model \cite{CAR74} and the symbols that of 
the Monte Carlo model. The open and closed symbols indicate layers oriented 
towards and away from the neutron source, respectively. The difference 
in the energy dependence for the two orientations is an effect of the
partial transfer of the linear  momentum. The thick red solid 
line shows the mean of all samples using the \textsuperscript{235}U mass per 
unit area $m_{\rm U}$ as weights.}
\end{figure}

   Figure~\ref{fig:FC-ZBEfficiency} shows the fragment detection efficiency 
$\epsilon_{\rm f}(E)$ as a function of neutron energy for all 
\textsuperscript{235}U layers used in the low-energy experiment.   
Results are shown for the neutron incident from the layer and backing side. 
The difference between the two orientations reflects the effect of the partial 
transfer of linear momentum. This effect almost cancels if an equal number of 
samples with either orientation and similar mass per unit area is used.
    
   The zero-bias efficiencies calculated using the Monte Carlo model are 
compared with Carlson's analytical formulas \cite{CAR74}. 
The good agreement confirms that the analytical approach can be used to 
propagate the uncertainties of the various influence quantities to the total 
uncertainty of the zero-bias efficiency.

\subsection{Data acquisition and analysis}
\label{subsec:FC-Analysis}

\begin{figure}[htb]
\centering
\includegraphics[width=7cm]{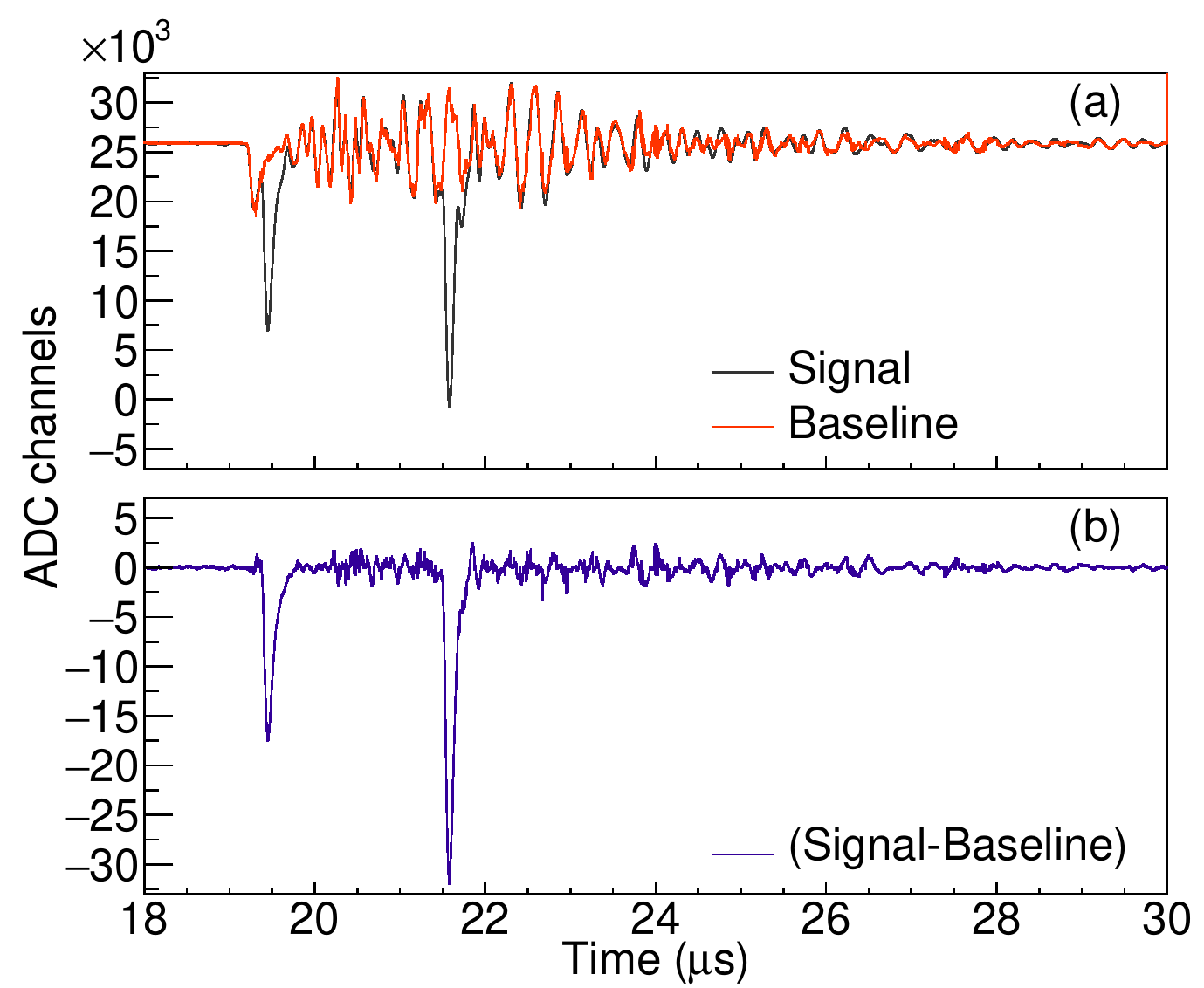}\qquad
\includegraphics[width=7cm]{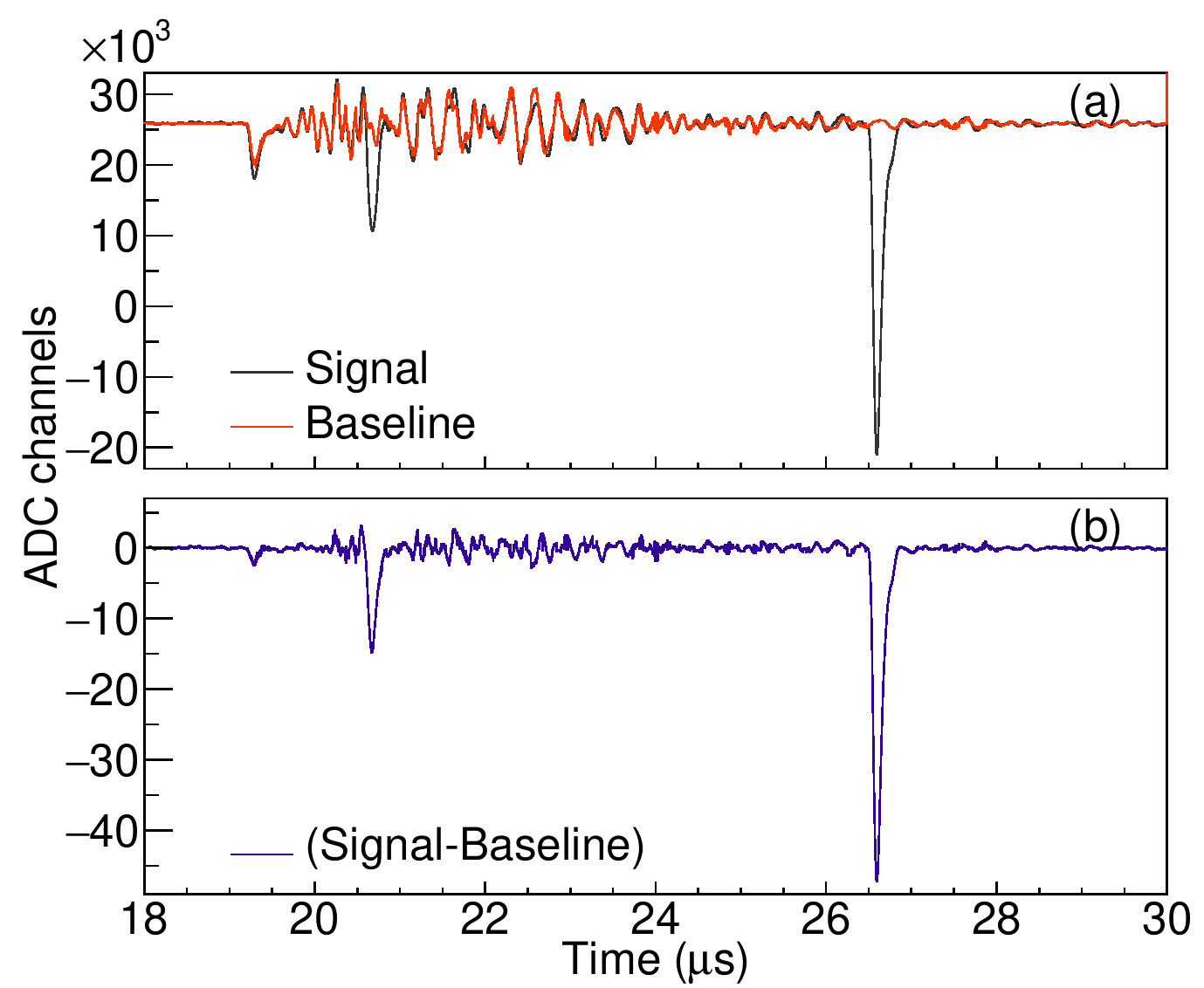}
\caption{\label{fig:FC-baseline} Two signals (frames of 12~{\micro s}) recorded 
for the target U1 of the PPFC. 
Panel~(a) shows the raw signal, with the gamma flash arriving with a timestamp 
of about 19.2~{\micro s}, followed by the ringing noise, which is overlapping to 
the detector signals. Panel~(b) shows the result of the baseline subtraction: 
the gamma flash is suppressed, and the noise under the two signals is strongly 
reduced. }
\end{figure}

The main problem that had to be solved via data analysis was the subtraction of
the ringing noise generated in the interaction with the gamma flash. 
The gamma flash signal was not very intense per se, however it was 
followed by strong oscillations which were hindering the discrimination of alpha 
particles (low amplitude signals) and fission fragments (high amplitude signals)
for about 5~{\micro s}. At 183~m distance from the neutron source, a time of 
flight of 5~{\micro s} corresponds to about 5~MeV in neutron energy, i.e. the 
disturbance was affecting the entire time interval of interest. 
The electromagnetic interference could be partially reduced by improving the 
grounding and the electric shielding of chamber and preamplifiers but,
as it can be seen from figure~\ref{fig:FC-baseline}(a), that was not 
sufficient to obtain a clean signal, and therefore the baseline distortion had 
to be corrected numerically during the waveform post-processing.

The amplitude of the oscillations was correlated with the intensity of the gamma
flash, the shape however was found to be independent from the beam and rather 
being a feature of the electronics, as it was always the same at every 
beam pulse. By summing the first 12~{\micro s} of consecutive (arbitrarily 
chosen) signals, it was possible to filter out the signals from low probability 
events (alpha particles from radioactive decay, fission fragments from 
high-energy neutrons) and therefore to determine, in average, the shape of the 
\lq distorted baseline\rq. The baseline subtraction could be then carried out as
usual with the PSA routine, with the sole caution of introducing a scale factor 
to take into account that noise amplitude was a function of the
gamma flash intensity.

The subtraction, as shown for example in figure~\ref{fig:FC-baseline}(b), did 
not completely cancel out the noise, but the pulse height of the residual 
fluctuations was similar or lower than that of the $\alphaup$-particle signals, 
which was enough for the analysis purposes. 
As it can be seen in the figure, the gamma flash signal was also suppressed
in this procedure, but that was not an issue as the PSA determined its timestamp
before proceeding with the baseline subtraction.
Beside the arrival time of the gamma flash ($t_{\rm \gamma}$), and the baseline
subtraction, the PSA determined also the timestamp of the pulses following the 
flash ($t$), the pulse height, and the pulse integral (later denoted as 
$A$, area), which is proportional to the charge induced on the anodes.

The time of flight $t_{\rm n}$ of the neutrons was then obtained as:
\begin{equation} 
    t_{\rm n} = t - t_{\rm \gamma} + \frac{L_{\rm FC}}{c}
\label{eq:FC-tof}
\end{equation}
where $c$ is the speed of light in vacuum, and $L_{\rm FC}$ is the flight path 
from neutron source to fission chamber. 
To take into account the spread in time caused by the moderation process inside 
the lead target following the neutron production, the conversion from 
time-of-flight $t_{\rm n}$ to 
neutron energy $E_{\rm n}$ was carried out following the procedure explained 
e.g. in \cite{GUE13}. The neutron velocity $v_{\rm n} = L_{\rm FC,eff} / t_{\rm n}$ 
was therefore calculated considering the effective flight path:
\begin{equation}
\label{eq:FC-EffectiveFP}
    L_{\rm FC,eff} = L_{\rm FC} + \lambda(E_{\rm n}) - \lambda({\rm 250~eV})
\end{equation}
where $\lambda(E_{\rm n})$ is the so-called neutron moderation distance,
determined in extensive Monte Carlo simulations of the n\_TOF spallation target.
Finally, the neutron energy $E_{\rm n}$ was calculated using:
\begin{equation}
\label{eq:NeutronEnergy}
    E_{\rm n} = m_{\rm n} c^2 \left( \frac{1}{\sqrt{1-{v_{\rm n}}^2/c^2}} - 1 \right)
\end{equation}
in an iterative procedure. Here $m_{\rm n}$ denotes the neutron mass.
The value of $L_{\rm FC} = \rm 182.61(5)~m$ in equation~\ref{eq:FC-tof} was 
calibrated using as reference the position of the resonances in the 
\textsuperscript{235}U fission cross section between 200~eV and 300~eV
reported in the ENDF/B-VIII.0 library.
Summing and subtracting $\lambda(E_{\rm n})$ and $\lambda({\rm 250~eV})$
in equation~\ref{eq:FC-EffectiveFP} was an expedient to avoid determining the 
relationship between $L_{\rm FC}$ and the scoring plane in the simulations.

\begin{figure}[htb]
\centering
\includegraphics[width=7cm]{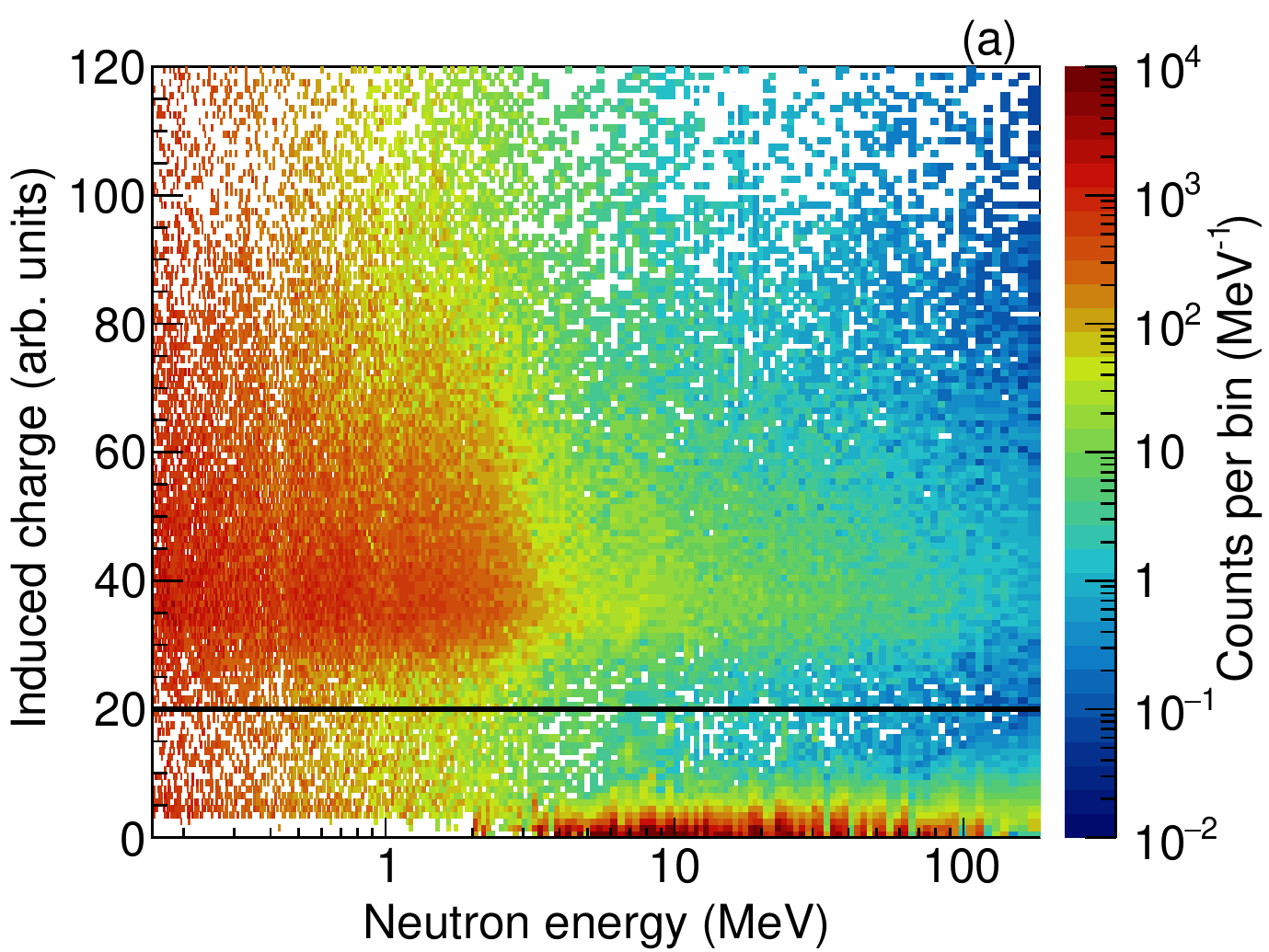}\qquad
\includegraphics[width=7cm]{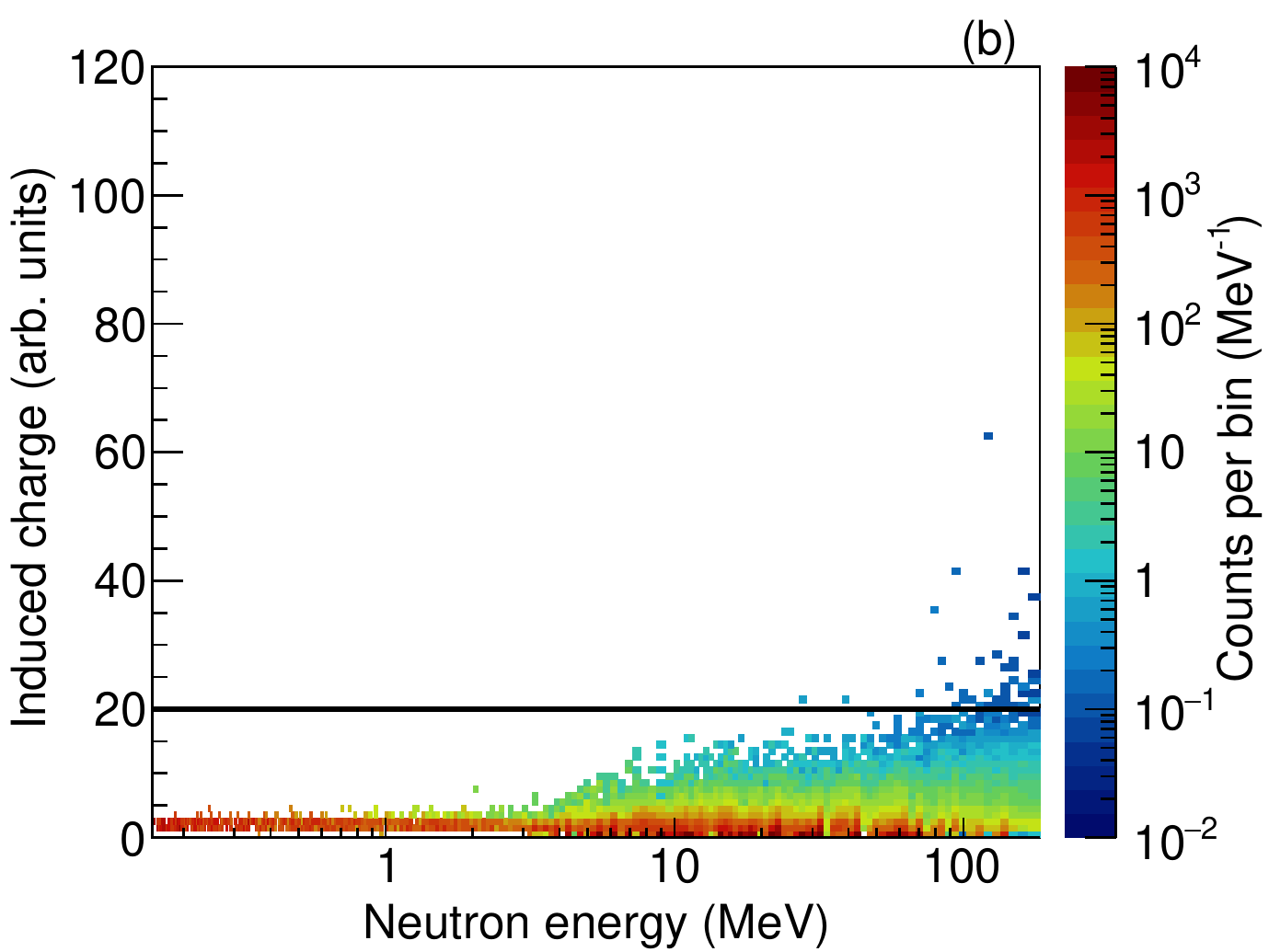}
\caption{\label{fig:FC-sort} Events recorded for a uranium target (figure (a)) 
and a blank target (figure (b)) are sorted according to the integral of the 
detector pulses (the induced charge) and the neutron energy determined using the
time-of-flight technique. The black lines represent the threshold used to 
separate real events (fission fragments or particles from n-Al reactions) from 
noise.}
\end{figure}

In figure~\ref{fig:FC-sort} the pulses recorded for the uranium target U1 and 
the blank target B1 are sorted by area, or induced charge, and neutron energy. 
The residual noise that the baseline subtraction did not completely suppress
is still noticeable, however the fission fragments are clearly separated. 
In case of the blank target, the events above threshold (the red line in the 
figure) are the product of neutron-induced light charged particle emission from
aluminum. The threshold that was chosen to separate noise from (n,f) events
(and light charged particles emitted in n-Al reactions) was different for every 
target as the noise level and the settings of the electronic chain were slightly 
different for each read-out channel.

The number of fission fragments $N_{{\rm f},i}^{\rm det}$ detected for a given 
uranium sample $i$ as a function of time-of-flight $t_{\rm n}$ or kinetic energy
$E_{\rm n}$ was obtained by summing the events with area above the noise 
threshold $A_{0,i}$, and subtracting the contribution from the blank targets: 
\begin{equation}
\label{eq:FC-counts}
    N_{{\rm f,}i}^{\rm det} (t_{\rm n}) = 
    k_{{\rm \tau,}i}(t_{\rm n})  \; C_i(t_{\rm n}) 
     - \frac {1}{2} \sum_{j=1,2} k_{{\rm \tau,}j}(t_{\rm n}) \;  B_j(t_{\rm n})
\end{equation}
where $C_i$ and $B_j$ denote the integrated number of events above the 
pulse-height threshold $A_{0,i}$ for the uranium sample $i$ and the two blank 
samples, respectively, and $k_{\rm \tau}$ is the dead-time correction.

\begin{figure}[htb]
\centering
\includegraphics[width=7cm]{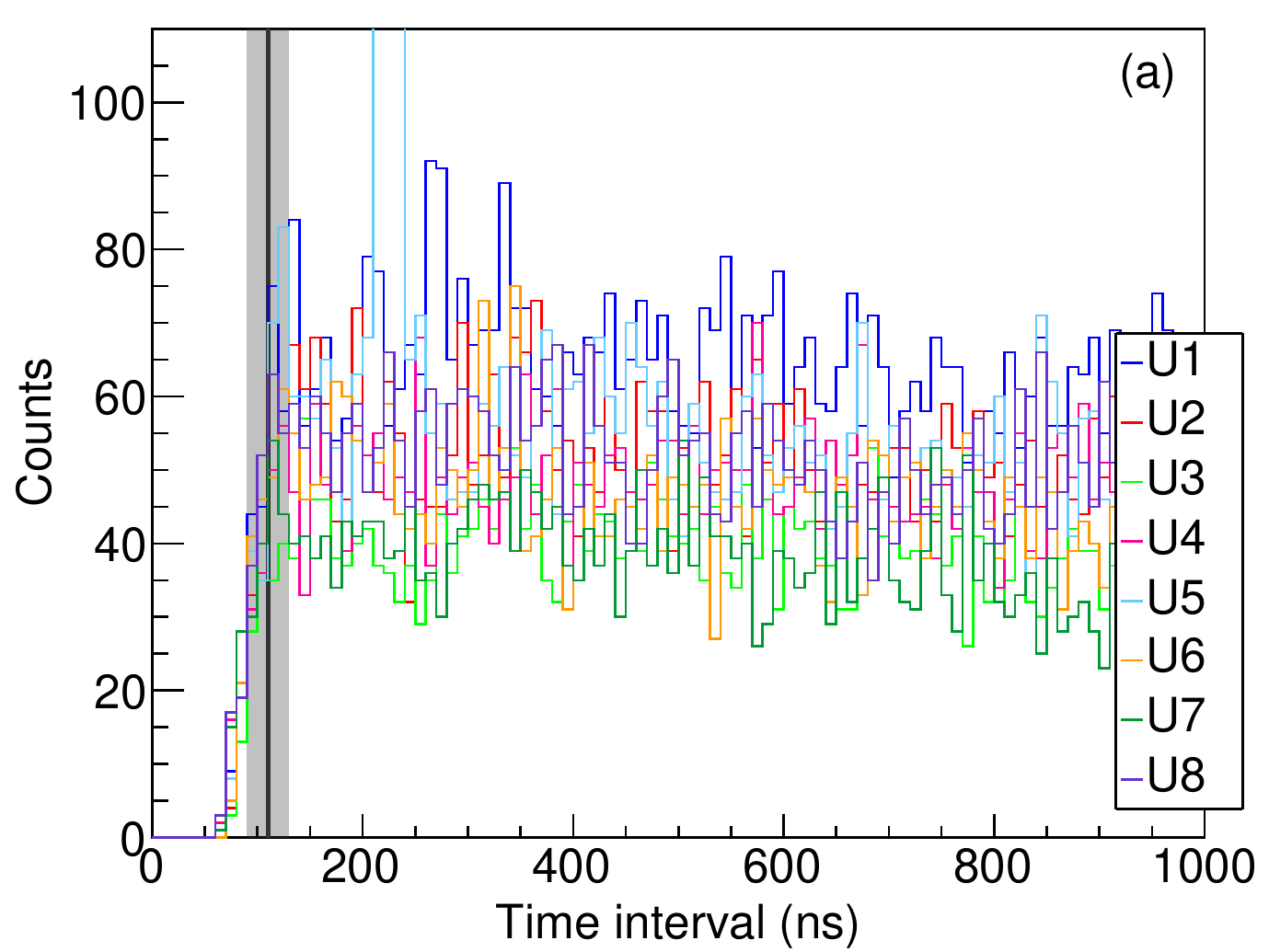}\qquad
\includegraphics[width=7cm]{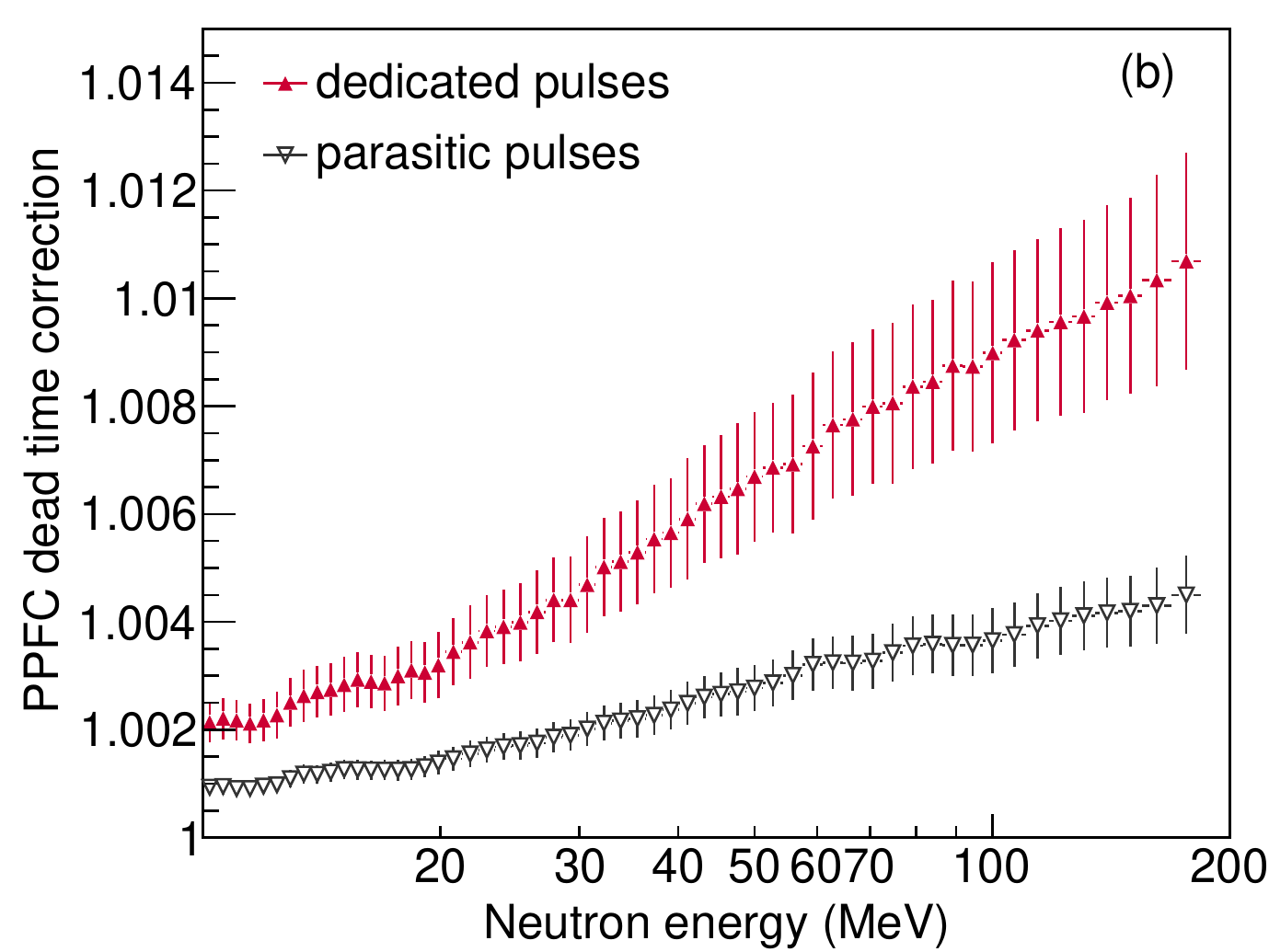}
\caption{\label{fig:FC-Deadtime} 
Panel (a): time interval distribution of the events recorded for each of the 
eight uranium samples mounted in the PPFC. The vertical line indicates the 
dead-time (and the shadow its uncertainty), defined as the shortest time for 
which the recorded distribution differs less than 10\,\% from the 
expected exponential behavior. 
Panel (b): dead time correction $k_{\tau}$ applied to the 
PPFC data; average for the eight uranium samples. The correction was 
calculated following the method of Whitten \cite{WHI91}.}
\end{figure}

The dead-time correction was determined as a function of the time of flight 
using the method outlined in \cite{WHI91}. It is based
on the assumption that all neutron pulses have the same intensity, which 
means that events produced by parasitic and dedicated PS proton pulses were 
treated separately.
The dead time of 110(20)~ns was obtained from the distribution of the time 
intervals between consecutive events, shown in figure~\ref{fig:FC-Deadtime}(a),
by taking the shortest time for which the recorded time-interval distribution 
differed less than 10\,\% from the expected exponential behavior.
There were slight variations from sample to sample, but they all fell within 
the uncertainty.
In figure~\ref{fig:FC-Deadtime}(b) the average correction for the eight uranium 
samples is shown for both dedicated and parasitic pulses. The effect ranges 
from 0.1\,\% at low energy up to 1.1\,\% at high energies, with an absolute 
uncertainty of 0.2\,\% at worst.

The fission reaction yield, normalized by the number of \textsuperscript{235}U 
atoms per unit area, was then calculated as:
\begin{equation}
\label{eq:FC-yield}
    Y_{\rm f}(t_{\rm n}) = \frac
    {\sum_{i=1}^{8}  k_{{\rm thr,}i}(t_{\rm n}) \; N_{{\rm f},i}^{\rm det}(t_{\rm n}) }
    {\sum_{i=1}^{8} \epsilon_{{\rm f,}i}(t_{\rm n}) \; n_{{\rm U,}i}}
\end{equation}
where $n_{{\rm U,}i} = k_{{\rm U,}i}\; m_{{\rm U,}i}\; N_{\rm A} / \mu_{\rm U} $. 
Here, $N_{\rm A}$ denotes the Avogadro number and $\mu_{\rm U}$ the atomic 
mass of \textsuperscript{235}U. The number of uranium atoms per unit area 
$n_{\rm U}$ and its correction $k_{\rm U}$ were discussed in 
section~\ref{subsec:FC-Targets}. The detection efficiency for fission fragments 
is the combination of two components, the zero-bias efficiency 
$\epsilon_{{\rm f,}i}$, introduced in section~\ref{subsec:FC-Efficiency}, and 
threshold-dependent correction factor to the efficiency $k_{{\rm thr,}i}$, 
which takes into account the fact that the fission fragments distribution 
extends also below the threshold used for the separation from the noise. 
The fission reaction yield $Y_{\rm f}$ of the eight deposits replaces the 
product $N_{\rm f} / (\epsilon_{\rm f} \; n_{\rm U} \; \bar{m}_{\rm U})$ in 
equation \ref{eq:Cross-section}.

To extrapolate the number of fission events below threshold, the Monte Carlo 
model for the transport of the fission fragments through the chamber 
(section~\ref{subsec:FC-Efficiency}) was used to obtain the shape of the pulse 
height distribution as function of the incident neutron energy. 
For each sample, and for each energy bin, the simulations were fit to the 
experimental histograms using only the events above the threshold. The results 
were then used to calculate the number of fission events below threshold and 
$k_{\rm thr}$.
The fit was carried out using ROOT-v6.18/04 \cite{BRU97} because it allows
to analyze large amount of data efficiently. To quantify the uncertainty,
however, the same fit was repeated, for a selected number of neutron energies, 
using also WinBUGS \cite{LUN00}. WinBUGS is Bayesian analysis software that uses 
Markov Chain Monte Carlo to fit statistical models, as for example measurement 
error models. The main advantage is that it produces the probability 
distributions of the model parameters, not only point estimates, which means 
that compared to analytical methods, the determination of the uncertainties is 
more robust. 

\begin{figure}[htb]
\centering
\includegraphics[width=7cm]{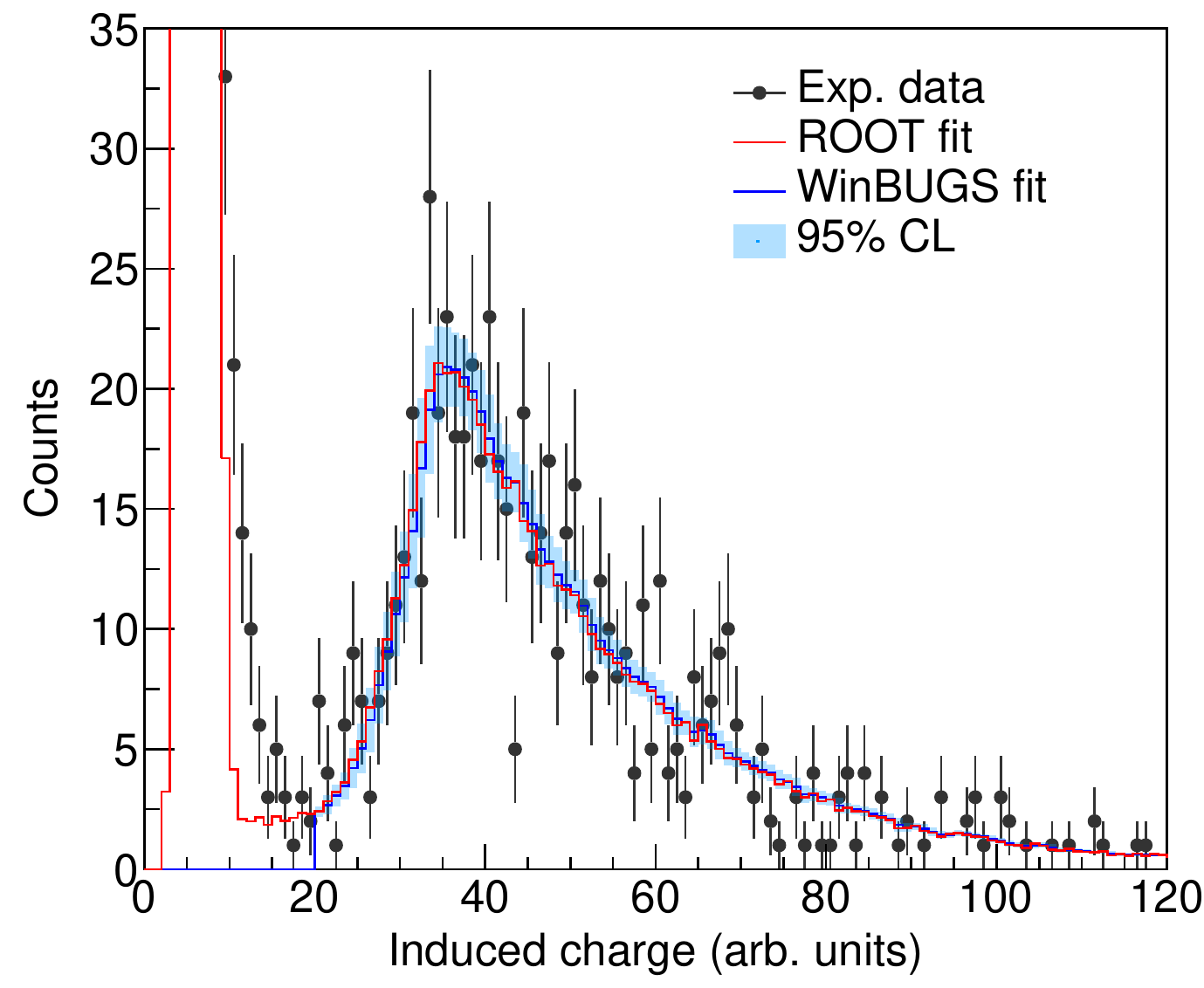}
\caption{\label{fig:FC-WinBUGS} Fit for one of the PPFC uranium targets of the 
pulse height distribution measured at the thermal point to that corresponding to
incident neutrons with energies from 103 to 110~MeV. The fit was performed with 
both Root and WinBUGS; the 95\,\% confidence level (CL) was determined with 
WinBUGS.}
\end{figure}

Figure~\ref{fig:FC-WinBUGS} shows for example the result of the fit
of the pulse height distribution measured with thermal neutrons to that 
corresponding to incident energies from 103~MeV to 110~MeV. 
The fit was performed with both Root and WinBUGS, and the experimental thermal 
distribution was used instead of simulated energy-dependent distributions 
for the sake of simplicity. The fit model included a 
normalization factor, to match the number of events above threshold, and a 
linear adjustment of the horizontal scale (gain and offset), which was 
introduced to estimate if the gamma-flash-induced noise had any effect on the 
shape of the distribution at high energies.

The fits performed with the two codes were always in agreement within the 95\,\% 
confidence level determined by WinBUGS. Moreover, within the uncertainties 
calculated with WinBUGS, there was no evidence of changes in gain or offset at 
all energies. The uncertainty on the normalization factor, which ultimately 
defines the uncertainty of the extrapolation of the events below threshold, was
found to range from 2\,\% to 4.5\,\%; on average it was determined to be 
about 3\,\%.

%-------------------------------------------------------------------------------

\section{Recoil proton telescope for flux measurements}
\label{sec:RPT}

\subsection{Design}
\label{subsec:RPT-Design}

   The design of the present RPTs was inspired by earlier experiments at 
\lq white\rq\ \cite{DOL86} and quasi-monoenergetic \cite{SCH99,DAN01} neutron sources. 
In the latter two experiments, the use of a 
triple-coincidence requirements was crucial to suppress background and provide a
clear signature for recoil proton events. For the present experiment, the 
surface barrier detectors, proportional counters and multiwire chambers used in 
 \cite{SCH99} and \cite{DAN01} were replaced by fast plastic scintillators
to reduce the sensitivity to photons from the gamma flash and increase the time 
resolution. 

   The transmission detectors consisted of square EJ~204 plastic scintillators. 
The lateral dimensions of the front transmission detector ($\Delta E_1$) were 
45~mm $\times$ 45~mm. The rear transmission detector ($\Delta E_2$) had lateral 
dimensions of 38~mm $\times$ 38~mm. 
Detectors with thicknesses of 0.5~mm, 1~mm, 2~mm and 5~mm were used for different energy
regions. The stop ($E$) detectors were cylindrical EJ~204 scintillators with a 
diameter of 80~mm and lengths of 50~mm, 75~mm, 100~mm and 150~mm.

   As shown in figure~\ref{fig:Setup}, the scintillators of the $\Delta E$ 
detectors were mounted in tent-like housing using 4~mm lucite pins to fix the 
scintillators to the triangular parts of the housing. The charged particles 
produced in the radiator traverse the detectors through the square parts of the 
housing which consist of 0.2~mm thick aluminum foils. The inner surfaces of the 
housings were painted with diffusively reflecting white EJ~510 paint. The 
bottom of the housings had a circular opening for conical light guides made 
from UV-transparent polymethylmethacrylate (UVT-PMMA). The distance between 
the surfaces of the light guides and the lower edge of the scintillator were 
7.5~mm ($\Delta E_1$) and 11~mm ($\Delta E_2$), respectively. The light guides 
were 45~mm long and tapered from 60~mm to 50~mm. On the narrow side they were 
coupled to Valvo XP2020Q photomultipliers (PMTs) using silicone grease.

   The cylindrical $E$ detectors were coupled to XP2020Q PMTs via conical light 
guides made from UVT-PMMA which tapered from 80~mm to 54~mm. The front face and
the curved surface of the cylinders as well as the upper part of the light 
guides were coated with EJ510 paint. The detectors were mounted in cylindrical 
thin-walled aluminum housings which had 0.2~mm aluminum front windows. 

   The optical transport of scintillation light in the detectors was 
investigated for all detectors using 511~keV $\gamma$-rays from an 
actively collimated \textsuperscript{22}Na source. The active collimation was 
achieved by requesting a coincidence with an event in a 2"$\times$2" 
BaF\textsubscript{2} detector about 1.5~m away from the source. 
Scans of the $\Delta E$ detectors along the cylindrical axis of the light guide
and perpendicular to the square surface of the scintillator showed that the 
light collection efficiency decreased linearly by about 12\,\% from the bottom 
of the scintillator to the top. 
Perpendicular to this direction and along the square surface the variation of 
the collection efficiency was 6\,\%. For the 100~mm $E$ detector, a scan along
the cylindrical axis showed a linear decrease of the collection efficiency of 
about 20\,\% from the front side to the PMT.
These data went into the Monte Carlo models used to calculate the efficiency of
the RPTs.

   The different detectors were combined to four RPTs for different energy
ranges (see table~\ref{tab:RPT-config}). The distance of the center of the 
$\Delta E_1$ and $\Delta E_2$ detectors from the center of the radiator were 
206~mm and 355~mm, respectively. The front surface of the $E$ detector was 
located 55~mm behind the center of the $\Delta E_2$ scintillator.
In all RPT configurations, the $\Delta E_2$ scintillator had the smallest 
cross-sectional area and determined the solid angle covered by the RPT when 
triple coincidences were requested. The geometric trajectories from any point on 
the radiator surface within the neutron beam profile to any point on the surface
of the $\Delta E_2$ scintillator were always hitting the surface of the 
$\Delta E_1$ scintillator, i.e. a loss of coincident events can only occur due to 
angular straggling which was accounted for in the Monte Carlo models of the 
RPTs. The large cross-sectional area of the $E$ detector compared with the 
$\Delta E_2$ detector helped to reduce incomplete energy deposition in the 
$E$ detector, especially at higher proton energies. The large spatial separation
between the $\Delta E_1$ and $\Delta E_2$ detector effected an angular 
acceptance of the RPTs of about $\pm$15~degree in the scattering plane which 
helped to suppress background originating from other parts of the experiment 
outside the radiator, e.g. from the PPFC or the two detectors of the 
high-energy experiment.

\begin{table}[htb]
\centering
\caption{\label{tab:RPT-config} RPT configurations used for the low-energy 
experiment. \\
\textsuperscript{1)} This configuration could not be employed for the final 
experiment because of time restrictions.}
\smallskip
\begin{tabular}{|l|l|ll|c|c|c|}
\hline
Config. & Energy range & \multicolumn{2}{c|}{Radiators} & $\Delta E_1$ & $\Delta E_2$ & $E$ \\
\quad   &\quad        & \multicolumn{2}{c|}{\quad}     & mm & mm & mm                       \\
\hline
1 & 25 MeV - 100 MeV & PE 1 mm,  & C 0.5 mm & 0.5 & 0.5 &  50 \\
2 & 35 MeV - 100 MeV & PE 2 mm,  & C 1 mm   & 1   & 1   &  50 \\
3 & 50 MeV - 150 MeV & PE 5 mm,  & C 2.5 mm & 2   & 2   & 100 \\
4 & 75 MeV - 200 MeV \textsuperscript{1)} 
                & PE 10 mm, & C 5 mm  & 2   & 5   & 150 \\
\hline
\end{tabular}
\end{table}

\begin{figure}[htb]
\centering
\includegraphics[width=7cm]{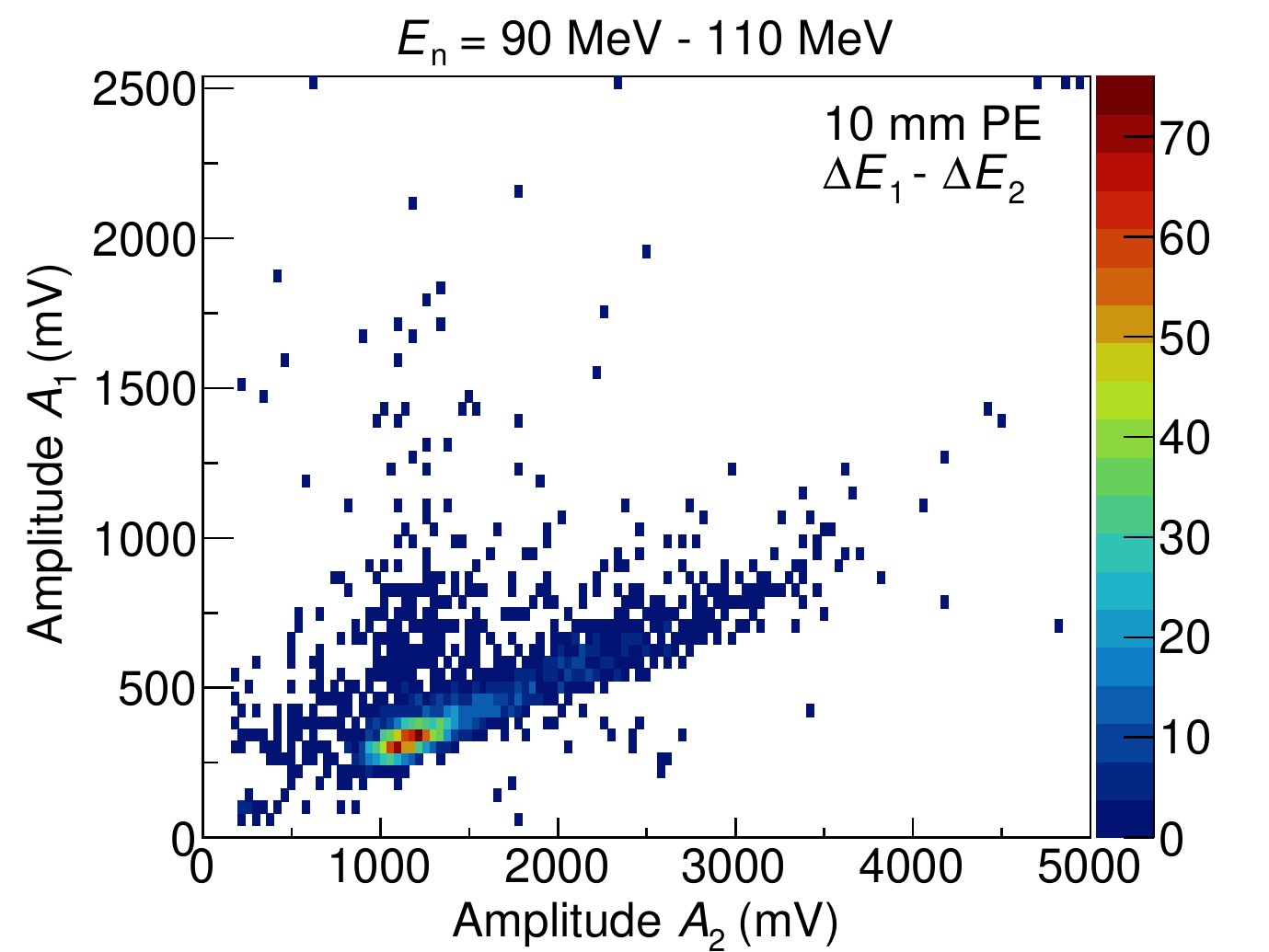}\qquad
\includegraphics[width=7cm]{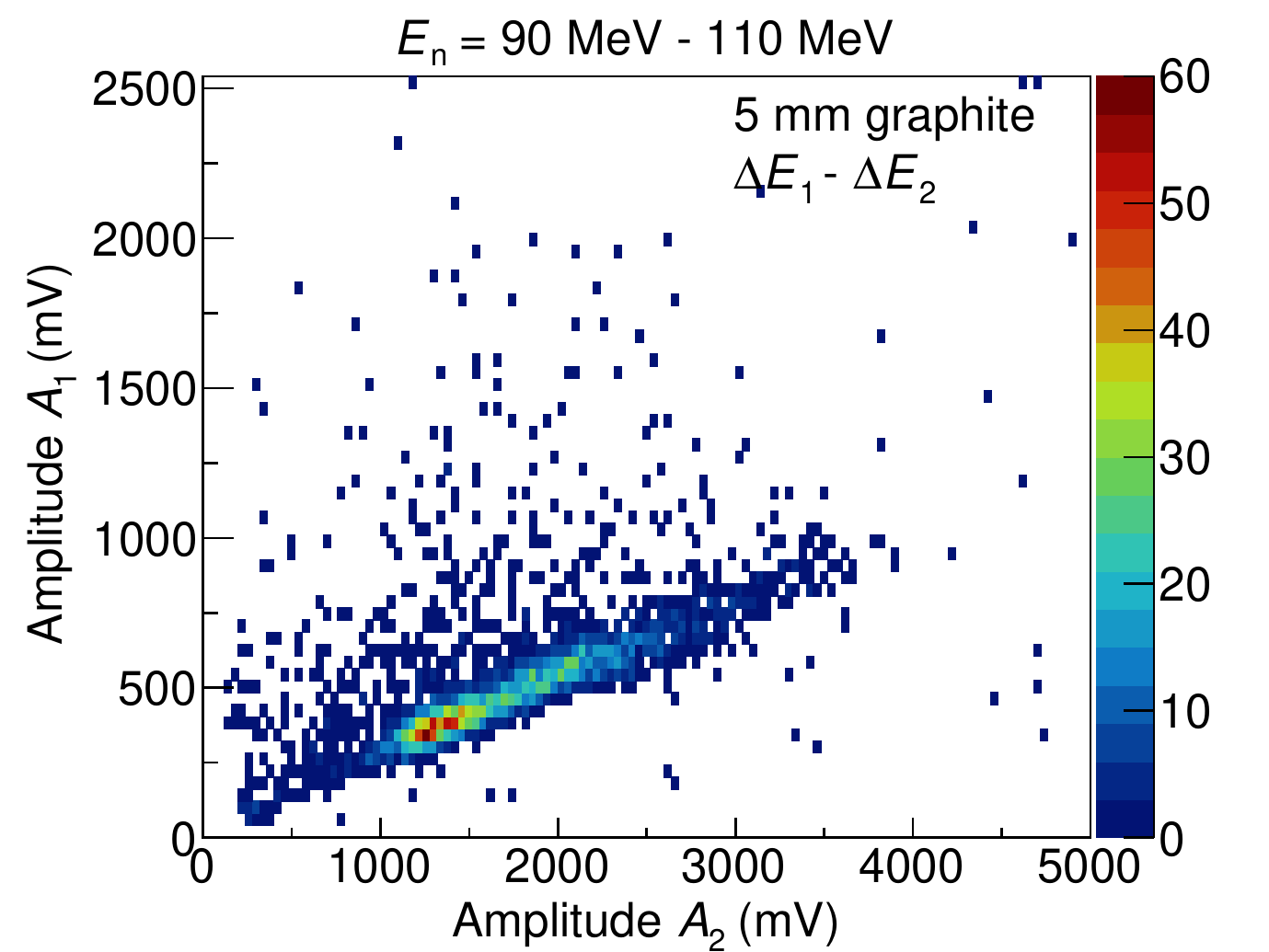}\qquad
\includegraphics[width=7cm]{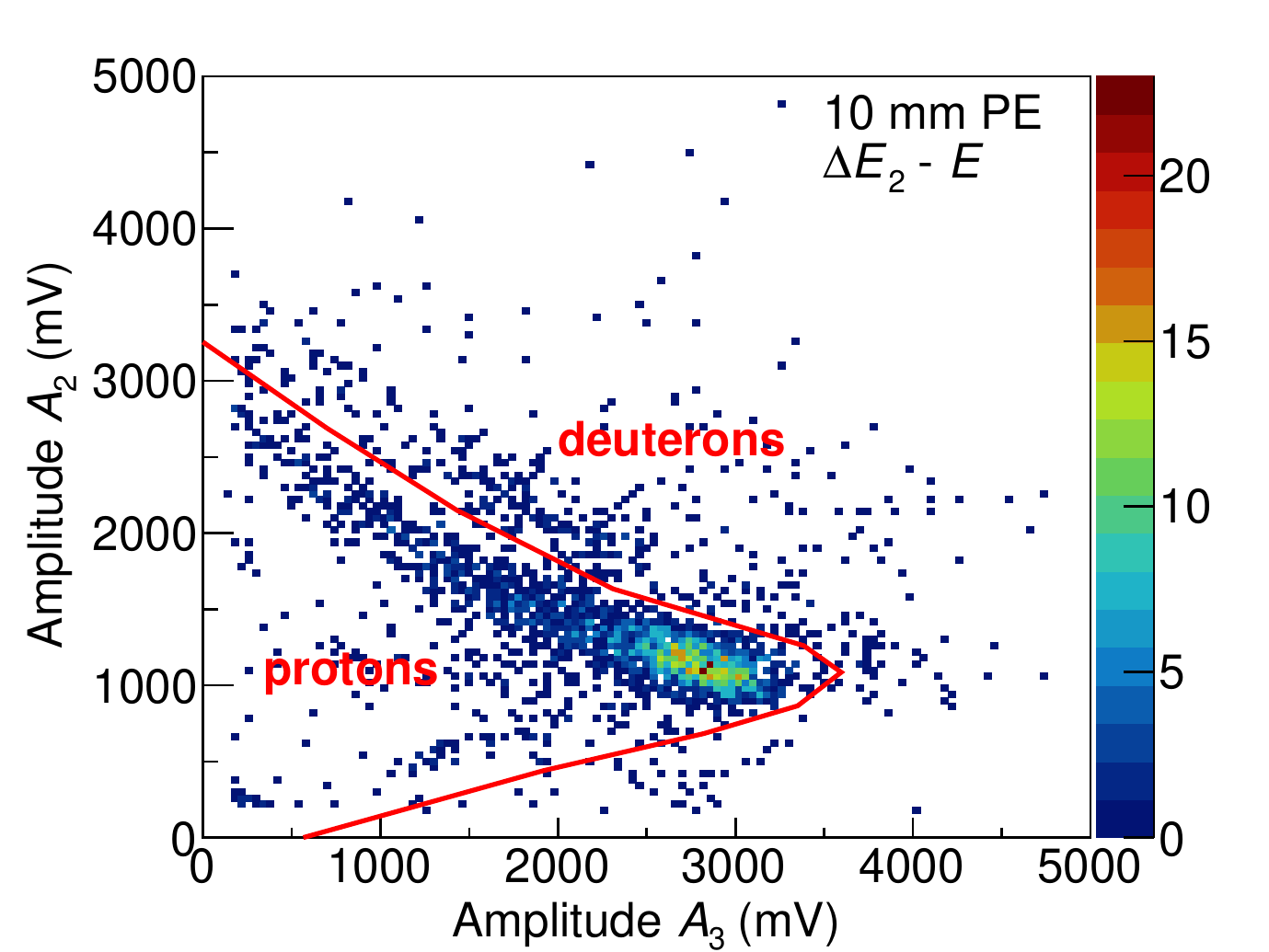}\qquad
\includegraphics[width=7cm]{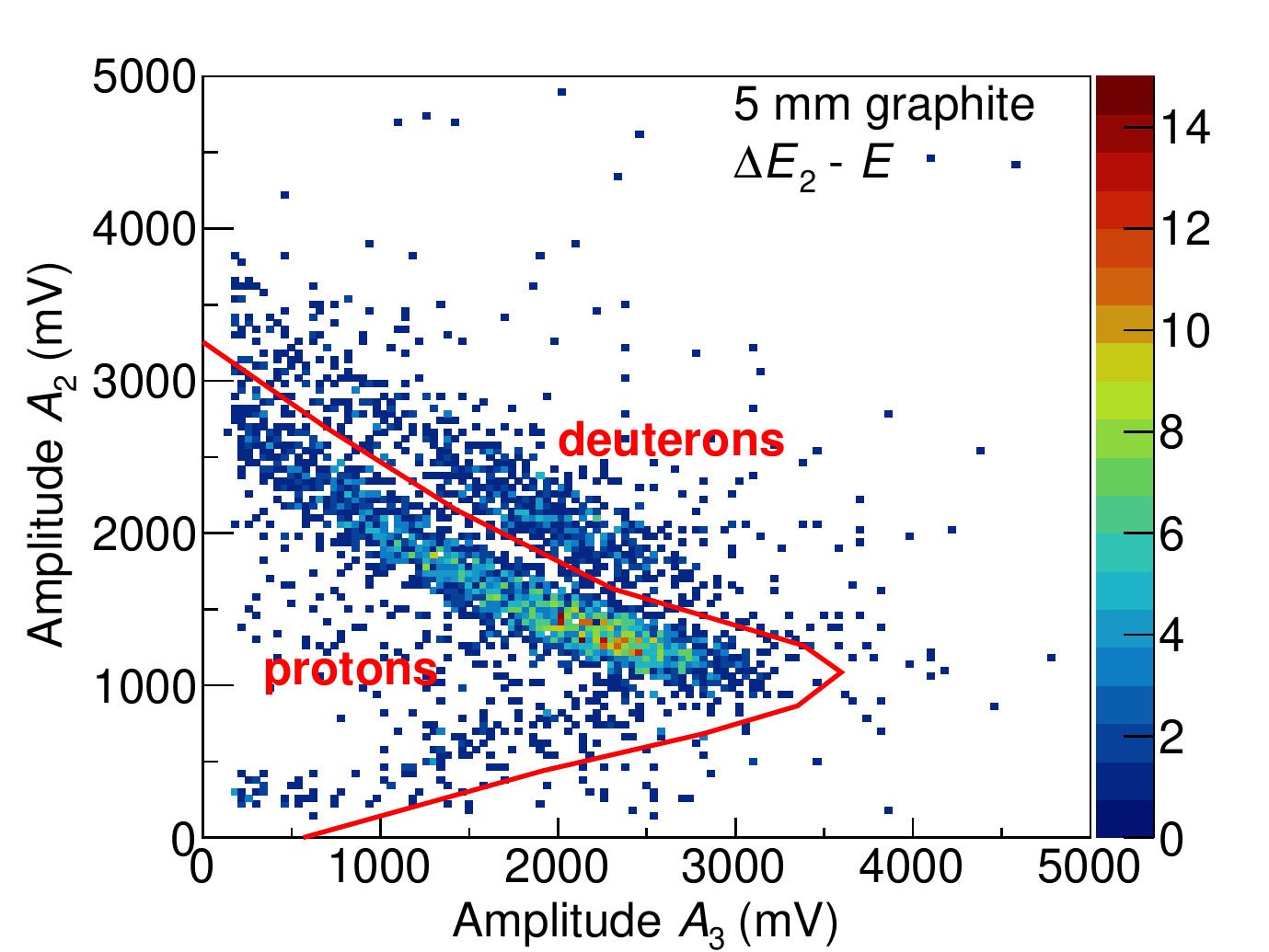}
\caption{\label{fig:RPT-E-DE-10mm} $\Delta E_1 - \Delta E_2$ and 
$\Delta E_2 - E$ event distributions for a 2~mm - 5~mm - 75~mm detector 
combination and 10~mm polyethylene (left panel) and 5~mm graphite (right panel) 
radiators. The neutron energy region extended from 90~MeV to 110~MeV. 
The red solid line shows the separation between proton- and deuteron-induced 
events.}
\end{figure}

   Figure~\ref{fig:RPT-E-DE-10mm} shows the $\Delta E_1-\Delta E_2$ and
$\Delta E_2-E$ event distributions obtained during the test experiments with a 
2~mm - 5~mm - 75~mm detector combination and 10~mm polyethylene and 5~mm 
graphite radiators. The neutron energy regions were restricted to windows from 
90~MeV to 110~MeV. The event distributions are clearly separated from the 
base lines, i.e. loss of events close to the noise level can be excluded. 
The ridges in the $\Delta E_2 - E$ distributions with proton events from n-p 
scattering and \textsuperscript{12}C(n,px) and deuteron events from 
\textsuperscript{12}C(n,dx) are clearly visible although the separation of 
protons from deuterons is not perfect due to the spatial variation of the 
light collection efficiency. Therefore, a conservative separation cut must 
ensure that no protons events are lost. The remaining deuteron events 
from \textsuperscript{12}C(n,dx) reactions leaking into the proton region must 
be subtracted together with the proton events from \textsuperscript{12}C(n,px) 
using the graphite sample.  

   A crucial aspect of for the calculation of the neutron detection efficiency 
of the RPTs is the definition of the solid angle using the $\Delta E_2$ 
detector. Due to edge effects, i.e. incomplete energy deposition by protons 
leaving the detector through the edges, such events could be lost below the 
pulse-height threshold. As a consequence, the effective cross-sectional area of 
the detector would be smaller than the geometrical one. Such effects were 
studied in more detail using single-ion proton beams at the PTB micro 
ion beam facility. 

\begin{figure}[htb]
\centering
\includegraphics[width=7cm]{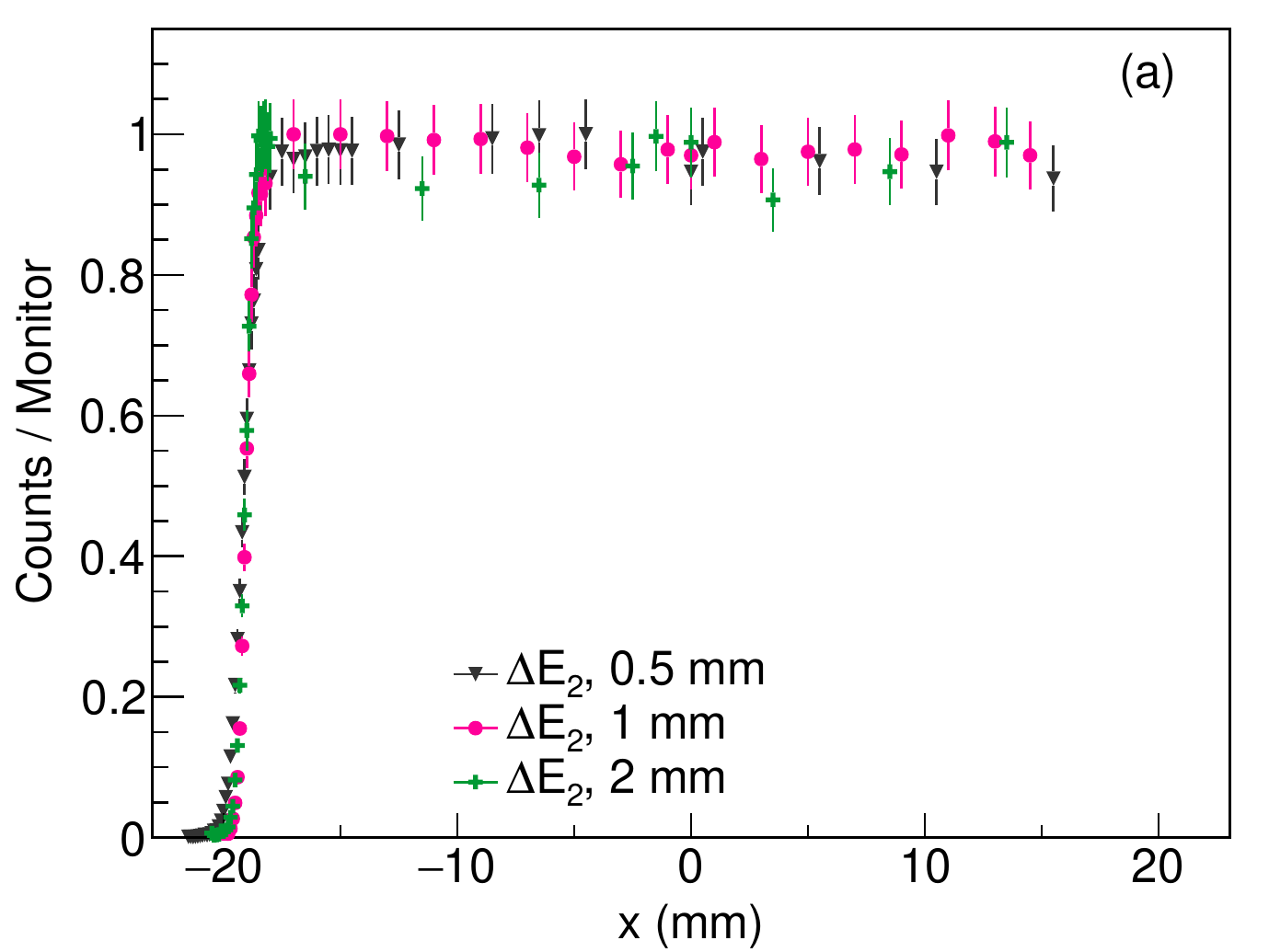}\qquad
\includegraphics[width=7cm]{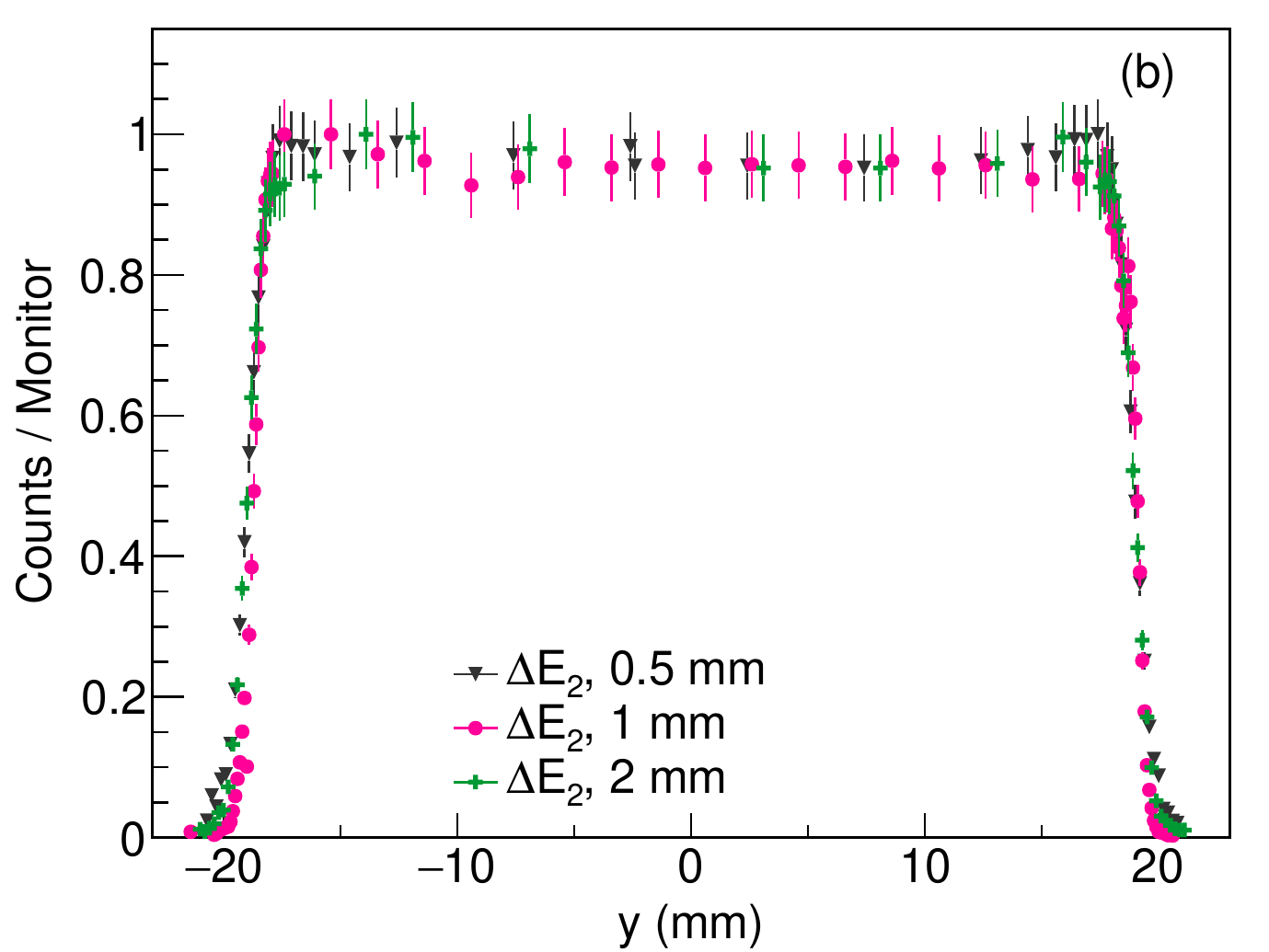}
\caption{\label{fig:RPT-DE2-profile} Measurements of the $\Delta E_2$ 
scintillator profiles performed at the PTB micro beam facility. The
detectors have lateral dimensions of 38~mm $\times$ 38~mm, and thickness
of 0.5~mm, 1~mm, or 2~mm (as written in the figure). The $x-y$ plane (the 
detector square face) was normal to the proton beam direction.
The $x$-axis (panel~(a)) corresponds to the axis of the PMT, with the PMT 
placed at $x > 19~\rm{mm}$; the $y$-axis (panel~(b)) is perpendicular to 
it. The measurements along the $x$-axis were not be performed edge-to-edge
but were stopped mid-way because otherwise the detector casing would have 
bumped against the beamline.}
\end{figure}

At this facility magnetic lenses are used to focus a 
charged particle beam, protons or $\alphaup$ particles, to a diameter of a few 
micrometers. To scan the $\Delta E_2$ scintillators, a proton beam of 
15~MeV was used. The detectors were positioned in air at the distance of 
a few centimeters from the exit window of the beamline. 
They were mounted on a motorized stand programmed to move in steps of 
10~{\micro m} along the axes $x$ and $y$ of the plane normal to the beam 
direction. Along the $x$-axis, which corresponded to the PMT axis, the scan 
was not performed edge-to-edge but had to be stopped mid-way because 
otherwise the detector housing would have hit the beamline.
The scintillator count rate was recorded, normalized by the beam 
monitor, as a function of the ($x$,$y$) coordinates. 
In figure~\ref{fig:RPT-DE2-profile}, the profiles along the $x$ and $y$ 
axes are shown for the $\Delta E_2$ scintillators of 0.5~mm, 1~mm, and 2~mm 
thickness. These were obtained by averaging the single measurement points at a 
given $x$ or $y$, excluding the corners of the detectors.
These results were therefore included in the model for the calculation of the RPT 
efficiency, as described in section~\ref{subsec:RPT-Efficiency}.

\subsection{Polyethylene radiators}
\label{subsec:RPT-Radiators}

Table~\ref{tab:RPT-Radiators} summarizes the main characteristics 
of the polyethylene and graphite samples used as radiators 
during the \textsuperscript{235}U(n,f) cross section measurement 
at the upstream position. 

\begin{table}
\centering
\caption{Summary of the properties of the RPT radiators at the upstream 
position. The samples are labeled with PE (for polyethylene) or C (for graphite) 
and their nominal thickness.
}
\label{tab:RPT-Radiators}
\begin{tabular}{|c|c|c|cc|}
\hline
Sample   & Thickness & Density    & \multicolumn{2}{c|}{Areal density} \\
\quad    & (mm)      & g/cm$^3$   & g/cm$^2$   & (rel. unc.)        \\
\hline
PE 1~mm  & 1.025(4)  & 0.9534(20) & 0.0978(4)  & (0.4\%) \\
PE 2~mm  & 1.824(11) & 0.9555(20) & 0.1743(11) & (0.6\%) \\
PE 5~mm  & 4.925(4)  & 0.9597(20) & 0.4726(11) & (0.2\%) \\
C 0.5~mm & 0.500(4)  & 1.7749(27) & 0.0887(8)  & (0.9\%) \\
C 1~mm   & 1.000(5)  & 1.7364(86) & 0.1736(12) & (0.7\%) \\
C 2.5~mm & 2.500(4)  & 1.7512(32) & 0.4378(11) & (0.3\%) \\
\hline
\end{tabular}
\end{table}

The polyethylene samples were characterized at PTB.
The density was measured by the Working Group \lq Solid State Density\rq\  
by hydrostatic weighing. The thickness of the samples was measured by the 
Working Group \lq Scientific Instrumentation\rq\ using two touch probes 
(manufactured by HEIDENHAIN) and a measuring stand (Mahr), which allowed to 
map the sample lateral profile by measuring it from both sides simultaneously.
Since the information of the orientation of the radiator during the beamtime was 
lost, the thickness values in table~\ref{tab:RPT-Radiators} were obtained by 
calculating the average thickness over the surface excluding the 
measurement points within 1~cm from the border. The uncertainty was obtained as 
the difference between maximum and minimum value.

The elemental composition and the stoichiometric relation of hydrogen to 
carbon atoms was independently measured by the ZEA-3 unit at the 
Forschungszentrum J{\"u}lich and the Institute for Inorganic and 
Analytical Chemistry at the Technische Universität Braunschweig.
In both cases, combustion analysis was used to determine the amount of carbon and 
hydrogen. Since the analysis is destructive, even if it requires only a few 
milligrams of material, it was performed only after the cross section
measurement.
The stoichiometric ratio between hydrogen and carbon was found to be 1.98(3) and
2.00(3), i.e. it is compatible with the nominal stoichiometry within an uncertainty
of about 1.5\,\%.

The graphite samples were only required to subtract those background protons 
not resulting from n-p scattering which were detected in the pulse-height 
region covered by the recoil proton peak. 
Hence, the uncertainty of the carbon mass per unit area had a less important 
impact on the uncertainty of the fission cross section.
Therefore, dimensional measurements using a caliper and weighing on a precision 
balance were considered sufficient to determine 
the mean mass per unit area for the graphite samples.

\subsection{Efficiency}
\label{subsec:RPT-Efficiency}

   The neutron detection efficiency of low-energy recoil telescopes with very 
thin radiators and detectors can be calculated semi-analytically by integrating 
over all geometric trajectories involved in the problem, neglecting the effect 
of angular straggling \cite{SLO82}. For high-energy RPTs this is not a viable 
option because thicker radiators must be used to compensate the decrease in the 
n-p scattering cross section. Moreover, the increased ranges of charged 
particles make edge effects much more important than at lower energies. 
Therefore, a detailed Monte Carlo model is required to account for such effects. 
For the present experiment, MCNPX ver.~2.7 \cite{PEL11} was chosen 
for simulating the transport of neutrons and charged particles in the RPT. 
Only protons and deuterons were transported in these simulations because alpha 
particles from \textsuperscript{12}C(n,$\alphaup$x) reactions cannot produce triple 
coincidence events in the present RPTs because of their low kinetic energy and 
short range. 
Nuclear data for neutron and proton-induced nuclear reactions were taken from 
the LA150 library \cite{CHA99} which extends up to 150~MeV and contains tabular 
data for all nuclides of the relevant materials in the present experimental 
setup. Above 150~MeV, nuclear models (Bertini intranuclear cascade plus 
pre-equilibrium model) were used. 

   For n-p scattering, the LA150 library includes the results from the 
phase-shift solution VL40 \cite{ARN87}. Although this solution is rather old and
certainly not the \lq best \rq\ solution, it is still recommended by the 
International Nuclear Data Committee (INDC) \cite{CAR97} to be used for the 
energy range from 20~MeV to 350~MeV, mostly to guarantee consistent results of 
measurements relative to the n-p scattering cross section. 

To be consistent with the ENDF-6 format system for nuclear data, MCNPX uses 
non-relativistic kinematics to sample the emission angle and energy of recoil 
particles in elastic scattering. In the energy range above 20~MeV, however, the 
deviation from the correct relativistic description cannot be ignored. For n-p 
scattering, this manifests itself in a deviation of the \lq experimental\rq\ 
differential proton emission cross section obtained by tallying recoil protons 
from the differential cross section calculated from the recommend differential 
n-p scattering cross section in the center-of-mass system using relativistic 
kinematics. 

\begin{figure}[htb]
\centering 
\includegraphics[width=7cm]{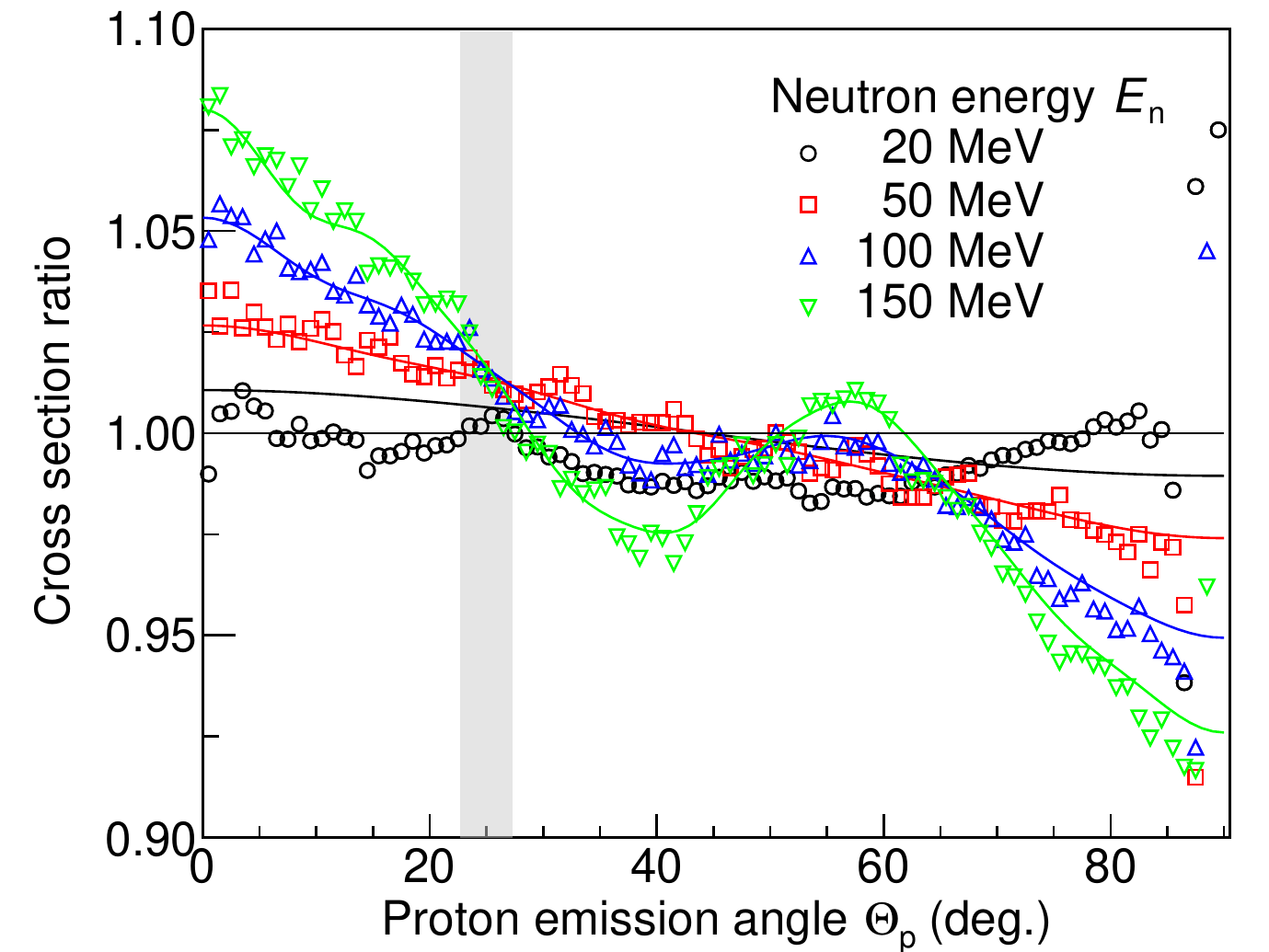} \qquad
\includegraphics[width=7cm]{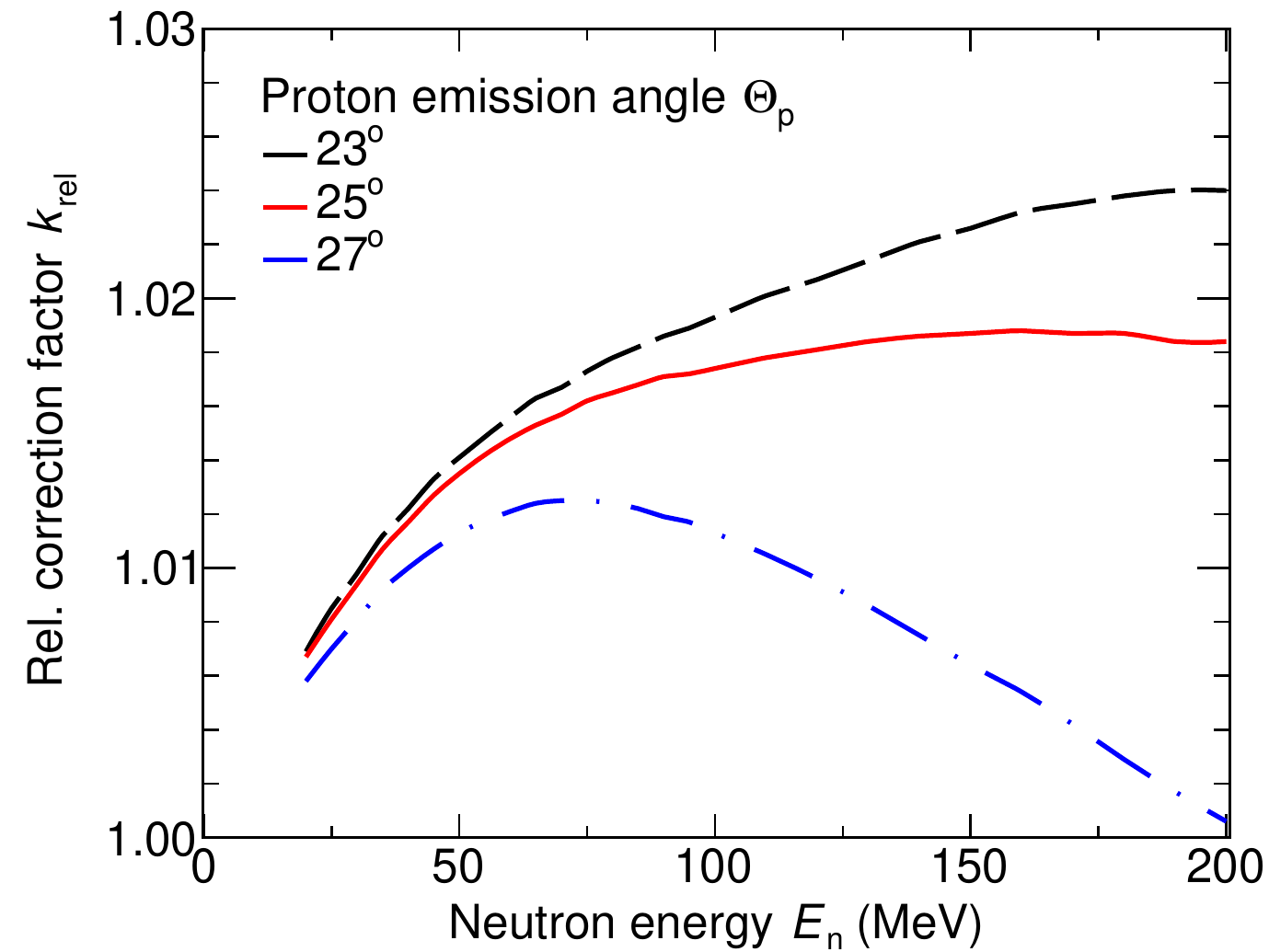} 
\caption{\label{fig:RelCor}  The left panel shows the ratio of proton emission 
cross section calculated relativistically and non-relativistically, starting 
from the VL40 cross section in the center of mass system (solid lines) or 
using MCNPX to determine the non-relativistic cross section from the tallied 
yield of recoil protons (circles). The shaded area shows the range of proton 
emission angles contributing in the RPTs. The right panel shows the corresponding 
correction factor $k_{\rm rel}$ for the MCNPX simulation (red curve) and for two 
angles corresponding to the FWHM of the distribution of scattering angles in the
RPTs.}
\end{figure}

   The left panel of figure~\ref{fig:RelCor} shows the ratio of the relativistic
differential recoil proton emission cross sections 
$({\rm d}\sigma_{\rm np}/{\rm d}\Omega_{\rm p})(\Theta_{\rm p})$ 
in the laboratory system to the non-relativistic calculation as a 
function of the proton emission angle $\Theta_{\rm p}$ in the laboratory system.
The solid lines show calculations starting from the differential VL40 cross 
section in the center-of-mass system. The symbols indicate the results of 
\lq numerical experiments\rq\ using MCNPX to determine the non-relativistic 
differential proton emission cross section from the yield of recoil protons in 
the laboratory system. It is evident that the kinematic calculation and the 
numerical experiments give consistent result. Moreover, at the laboratory proton
emission angle $\Theta_{\rm p} = 25$\textdegree, this ratio is very close to 
unity and depends only weakly on the neutron energy. The right panel shows this 
cross-section ratio as a function of neutron energy for several angles around 
25\textdegree. The data around $\Theta_{\rm p} = 25$\textdegree\, were used as 
a correction factor 
\begin{equation}
\label{eq:k_rel}
k_{\rm rel}(\Theta_{\rm p}) = \frac
     { \bigl( {\rm d}\sigma_{\rm np} / {\rm d}\Omega_{\rm p}^{\rm rel.} \bigr) (\Theta_{\rm p}) }
     { \bigl( {\rm d}\sigma_{\rm np} / {\rm d}\Omega_{\rm p}^{\rm non-rel.} \bigr) (\Theta_{\rm p}) }
\end{equation}
for the Monte Carlo simulations used to analyze the PRT data (see 
section~\ref{subsec:RPT-Analysis}). More details on this problem and a 
comparison of  MCNPX and Geant4 simulations will be presented in a forthcoming 
publication.
 
%\sidenote{check reasons for `bumps' in the solid lines of the left panel!} 

\begin{figure}[htb]
\centering 
\includegraphics[width=7cm]{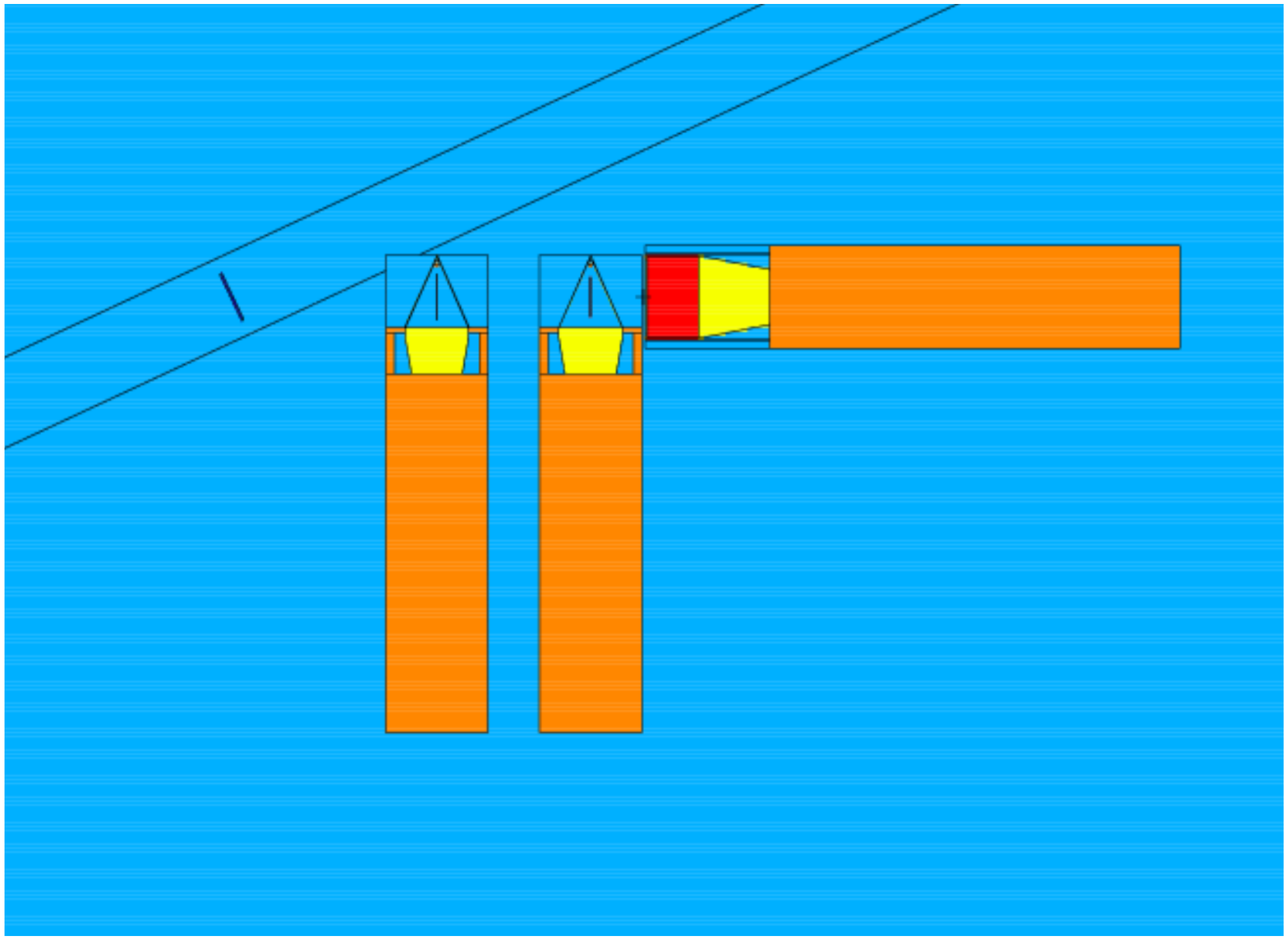} \qquad
\includegraphics[width=7cm]{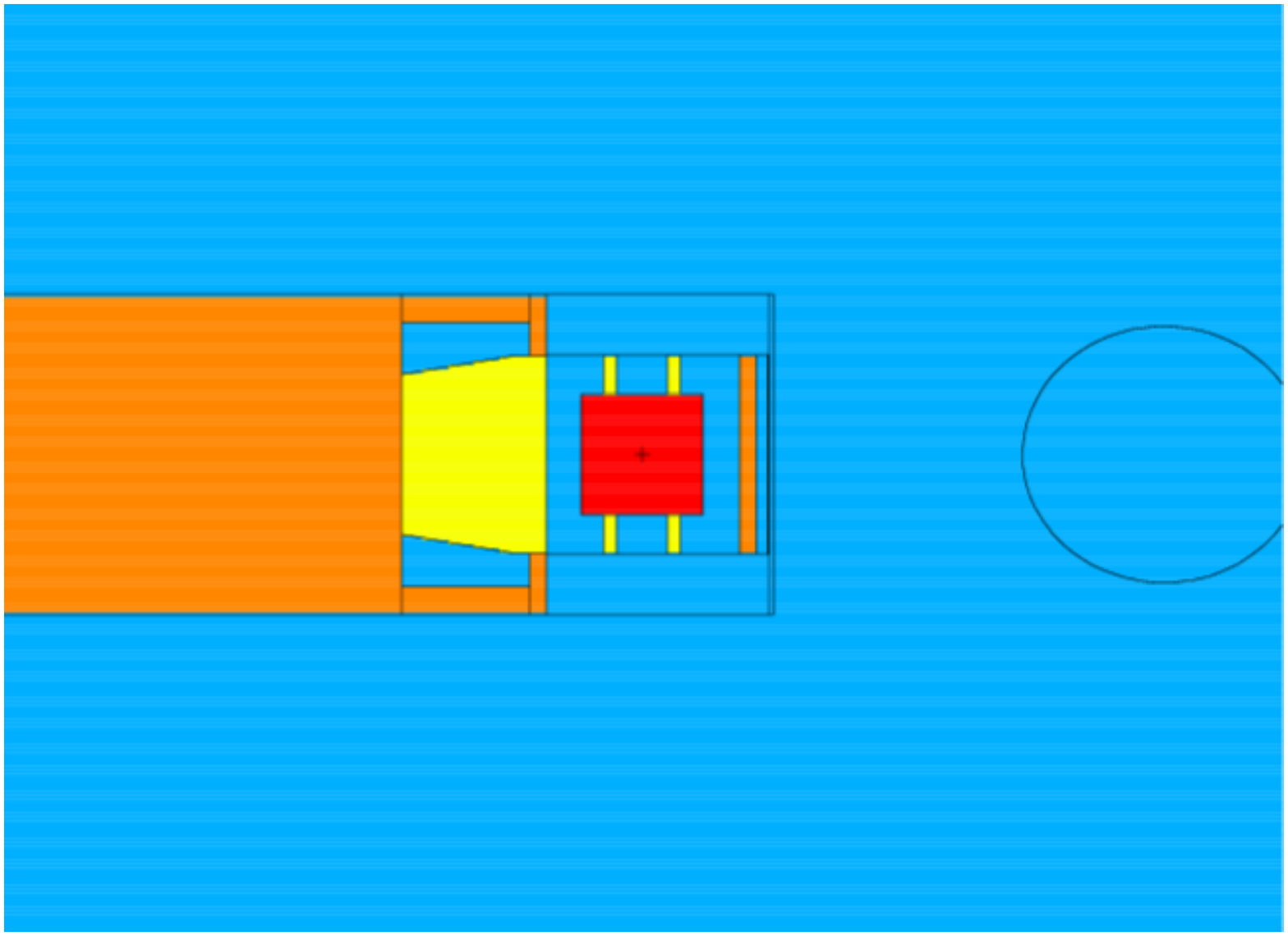} 
\caption{\label{fig:MCNPX_model} MCNPX model of the RPTs. The left panel shows 
the full PRT setup in the scattering plane. The right panel exhibits a cut 
through the $\Delta E_2$ detector perpendicular to the scattering plane. 
The black lines show cell boundaries and the colors encode different materials:
air (blue), EJ204 scintillator (red), lucite (yellow), aluminum (orange).}
\end{figure}

   Figure~\ref{fig:MCNPX_model} shows the full MCNPX model of the RPTs and 
details of the $\Delta E_2$ detector. The model contains detailed descriptions 
of those parts of the experiments which are hit by charged particles 
(radiator samples, scintillators, reflective housings and light guides of the 
$\Delta E$ and $E$ detectors) and more schematic models of the other parts. 
The simulation starts with a parallel beam of neutrons. 
The energy-dependent lateral fluence distribution of the beam is taken from
the PPAC data of the high-energy experiment. When the neutrons hit the radiator, 
an interaction is forced and the weight $w$ of the secondary particles reduced 
accordingly from unity to $w = \exp(-\Sigma_{\rm r} d_{\rm r})$, where 
$\Sigma_{\rm r}$ and $d_{\rm r}$ denote the macroscopic total cross section of 
the radiator material (polyethylene or graphite) and the thickness of the 
radiator layer traversed by the neutrons, respectively. To conserve the number of particles, 
a second \lq uncollided\rq\ neutron is produced with the same momentum vector as
the incident neutron and with the weight 
$w = 1 - \exp(-\Sigma_{\rm r} d_{\rm r})$. 
These \lq uncollided\rq\ neutrons can create secondary particles in the air 
behind the radiator and eventually produce rare events with a very high 
statistical weight but very large variance.

\begin{figure}[htb]
\centering 
\includegraphics[width=7cm]{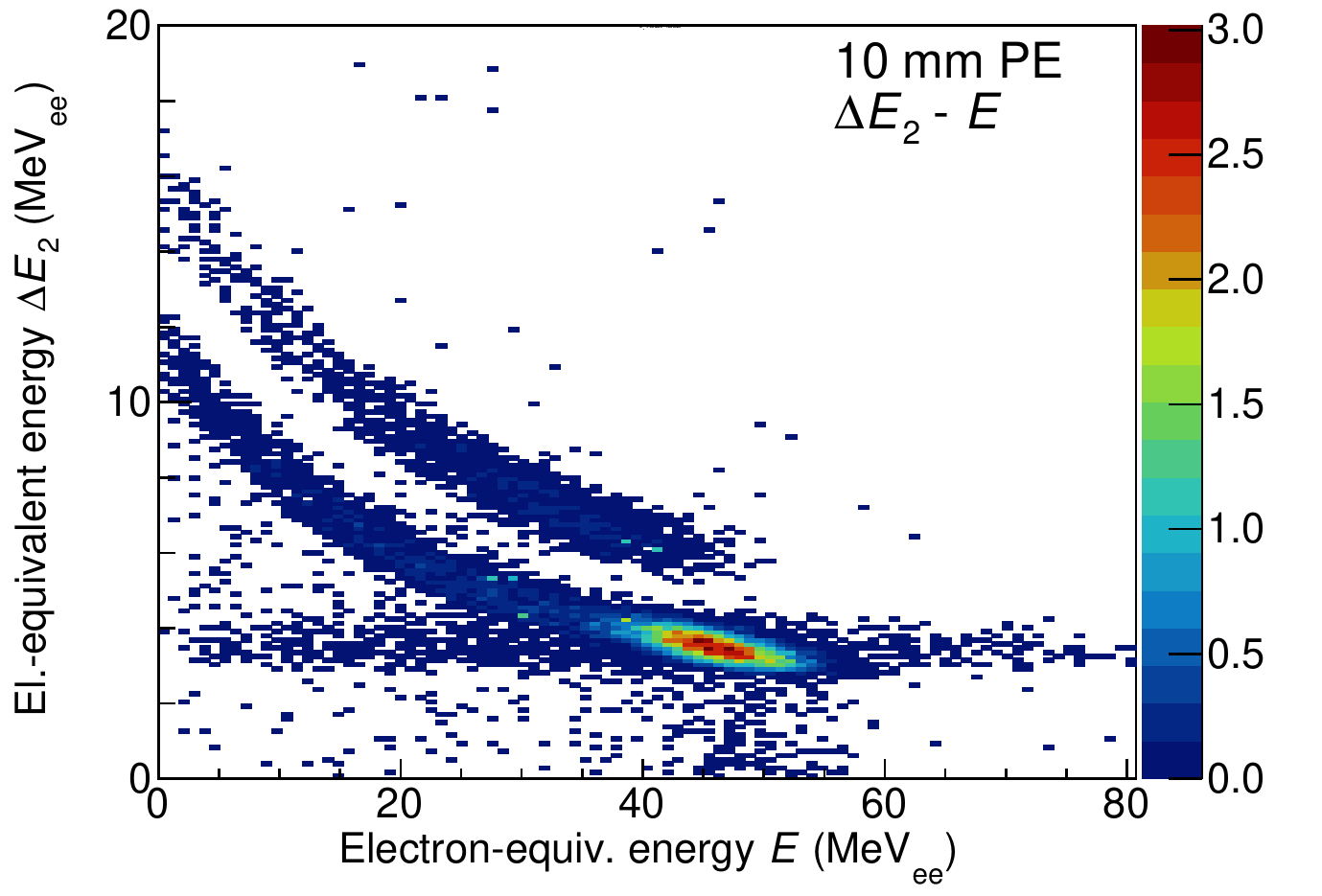}\qquad
\includegraphics[width=7cm]{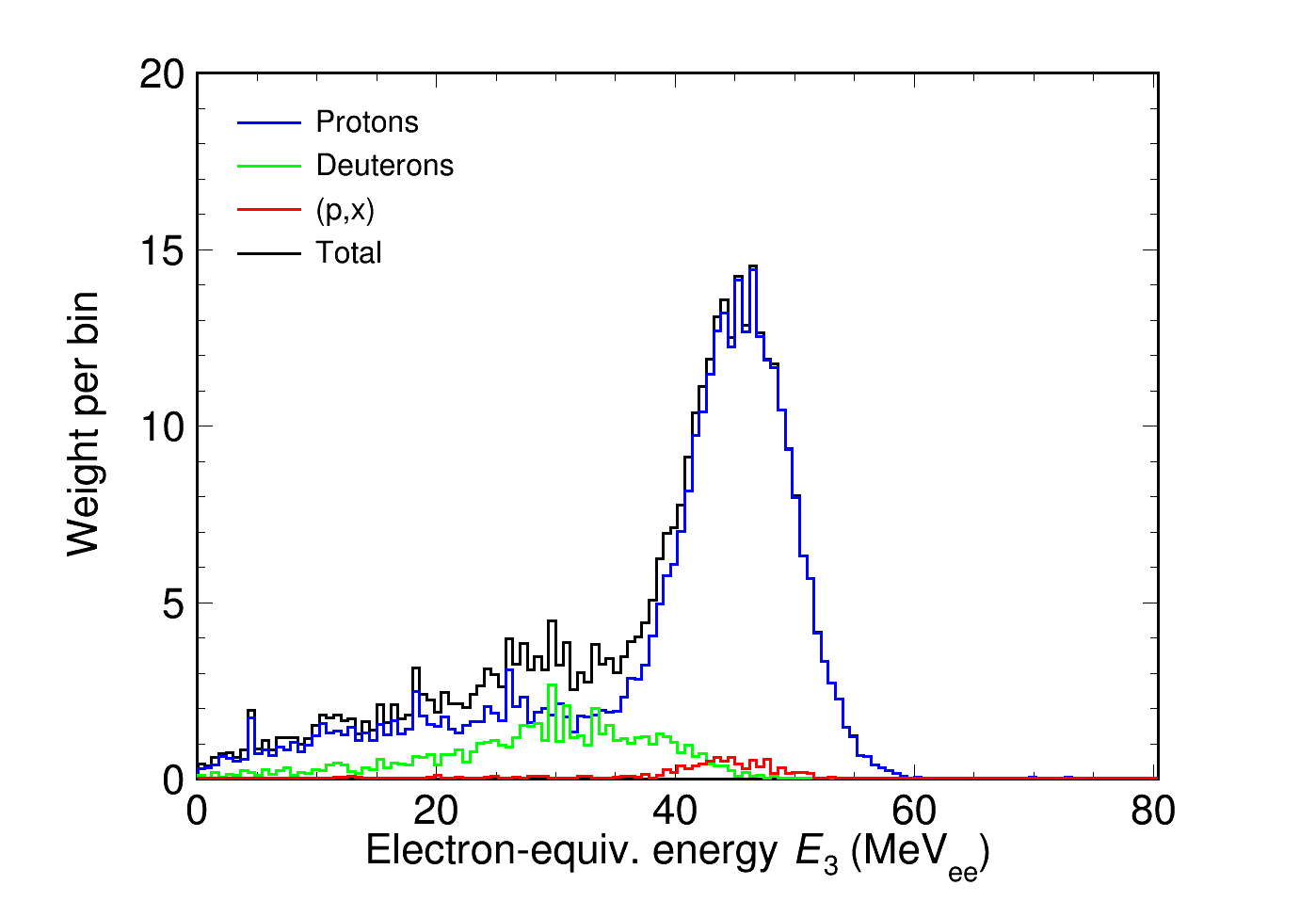} 
\caption{\label{fig:RPT_Sim_100MeV}  Simulated $\Delta E_2$-$E$ event 
distribution (left) panel and pulse-height distribution in the $E$ detector 
(right panel) for a 2~mm - 5~mm - 75~mm RPT and  a 10~mm polyethylene radiator. 
The neutron energy interval extended from 95~MeV to 105~MeV. The amount of 
scintillation light is measured in units of electron-equivalent energy using the
non-linear light output function for organic scintillators like EJ~204.}
\end{figure}

   The secondary particles are followed through the full geometry. 
The particle tracking capability of MCNPX is used to write the complete history 
of a source neutron to a \lq PTRAC\rq\ file \cite{PEL11} if at least one charged 
secondary particle hits one of the sensitive volumes of the scintillators. The PTRAC 
file is analyzed using a dedicated code which determines the coincidence 
pattern, the energy deposition and the amount of scintillation light produced in
the scintillators by each of these source neutrons. It also identifies the 
type of nuclear interaction in the radiator and tracks proton-induced nuclear 
reactions effecting a loss of valid events. The information on the spatial 
inhomogeneity of light collection and on the edge effects in the scintillator 
(see section~\ref{subsec:RPT-Design}) are also included at this stage. Finally, 
histograms of the distribution of scintillation light are incremented by the 
weight $w$ for valid triple-coincidence events and compared with the 
experimental pulse-height distributions of triple coincidence events in the $E$ 
detector (see \ref{subsec:RPT-Analysis}). As an example, 
figure~\ref{fig:RPT_Sim_100MeV} shows a $\Delta E_2$-$E$ event distribution and 
a pulse-height distribution for the $E$ detector for the same RPT and neutron 
energy bin as the experimental data shown in figure~\ref{fig:RPT-E-DE-10mm}.  
 
   It should be noted that the simulated data show a small number of events 
above the recoil peak resulting from n-p scattering. This is an artifact from 
the tabular sampling technique used by MCNPX for cross section data from nuclear
data libraries when more than one particle is emitted in a nuclear reaction. 
In such cases MCNPX does not conserve energy on an event-by-event basis 
but only on average \cite{PEL11}. 

\subsection{Data analysis}
\label{subsec:RPT-Analysis}

The n\_TOF DAQ did not offer the possibility of on-board coincidences; the
read-out channels however were all triggered using the same external 
signal coming from the PS. The three RPT detectors were therefore effectively 
synchronized and the time difference between events could be used to identify 
triple coincidences.

\begin{figure}[htb]
\centering
\includegraphics[width=7cm]{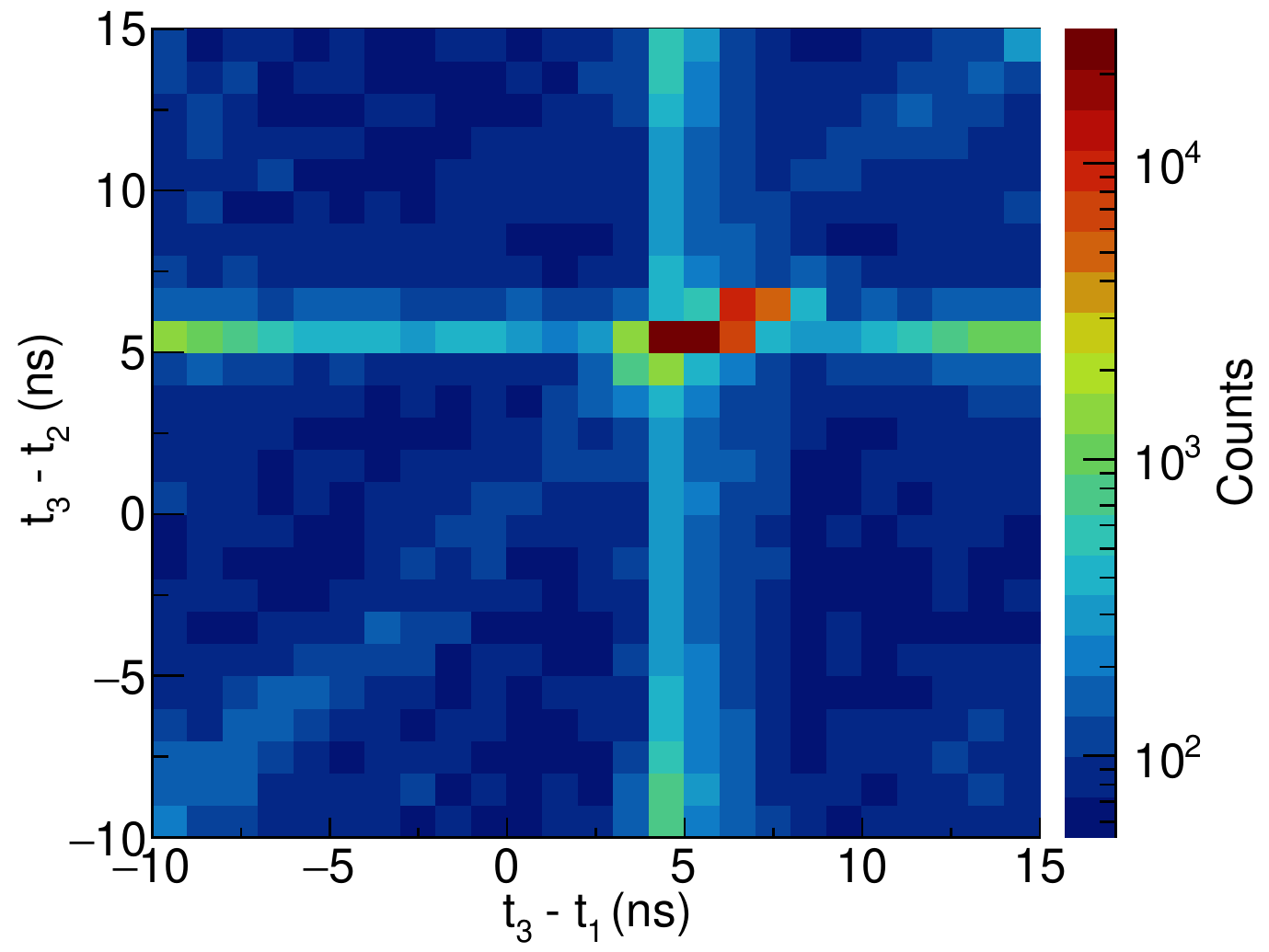}
\caption{\label{fig:RPT-dt} Time difference $t_3-t_1$ and $t_3-t_2$ between 
signals from detector $E$ and detectors $\Delta E_1$ and $\Delta E_2$, 
respectively. These events were recorded during the run with the 2-mm 
polyethylene radiator. }
\end{figure}

The coincidence filter worked in the following manner: for every neutron pulse,
the events of the three scintillators were listed and all possible 
combinations were analyzed. The time difference between the transmission 
detectors and the stop detector were calculated, and among all triplets only 
that with the smallest time difference was selected. The result of this procedure
is shown in figure~\ref{fig:RPT-dt}, where $t_1$, $t_2$, and $t_3$ are used to 
indicate the timestamp of detector $\Delta E_1$, $\Delta E_2$, and $E$, 
respectively. The \lq real\rq\ coincidences form a peak in the 
$t_3-t_1$-vs.-$t_3-t_2$ two-dimensional histogram in correspondence of the time
matching the cable delay between detectors. The peak is surrounded by a 
background of random coincidences which is orders of magnitude lower.
The events in the peak were therefore selected for further analysis, while the 
background counts were subtracted.

The time of flight of the neutrons incident on the PE radiators was
calculated using as reference the timestamp $t_3$ of the $E$ detector.
The event arrival time at the stop detector includes the time of flight of the 
neutron from the source to the radiator, and the time of flight of the recoil 
proton from the radiator to the detector. The proton time of flight 
$t_{\rm p}$ can be determined using the MCNPX model of the RPT, and thus the 
neutron time of flight $t_{\rm n}$ was calculated as:
\begin{equation}
\label{eq:RPT-tof}
    t_{\rm n} = t_3 - t_{\rm 3,\gamma} - t_{\rm p} + (L_{\rm PE} + L_3)/c.
\end{equation}
Here $t_{\rm 3,\gamma}$ is the time of arrival of the gamma flash in the stop
detector,
$L_{\rm PE} = \rm 183.55(5)~m$ is the distance from the neutron source to the 
polyethylene sample, obtained by adding to $L_{\rm FC}$ the distance between 
PPFC and radiator, and 
$L_3$ is the distance from the center of the PE sample to the $E$ detector.
The incident neutron energy was then calculated using 
equation~\ref{eq:NeutronEnergy}, with $v_{\rm n} = L_{\rm PE,eff} / t_{\rm n}$ and
\begin{equation}
\label{eq:RPT-EffectiveFP}
    L_{\rm PE,eff} = L_{PE} + \lambda(E_{\rm n}) - \lambda({\rm 250~eV}).
\end{equation}

\begin{figure}[htb]
\centering
\includegraphics[width=7cm]{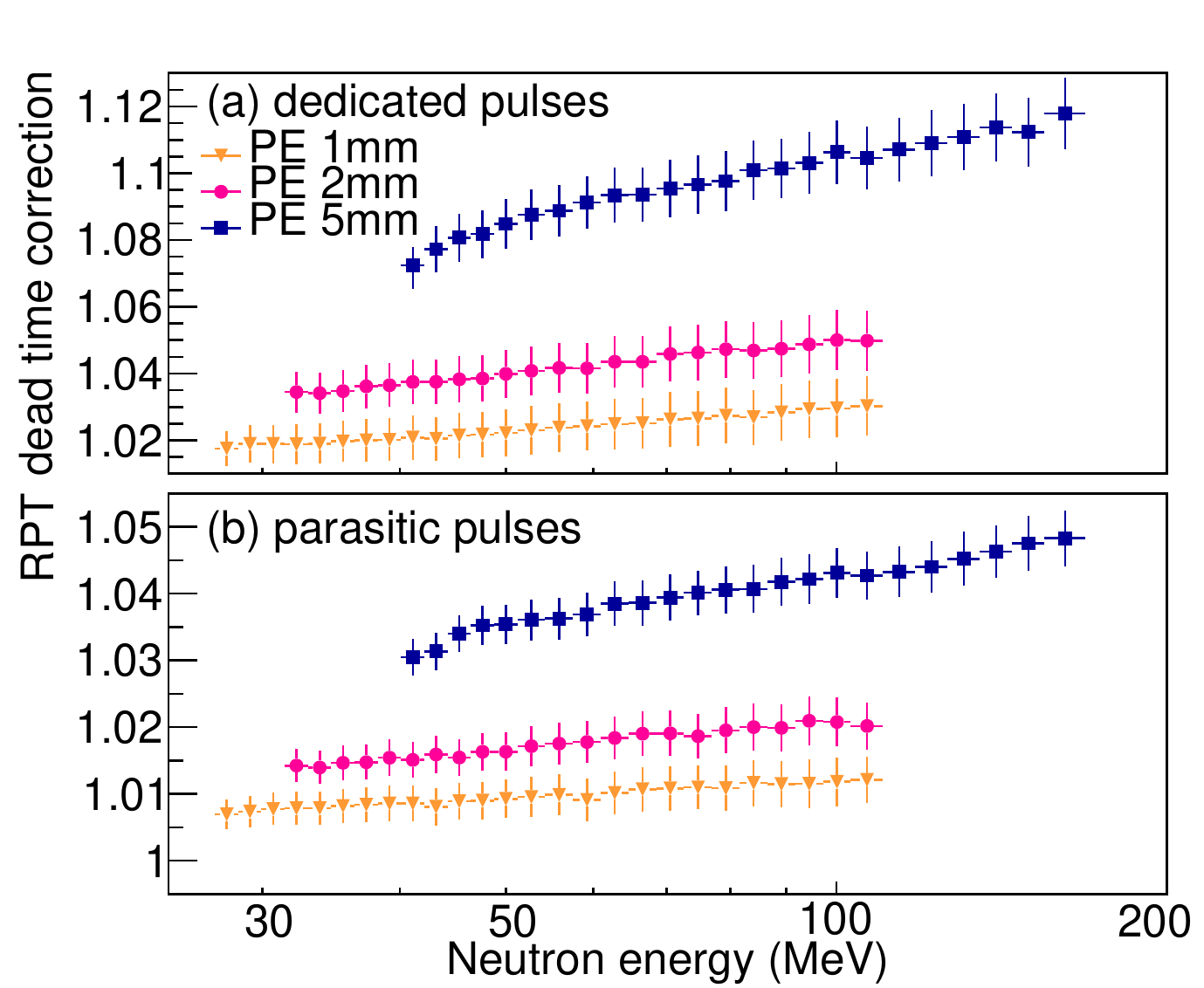}
\caption{\label{fig:RPT-deadtime} Dead time correction applied to the triple 
coincidences detected by the recoil proton telescopes coupled to the 
polyethylene radiators of 1, 2 and 5~mm thickness. The analysis was performed 
separately for events acquired during dedicated (a) and parasitic PS pulses 
(b).}
\end{figure}

Similarly as for the fission chamber, the dead time correction for the RPT was
calculated following the method of Whitten \cite{WHI91}. The formula reported 
in the paper however was modified to take into account the fact that the final 
count rate does not depend on a single detector but it is given by the 
coincidence of three independent detectors. 
The following expression was derived:
\begin{equation}
\label{eq:RPT-deadtime}
    k_{\rm \tau,RPT}(i) = -\frac{N_{\rm c}}{N_{\rm t,0}(i)} \ln 
    \left\{ 1 - \frac{ N_{\rm t,0}(i)/N_{\rm c} }{ 
    \prod_{d=1}^3 \left[ 1-\sum_{k=i-\tau_{\rm d}}^{i-1} N_{\rm d,0}(k)/N_{\rm c} \right] }
    \right\}
\end{equation}
where $k_{\rm \tau,RPT}$ is the correction applied to a given RPT configuration,
$i$ indicates the index of the time-of-flight bin,
$\tau_{\rm d}$ is the dead time of the detector $d$ ($d=1,2,3$) expressed in  
time-of-flight bins,
$N_{\rm d,0}$ is the number of single events per bin recorded by each 
detector,
$N_{\rm t,0}$ is the number of the detected triple coincidences,
$N_{\rm c}$ is the number of the accelerator repetition cycles completed during 
the measurement run.
The dead time $\tau_{\rm d}$ was determined by analyzing the time interval 
distributions obtained for the single scintillators before applying the 
coincidence filter. It was found to range from 5(2)~ns for the
thinnest $\Delta E$ detector, to 19(2)~ns for the thickest $E$ detector.
The correction $k_{\rm \tau,RPT}$ obtained then with 
equation~\ref{eq:RPT-deadtime} is shown in figure~\ref{fig:RPT-deadtime}. 
Also in this case, the analysis of events acquired during dedicated and 
parasitic pulses was performed separately to comply with the underlying assumption
that all neutron pulses have the same intensity.
The absolute uncertainty on $k_{\rm \tau,RPT}$ was obtained by varying the values of 
$\tau_{\rm d}$ within their uncertainty, and it was found to be 1\,\% at worst.

\begin{figure}[htb]
\centering
\includegraphics[width=7cm]{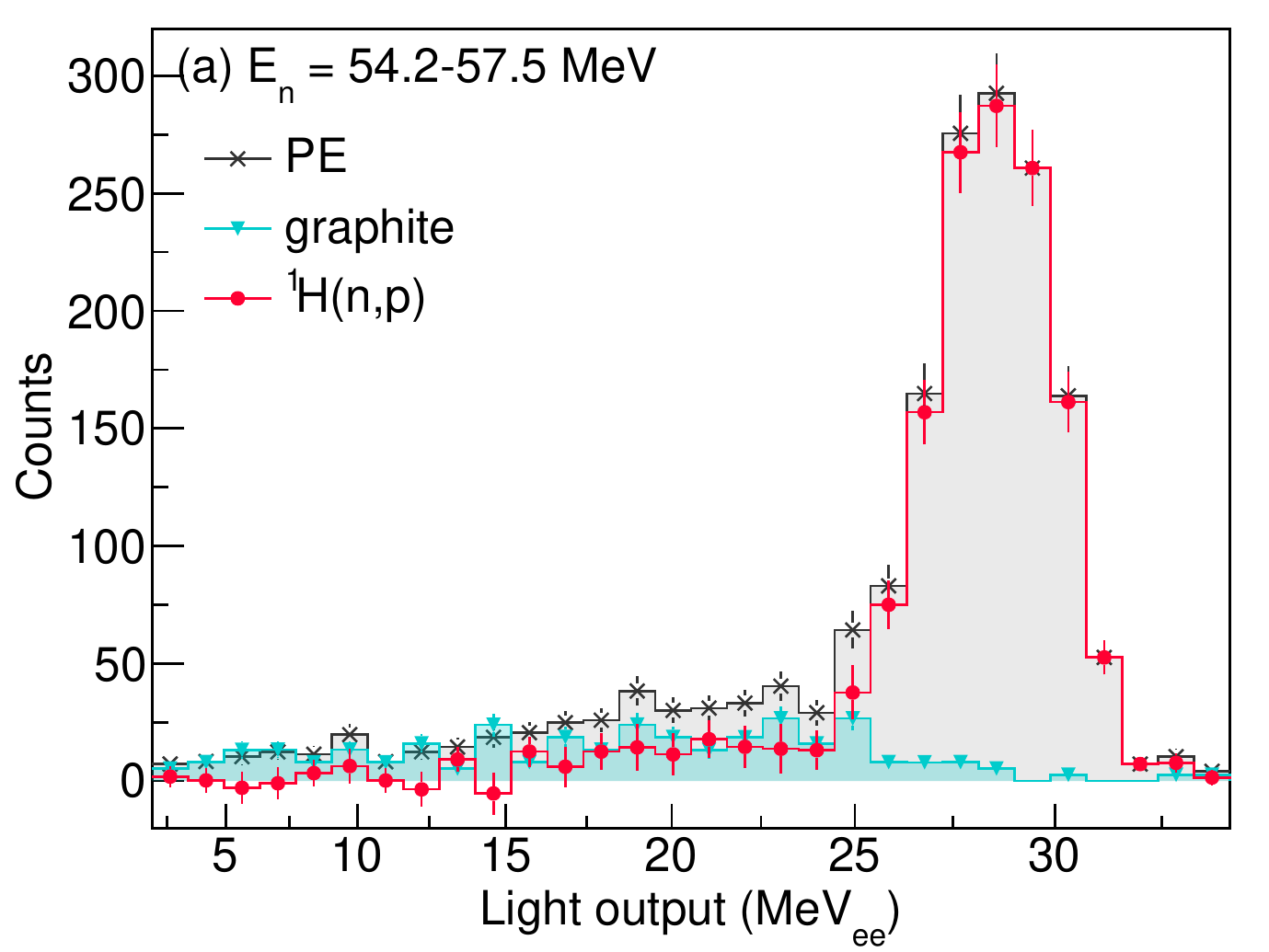}\qquad
\includegraphics[width=7cm]{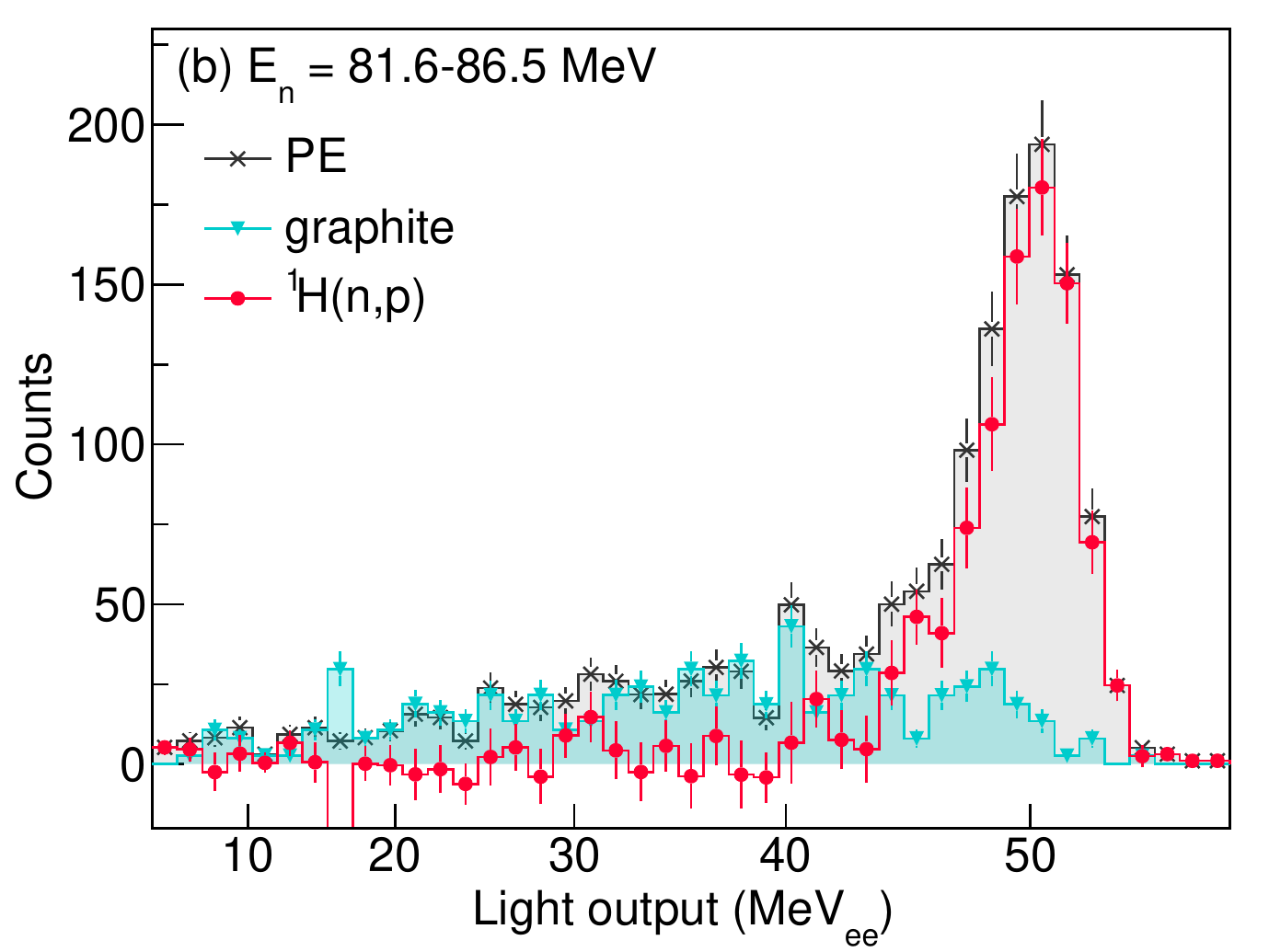}
\caption{\label{fig:RPT-Carbon-subtraction} Light-output distributions produced 
by protons in the stop detector. The events were collected during 
the runs with the PE sample of 2~mm and the C sample of 1~mm. The histogram
labeled \lq \textsuperscript{1}H(n,p)\rq\ was obtained after the subtraction of
the graphite data from the polyethylene data. Figure (a) corresponds to the 
events produced by neutrons of energy from 54.2~MeV to 57.5~MeV, figure (b) 
from 81.6~MeV to 86.5~MeV. }
\end{figure}

Protons produced in \textsuperscript{1}H(n,p) and \textsuperscript{12}C(n,p) 
reactions were separated from deuterons from \textsuperscript{12}C(n,dx) 
reactions by analyzing the light output distributions produced in the stop 
detector $E$ and in the second transmission detector $\Delta{}E_2$, using the 
$\Delta{}E-E$ technique.  
The resulting two-dimensional distributions were already shown for 
the PE sample of 10~mm in figure~\ref{fig:RPT-E-DE-10mm} in 
section~\ref{subsec:RPT-Design}.
The contribution of \textsuperscript{12}C(n,p) was determined by scaling 
the measurements with the graphite sample to PE. The scaling factor was 
calculated as the ratio of the carbon areal density in both samples, 
times the ratio of number of protons on target during the two runs.
In figure~\ref{fig:RPT-Carbon-subtraction}, the light output distribution in 
the stop detector is shown before and after the subtraction for different
neutron energies.

Finally, the number of neutron incident on the PE radiators was determined by
fitting the results of the simulations, the light output distributions produced
in the stop detector, to the experimental data, i.e. the 
\textsuperscript{1}H(n,p) light output distributions shown in 
figure~\ref{fig:RPT-Carbon-subtraction}. The numerical procedure mimicked the
experimental one: for each neutron energy interval, two simulations, one for the
PE sample and one for graphite, were run; then the carbon data were subtracted 
from the polyethylene data, using the ratio of the carbon number densities as 
scaling factor.

\begin{figure}[htb]
\centering 
\includegraphics[width=7cm]{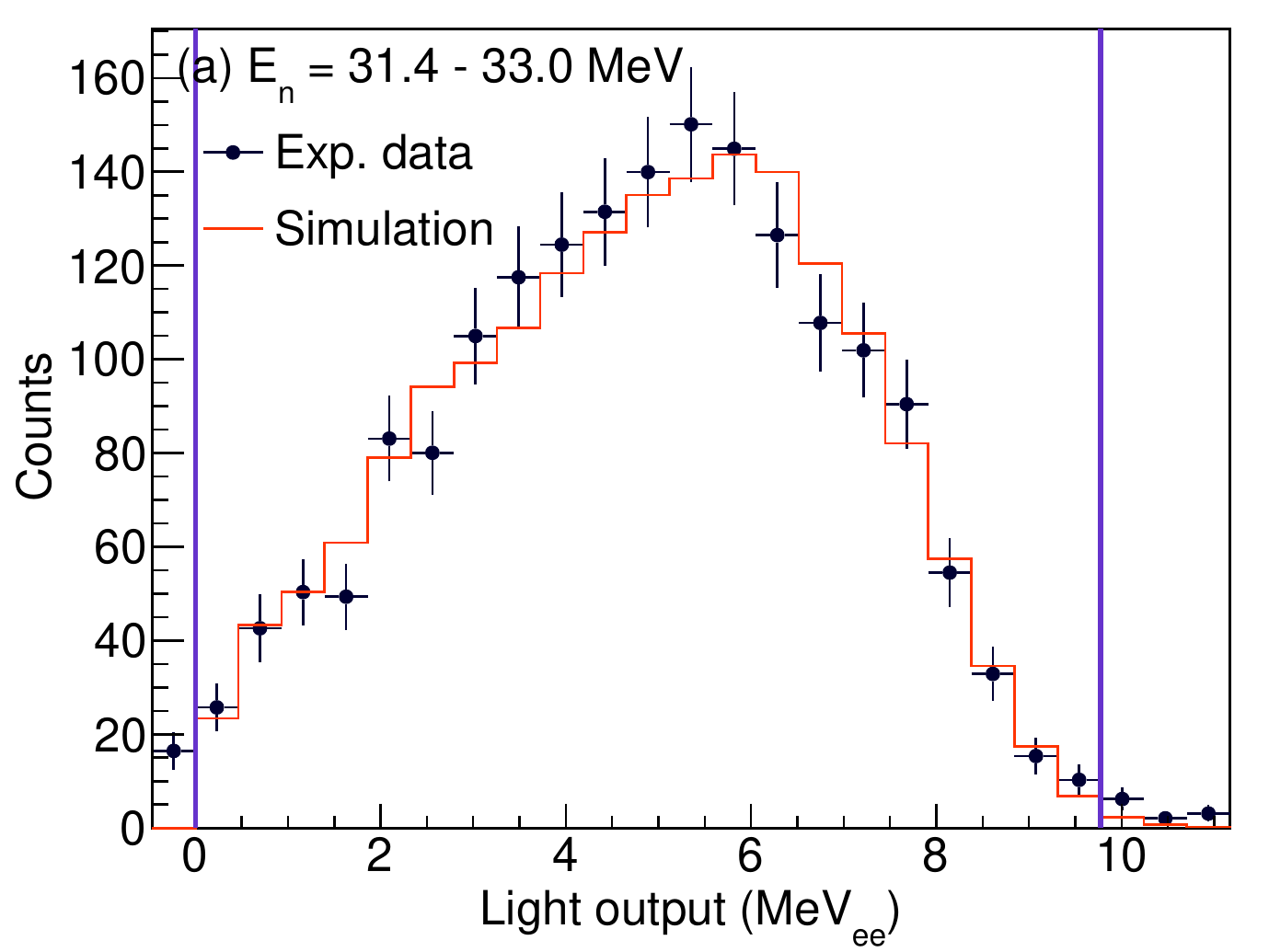}\qquad
\includegraphics[width=7cm]{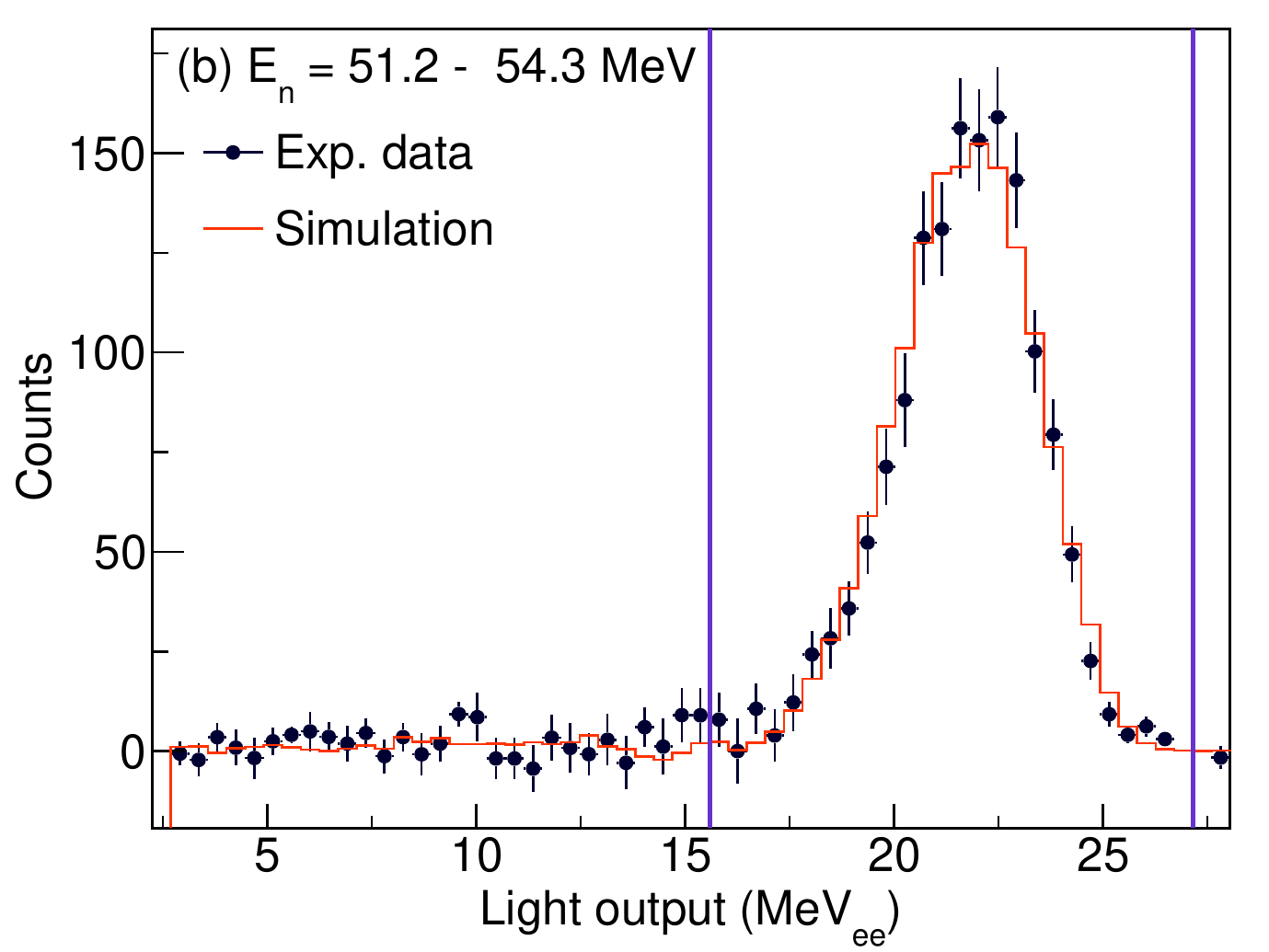}\qquad
\includegraphics[width=7cm]{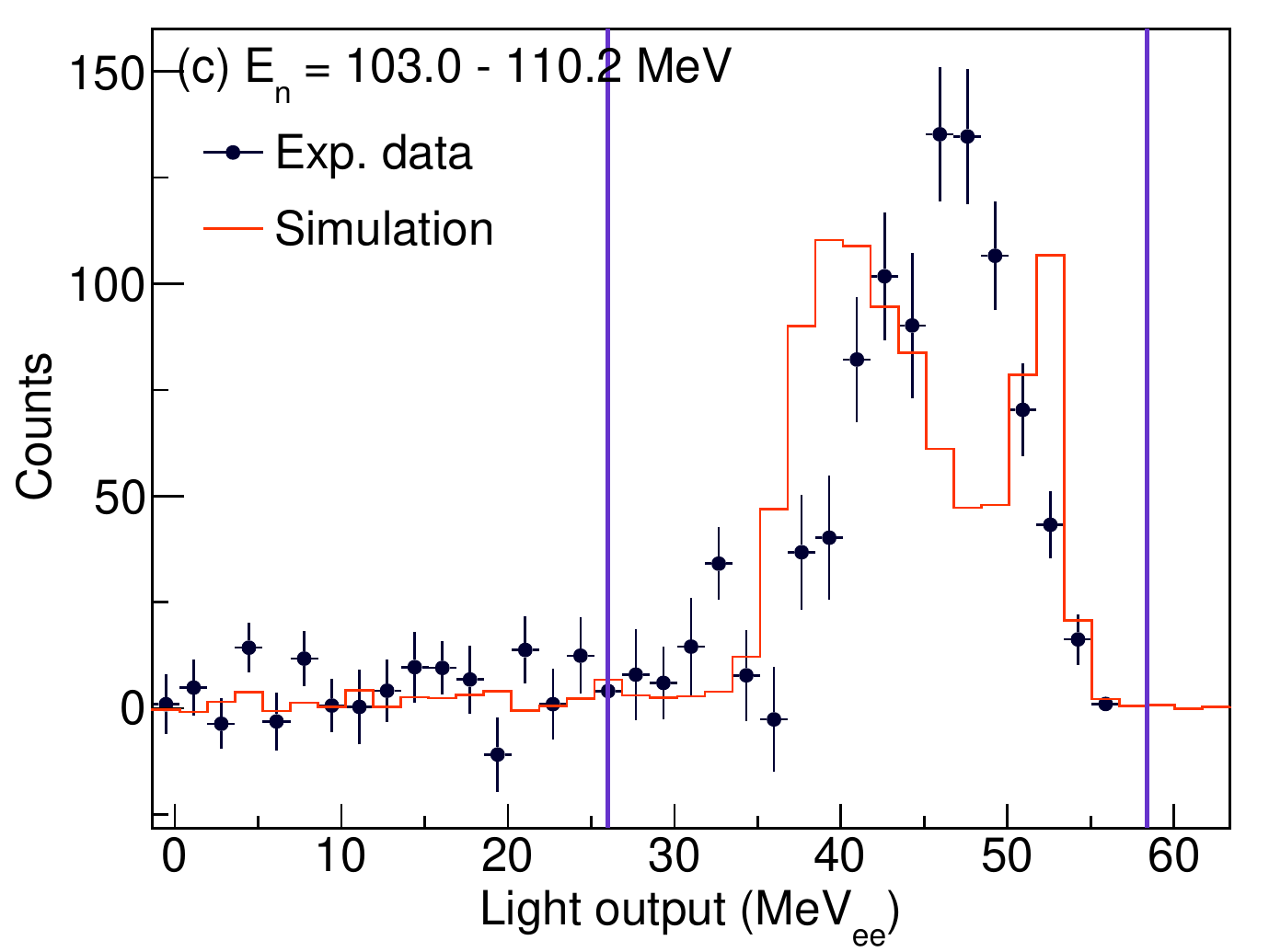}
\caption{\label{fig:RPT-fit} Results of the fit of the simulated proton light 
output distributions in the stop detector to the experimental data collected 
with the 2 mm PE sample and 1 mm graphite sample 
(configuration 2 from table \ref{tab:RPT-Radiators})
for three incident neutron energy intervals. 
The light output is expressed in units of electron-equivalent energy.
The black points are the experimental data, the red line the MCNP simulation
after the subtraction of the carbon contribution, the vertical bars show 
the limits of the fit interval.
The protons produced by neutrons with energy above 103~MeV (figure c) have 
enough kinetic energy to punch through the stop detector; when this happens, the
MCNP model fails to properly represent the light output distribution. }
\end{figure}

In figure~\ref{fig:RPT-fit} the results of the fit are shown for three incident 
neutron energy intervals. 
The fit is successful only for protons with sufficient energy to produce
a clear signal in the stop detector, but not enough energy to 
\lq punch through it\rq, i.e. when the protons fully deposit their energy in
the active volume of the detector. Such cases are shown in 
figures~\ref{fig:RPT-fit}(a) and (b), where the centroid and the width of the 
proton peak can be defined unambiguously. In contrast, 
figure~\ref{fig:RPT-fit}(c) shows an example where the proton is not 
completely stopped by the plastic scintillator but manages to escape: the 
issue is not that the protons do not produce a clear signal, but it is rather 
MCNPX that fails to describe the measurements, so the fit fails.
These limitations on the proton energy and the fit are what ultimately defined 
the energy range where it was possible to reconstruct the incident neutron 
fluence.

The reason why MCNPX fails to represent light-output distribution when punch-through
occurs might lie in the fact that the model does not include the light output 
production processes and the optical transport. 
This might explain why the two peaks in the modeled distribution,
obtained by considering just nuclear interactions, stopping power data, and an 
empirical parametrization of the light-output function, are merged in
a \lq single-peak\rq\ in the experimental data. However, as 
developing an ad-hoc simulation code for the proper treatment of optical 
processes would have been an effort beyond the main objective, this issue has 
not been tackled yet.

To determine the uncertainty of the fit procedure, the fit was repeated 
changing the interval limits, first by narrowing them down around the proton 
peak, then enlarging them to include the low energy tail.
It was found that by changing the fit interval the reconstructed fluence would
change by 2.5\,\% in the worst case.

%-------------------------------------------------------------------------------

\section{Conclusions and outlook} 
\label{sec:Conclusions}

One of the main strengths of the experimental campaign for the measurement 
of the \textsuperscript{235}U(n,f) cross section was the availability of 
different detector setups measuring simultaneously. 
The comparison of the PPFC and the PPAC, and of the three RPTs, is what
ultimately was also used to qualify the performance of the detectors.
Its analysis and the reconstruction of the neutron fluence, however, are 
tightly connected to the cross section results and its final uncertainty, and 
therefore they will be discussed in dedicated publications
(in preparation). 

The uncertainties that could be estimated already during the technical 
development phase are summarized in tables~\ref{tab:FC-Budget} for the PPFC 
and \ref{tab:RPT-Budget} for RPT. These values are an indication of the expected
systematic uncertainty of the \lq low-energy arm\rq\ of the $\rm ^{235}U(n,f)$ 
cross section measurement.

\begin{table}[htb]
\centering
\caption{\label{tab:FC-Budget} Systematic uncertainties affecting the 
fission-rate measurements with the PPFC. They were calculated both for each 
uranium target separately (\lq{}single deposit\rq{}) and for the average.}
\smallskip
\begin{tabular}{|l|c|c|}
\hline
Contribution & Uncertainty (average) & Single deposit \\
\hline
$^{235}$U mass fraction                          &  0.0014\,\%   & 0.0014\,\%    \\
$^{235}$U mass per unit area                     &  0.2\,\%      & 0.6\,\%       \\
$^{235}$U effective density correction $k_{\rm U}$  &  0.6\,\%   & 1-2.5\,\%     \\
Zero-bias efficiency                             &  1.3\,\%      & 1.1-1.3\,\%   \\
Efficiency, extrapolation below thr.             &  3\,\%        & 2-4.5\,\%     \\
Dead-time correction $k_{\rm \tau}$              &  0.2\,\%      & 0.04-0.2\,\%  \\
\hline
\end{tabular}
\end{table}

\begin{table}[htb]
\centering
\caption{\label{tab:RPT-Budget} Systematic uncertainties affecting the 
neutron fluence measurement with the RPT. The values correspond to the 
uncertainty on the detection efficiency.}
\smallskip
\begin{tabular}{|l|c|}
\hline
Contribution & Uncertainty \\
\hline
Beam transmission through PPFC, PPAC                                     &  0.5\,\%        \\
Isotopic composition of PE                                               &  1.5\,\%        \\
Areal density of PE sample                                               &  0.2-0.6\,\%    \\
Areal density of C sample                                                &  0.2-0.9\,\%    \\
Cuts the $\Delta E$-$E$ matrix for selecting proton events               &  0.5\,\%        \\
Fit of MCNPX simulations to the experimental light-output distributions  &  $\leq$2.5\,\%  \\
Effective area of the $\Delta E_2$ detector                              &  0.5\,\%        \\
Distance of the detectors from the PE or C sample                        &  0.8\,\%        \\
Angle relative to the neutron beam                                       &  0.1-0.6\,\%    \\
Dead-time correction                                                     &  0.5-1.0\,\%    \\
\hline
\end{tabular}
\end{table}

Overall, the detector development can be considered successful since
the observables of interest, the fission fragment yield in the case of the PPFC 
and the recoil proton yield for the RPT, could be measured up to a neutron energy of 150~MeV. 
The 200-MeV energy mark was not reached due to run time 
constraints that did not allow to use every available RPT configuration, 
however it was theoretically within the technical capabilities. 

The main difficulty in the case of the fission chamber was the electromagnetic 
interference, which was nevertheless handled in the analysis.
If further measurements will be necessary in future, it would be anyway 
advantageous to study a construction with better grounding and shielding to 
reduce the noise interference. Trying to extend the PPFC measurement range to 
higher energies, at the expense of its simplicity, is unnecessary as there 
already are detectors that are better suited for it (e.g. the PPACs). 
For the particle telescope, however, the 
extension to higher energies might be desirable e.g. for measurements of 
interest for hadrontherapy. This could be achieved for example by improving the 
numerical models to possibly extend the analysis to particles not completely 
stopped by the last detector. Other lines of development which are currently 
under study \cite{infomercial}, include for example the use of gated PMT and 
preamplifiers to avoid saturation by signals induced by the gamma flash, and the 
use of inorganic scintillators as stop detectors to stop more efficiently 
energetic charged particles.

%-------------------------------------------------------------------------------

\acknowledgments

The authors would like to thank the colleagues of PTB for the technical support,
in particular A.~Eckert, F.~Langner, S.~L{\"o}b and M.~Thiemig of the Department 
\lq Neutron Radiation\rq , and M.~Ehlers of the Department \lq Radioactivity\rq. 
They are grateful to M.~Reginatto for the assistance with the data analysis
and the insights on Bayesian methods. 
They wish to thank the n\_TOF local team for the assistance provided
during and after the experimental campaign, especially 
O.~Aberle, M.~Bacak, and D.~Macina.
The very useful discussions with G.~Sibbens of JRC Geel on the composition
of the \textsuperscript{235}U samples were sincerely appreciated.
The support provided by the PTB Working Groups \lq Solid State Density\rq\ and 
\lq Scientific Instrumentation\rq, ZEA-3 unit at the Forschungszentrum 
J{\"u}lich, and the Institute for Inorganic and 
Analytical Chemistry at the TU Braunschweig, 
for the characterization of the polyethylene radiators, is
also gratefully acknowledged.
This project has received funding from the Euratom research and training 
programme 2014-2018 under grant agreement No 847594.

\bibliographystyle{JHEP}
\bibliography{bibliography}

\end{document}